\documentclass[11pt,a4paper]{emulateapj}

\usepackage{color}
\usepackage{rotating}
\usepackage{ulem}
\usepackage{graphicx}
  \usepackage{epsfig}

\usepackage{amsmath}
\usepackage{subfigure}
\DeclareGraphicsExtensions{ps,eps,ps.gz,frh.eps,ml.eps}
\usepackage{natbib}
\newcommand{\sako}{S11}
\newcommand{\kesslera}{K10a}
\newcommand{\kesslerb}{K10b}

\newcommand{\prepSako}{Sako et al. (in prep.)}

\newcommand{\prepBetoule}{Betoule et al. (in prep.)}

\newcommand{\notargeted}{3761}  
\newcommand{\notargetSN}{2781}
\newcommand{\notargetbias}{980}
 
\newcommand{\nogood}{3323}
\newcommand{\nogoodSDSS}{3500} 
\newcommand{\noSDSS}{177} 
\newcommand{\nogoodSN}{2382}
\newcommand{\nogoodbias}{941}
\newcommand{\nofinal}{752}
\newcommand{\nofinalSpec}{208}
\newcommand{\nofinalnoSpec}{544}
\newcommand{\nofinalphotoSpec}{115}

\newcommand{\finalcontam}{3.9$\%$}
\newcommand{\finaleff}{70.8$\%$}

\begin{document}

\title{Cosmology with Photometrically-Classified Type Ia Supernovae from the SDSS-II Supernova Survey}

\author{Heather~Campbell\altaffilmark{1},
	   Chris~B~D'Andrea\altaffilmark{1},
        Robert~C.~Nichol\altaffilmark{1},
           Masao~Sako\altaffilmark{2},
           Mathew~Smith\altaffilmark{3,1},
        Hubert~Lampeitl\altaffilmark{1},
        Matthew~D.~Olmstead\altaffilmark{8},
	Bruce~Bassett\altaffilmark{4,5,6},
	Rahul~Biswas\altaffilmark{7},
	Peter~Brown\altaffilmark{8,9},
	David~Cinabro\altaffilmark{10},
	Kyle~S.~Dawson\altaffilmark{8},
	Ben~Dilday\altaffilmark{11,12},
	Ryan~J.~Foley\altaffilmark{14,15},
Joshua~A.~Frieman\altaffilmark{16,17,18},
Peter~Garnavich\altaffilmark{19},
Renee~Hlozek\altaffilmark{20},
Saurabh~W.~Jha\altaffilmark{13},
Steve~Kuhlmann\altaffilmark{7},
Martin~Kunz\altaffilmark{6,21}, 
John~Marriner\altaffilmark{18},
Ramon~Miquel \altaffilmark{22,23},
Michael~Richmond\altaffilmark{24},
Adam~Riess\altaffilmark{25,26},
Donald~P.~Schneider\altaffilmark{27,28},  
Jesper~Sollerman\altaffilmark{29},
Matt~Taylor\altaffilmark{10},
Gong-Bo~Zhao \altaffilmark{1,30}
}

\affil{$^{1}$Institute of Cosmology and Gravitation, University of Portsmouth, Portsmouth, PO1 3FX, UK}
\affil{$^{2}$ Department of Physics and Astronomy, University of Pennsylvania, 209 South 33rd Street, Philadelphia, PA 19104, USA}
\affil{$^{3}$Department of Physics, University of Western Cape, Bellville 7530, Cape Town, South Africa}
\affil{$^{4}$ Mathematics Department, University of Cape Town, Rondebosch, Cape Town, South Africa}
\affil{$^{5}$ South African Astronomical Observatory, Observatory, Cape Town, South Africa }
\affil{$^{6}$ African Institute for Mathematical Sciences, Muizenberg, 7945, Cape Town, South Africa} 
\affil{$^{7}$ Argonne National Laboratory, 9700 South Cass Avenue, Lemont, IL 60439, USA}
\affil{$^{8}$ Department of Physics \& Astronomy, University of Utah, 115 South 1400 East 201, Salt Lake City, UT 84112, USA}
\affil{$^{9}$George P. and Cynthia Woods Mitchell Institute for Fundamental Physics \& Astronomy, Texas A. \& M. University, Department of Physics and Astronomy, 4242 TAMU, College Station, TX 77843, USA}
\affil{$^{10}$ Wayne State University, Department of Physics and Astronomy, Detroit, MI 48126, USA}
\affil{$^{11}$ Las Cumbres Observatory Global Telescope Network, 6740 Cortona Dr., Suite 102, Goleta, California 93117, USA}
\affil{$^{12}$ Department of Physics, University of California, Santa Barbara, Broida Hall, Santa Barbara, California 93106-9530, USA}
\affil{$^{13}$ Dept. of Physics and Astronomy, Rutgers University, 136 Frelinghuysen Road, Piscataway NJ 08854, USA}
\affil{$^{14}$Harvard-Smithsonian Center for Astrophysics, 60 Garden Street, Cambridge, MA 02138 USA}
\affil{$^{15}$ Clay Fellow.}
\affil{$^{16}$ Department of Astronomy and Astrophysics, The University of Chicago, 5640 South Ellis Avenue, Chicago, IL 60637, USA }
\affil{$^{17}$ Kavli Institute for Cosmological Physics, The University of Chicago, 5640 South Ellis Avenue Chicago, IL 60637, USA }
\affil{$^{18}$ Center for Particle Astrophysics, Fermi National Accelerator Laboratory, P.O. Box 500, Batavia, IL 60510, USA}
\affil{$^{19}$ Department of Physics, University of Notre Dame, Notre Dame, IN 46556, USA}
\affil{$^{20}$Department of Astrophysical Sciences, Priceton University, Princeton, New Jersey 08544, USA } 
\affil{$^{21}$ D\'epartement de Physique Th\'eorique and Center for Astroparticle Physics, Universit\'e de Gen\`eve, Quai E.\ Ansermet 24, CH-1211 Gen\`eve 4, Switzerland}
\affil{$^{22}$ Institut de F\'{\i}sica d'Altes Energies, Universitat Aut\`onoma de Barcelona, E-08193 Bellaterra (Barcelona), Spain.} 
\affil{$^{23}$ Instituci\'o Catalana de Recerca i Estudis Avan\c{c}ats, E-08010 Barcelona, Spain} 
\affil{$^{24}$ Physics Department, Rochester Institute of Technology, Rochester, NY 14623, USA}
\affil{$^{25}$ Space Telescope Science Institute, 3700 San Martin Drive, Baltimore, MD 21218, USA}
\affil{$^{26}$ Department of Physics and Astronomy, Johns Hopkins University, 3400 North Charles Street, Baltimore, MD 21218, USA}
\affil{$^{27}$ Department of Astronomy and Astrophysics, The Pennsylvania State University, University Park, PA
16802, USA} 
\affil{$^{28}$ Institute for Gravitation and the Cosmos, The Pennsylvania State University,  University Park, PA 16802, USA} 
\affil{$^{29}$ The Oskar Klein Centre, Department of Astronomy, AlbaNova, Stockholm University, 106 91 Stockholm, Sweden }
\affil{$^{30}$ National Astronomy Observatories, Chinese Academy of Science, Beijing, 100012, P.R.China}

\email{Heather.Campbell@port.ac.uk}

\begin{abstract}

We present the cosmological analysis of \nofinal\ photometrically--classified Type Ia Supernovae (SNe~Ia) obtained from the full Sloan Digital Sky Survey II (SDSS-II) Supernova (SN) Survey, supplemented with host--galaxy spectroscopy from the SDSS-III Baryon Oscillation Spectroscopic Survey (BOSS). Our photometric--classification method is based on the SN typing technique of \citet{Sako:2011}, aided by host galaxy redshifts ($0.05<z<0.55$).  SNANA simulations of our methodology estimate that we have a SN~Ia typing efficiency of 70.8\%, with only 3.9\% contamination from core-collapse (non-Ia) SNe. We demonstrate that this level of contamination has no effect on our cosmological constraints. We quantify and correct for our selection effects (e.g., Malmquist bias) using simulations. When fitting to a flat $\Lambda$CDM cosmological model, we find that our photometric sample alone gives $\Omega_m= 0.24^{+0.07}_{-0.05}$ (statistical errors only). If we relax the constraint on flatness, then our sample provides competitive joint statistical constraints on $\Omega_m$ and $\Omega_{\Lambda}$, comparable to those derived from the spectroscopically-confirmed  three-year Supernova Legacy Survey (SNLS3). Using only our data, the statistics--only result favors an accelerating universe at 99.96\% confidence. Assuming a constant $w$CDM cosmological model, and combining with $H_0$, CMB and LRG data, we obtain $w=-0.96^{+0.10}_{-0.10}$, $\Omega_m = 0.29^{+0.02}_{-0.02}$ and $\Omega_k = 0.00^{+0.03}_{-0.02}$ (statistical errors only), which is competitive with similar spectroscopically confirmed SNe~Ia analyses. Overall this comparison is reÐassuring, considering the lower redshift leverage of the SDSS-II SN sample ($z < 0.55$) and the lack of spectroscopic confirmation used herein. These results demonstrate the potential of photometrically--classified SNe~Ia samples in improving cosmological constraints.

\end{abstract}

\section{Introduction}
\label{sec:intro}
Supernovae (SNe) have historically been classified based on their optical spectroscopic properties (e.g., \citealp{Filippenko:1997}).  Type Ia supernovae (SNe~Ia) are distinguished from other classes of SNe by their lack of Hydrogen and Helium spectral features, and the presence of other spectral features such as Si II absorption at rest--frame wavelength 6150~\AA. This particular optical classification is unique, as it efficiently separates two distinct SN physical processes. The progenitors of SNe~Ia are White Dwarfs (WDs), unlike other common categories of SNe, which result from the core collapse of stars with initial mass M $\gtrsim8$ M$_{\odot}$. As the progenitors for core--collapse (non-Ia) SNe, such as Type II SNe, are more luminous, the progenitor star has been identified from archival images on multiple occasions \citep{Smartt:2009}, while there have not yet been any direct observations of the WD progenitor of SNe~Ia (though deep limits do exist on SN~2011fe; \citealp{li:2011}). 

However, the lack of H and He in the spectrum, the composition of the ejecta, and the energy released in the explosion all strongly indicate that SNe~Ia are the visible manifestation of a thermonuclear runaway explosion in a carbon-oxygen WD as its mass approaches the Chandrasekhar limit \citep{hoyle:1960,Nomoto:1982,Iben:1984}. The exact nature of the binary progenitor system (a single degenerate object accreting mass from a companion, or the merger of two WDs) has long been an open question.  Recent observations have shown that both channels can lead to SNe~Ia \citep{nugent:2011a,bloom:2011a,dilday:2012a}, making the relative prevalence of each channel of primary concern amongst progenitor studies.

For over two decades SNe~Ia have been of interest to cosmologists as distance indicators, as they display less dispersion at their peak magnitude than other classes of supernovae and have high optical luminosities ($>10^9$~L$_{\odot}$). However, SNe~Ia are not simple ``standard candles", and the usefulness of SNe~Ia are greatly enhanced by our ability to standardize their magnitudes. The discovery that photometric properties of SNe~Ia, such as the light-curve width \citep{phillips:1993a} and color \citep{Riess:1996,tripp:1997a}, are correlated with the absolute magnitude at peak allowed for accurate distance measurements using SNe~Ia, reducing the dispersion in the measured distance modulus ($\mu$) to $\sim$0.14 mag. This work led directly to the discovery that high-redshift SNe~Ia appear fainter than expected unless the expansion of the universe is accelerating \citep{riess:1998a, perlmutter:1999a}. 

Over the last decade SN surveys have evolved, spanning greater fractions of the sky and discovering SNe at higher redshifts, with better photometric calibration and an improved understanding of systematic uncertainties.  Due to the differing observational requirements necessary for monitoring SNe~Ia over a wide range of redshifts (survey depth, area, wavelength coverage, etc.), the redshift--distance relationship, or ``Hubble diagram", for SNe~Ia is comprised of data from a number of surveys.  The Hubble Space Telescope Program \citep[GOODS SN sample; ][]{riess:2004a,riess:2007a}, Supernovae Cosmology Project \citep[SCP or HST Cluster SN sample;][]{knop:2003a, kowalski:2008a, amanullah:2010a, suzuki:2011a}, the Supernova Legacy Survey \citep[SNLS;][]{Guy:2010}, and Equation of State: SupErNovae trace Cosmic Expansion \citep[ESSENCE;][]{miknaitis:2007a} provide nearly all SNe~Ia measurements at $z>0.4$, with the first two surveys dominating at $z>0.9$. 

The Sloan Digital Sky Survey II Supernova Survey \citep[SDSS-II SN survey;][]{Frieman:2008,kessler:2009a} has populated the Hubble diagram at intermediate redshifts ($0.1<z<0.4$). Panoramic Survey Telescope \& Rapid Response System (PanSTARRS) is currently finding thousands of SNe~Ia in the same redshift range. The largest of the low redshift surveys include the Harvard-Smithsonian Center for Astrophysics SN group \citep[CfA;][]{hicken:2009a}, the Nearby Supernova Factory \citep[SNfactory;][]{aldering:2002a}, the Carnegie Supernova Project \citep[CSP;][]{hamuy:2006a,folatelli:2010a}, the Lick Observatory Supernova Search \citep{ganeshalingam:2010a}, the Palomar Transient Factory \citep{law:2009a}, and the Catalina Real-Time Transient Survey \citep[CRTS;][]{drake:2009a}. Using a combination of data from different SN surveys, spanning the full range of redshifts available, the most recent cosmological analysis using only SNe~Ia finds $\Omega_{\rm m}\,=0.18^{+0.08}_{-0.10}$(stats) $\pm{0.06}$(sys) and $w=\,-0.90^{+0.16}_{-0.20}$(stats) $^{+0.07}_{-0.14}$(sys) \citep{conley:2011a}, under the assumption of a flat Universe and a constant equation--of--state of dark energy ($w$).

A common theme across most previous SN cosmology surveys is that SN spectroscopy is used only to classify the SN event and obtain a redshift (with the exception of the SNfactory and, in some cases, the SNLS \citep[]{bronder:2007a,walker:2010a,ellis:2008a}). The distance to each object classified as a SN~Ia is then determined from the multi-color time series photometry using one or more light-curve fitting models, e.g., SALT2 \citep{guy:2007a}, MLCS2k2 \citep{Riess:1996,Jha:2007}, and SiFTO \citep{Conley:2008}.  Methods for constraining the absolute magnitude of an observed SN based on spectroscopic line ratios have also been developed and applied to nearby SNe \citep{foley:2008a,Bailey:2009,Chotard:2011,Nordin:2011, blondin:2012a,silverman:2012a}, though as of yet they have not been tested over a cosmologically interesting redshift range.

In the future, however, taking spectra of a large number of high-redshift SNe for classification purposes will be challenging, and could potentially limit the size and usefulness of  any new survey. This will be the case for the Large Synoptic Survey Telescope \citep{LSST}, where the complete spectroscopic classification of all its SN candidates will be simply impossible, thus other methods need to be employed to utilize these huge SN programs. This challenge is already being confronted by the latest SN surveys, such as the PanSTARRS, who have devised new and innovative techniques for classifying their SN candidates using only photometric imaging data \citep[thus far tested on a small subset; see][]{scolnic:2009a}. The SN survey from the Dark Energy Survey (DES) will probably not be able to spectroscopically classify all of their expected $\simeq4000$ high-quality, high--redshift SNe~Ia \citep{Bernstein:2011}.

An alternative is to use photometric-only classification techniques. This idea is not new -- \citet{pskovskii:1977a} proposed classifying SNe based on their observed decline rate -- and even early SN~Ia cosmology results included a significant fraction of high-redshift events that lacked spectroscopic identification \citep{riess:1998a,perlmutter:1999a,tonry:2003a,riess:2004a}. One of the primary scientific drivers for developing photometric classifiers has been to aid in the spectroscopic follow-up of SN surveys, thus allowing them to use their spectroscopic resources more efficiently \citep{sullivan:2006a,sako:2008a}. However, making a Hubble diagram solely from photometrically--classified SNe requires a lower false-positive rate (i.e., contamination by non-Ia~SNe) than does spectroscopic target selection from photometrically--classified SNe.  Photometry-only Hubble diagrams were introduced by \citet{barris:2004a}, and have been presented more recently by \citet{rodney:2010a} and \citet{Bazin:2011}.  

Most photometric classification methods fit observed light-curves to templates of different SN types and determine the likelihood of each class.  These methods \citep[e.g.,][]{poznanski:2002a,sullivan:2006a,johnson:2006a,poznanski:2007a,kuznetsova:2007a,kunz:2007a,rodney:2009a,gong:2009a,falck:2010a,Sako:2011} typically remove SNe which resemble non-Ia~SN templates based on their likelihoods, although there is considerable variety in the details. While \citet{hlozek:2011a} (based on \citealp{kunz:2007a} and further developed by \citealp{newling:2011a,knights:2012a}) uses templates to compute likelihoods for the type of each SN, they do not remove any objects from their cosmological fit. Rather, they use the likelihoods of each SN as a weight, retaining all possible information while computing an unbiased cosmology in a Bayesian manner.  Higher level statistical analyses have also been applied to this problem, with a goal of finding a lower dimensional parameter space where there exists a cleaner separation between the different types, thus simplifying the classification problem.  Examples of these approaches include semi--supervised learning techniques such as diffusion maps \citep{richards:2011a}, and kernel Principal Component Analysis (PCA) applied to SN light curves \citep{ishida:2012a}. 

In anticipation of the SN typing requirements to be encountered by DES, \citet[hereafter \kesslera]{kessler:2010a} issued the ``Photometric Supernova Classification Challenge", providing simulated light-curves of different SN types based on a realistic DES-like SN survey and a training sample where the true SN type was given.  The results of this challenge are presented in \citet[hereafter \kesslerb]{kessler:2010b}, and provide some interesting insights into the relative performance of different SN classifiers.  Overall, several different classification strategies produce similarly high scores in terms of both efficiency and contamination, but all proved subject to significant level of contamination ($\sim20$\%). A problem which is common for many methods is that they require a training set of known SN types. If this set is biased and not representative of the whole sample then the classification will be biased as well; an effect that was seen in the \kesslera\ challenge.

In this paper we build upon the photometric--classification algorithm of \citet[hereafter \sako]{Sako:2011}, which obtained the highest overall Figure--of--Merit in \kesslerb.  We use here the full three-year data-set from SDSS-II SN Survey, including a new collection of host-galaxy redshifts obtained by the SDSS-III Baryon Oscillation Spectroscopic Survey \citep[BOSS;][]{Eisenstein:2011,dawson:2012a}. 

We optimize selection cuts and determine the biases of our new method with extensive simulations using the SuperNova ANAlysis \citep[SNANA;][]{kessler:2009b} software package, and apply redshift-dependent corrections to our data.  We show that photometric classification can provide SN~Ia samples with low contamination and well-understood biases, and present cosmological constraints that are competitive with those derived from existing spectroscopic samples.

The paper is organized as follows.  In Section~\ref{sec:Data} we detail the SN and host-galaxy data that we analyse in this paper.  In Section~\ref{sec:algorithms} we discuss our SN classifier, with emphasis on the light-curve fitter and selection criteria. We perform a rigorous analysis of and derive corrections for biases introduced by our selection criteria, shown in Section~\ref{sec:bias}.  In Section~\ref{sec:results} we present our full photometric Hubble diagram and consistency checks.
Cosmological constraints from our photometrically derived sample are in Section~\ref{subsec:cosmo_anal} along with comparisons to other spectroscopic cosmological fits. In Section~\ref{sec:discuss} we discuss our results, how this analysis could be improved, and how our work applies to upcoming large-scale SN surveys. Finally in Section~\ref{sec:conclusion} we detail the main conclusions of this paper.

\section{Data}
\label{sec:Data}

\subsection{The SDSS-II Supernova Survey}
\label{subsec:SDSSSN}
The SDSS-II SN Survey is a dedicated search for intermediate-redshift SNe from repeated scans of the equatorial ``Stripe 82" region (covering $\simeq$300 deg$^2$) of the original SDSS \citep{York:2000,Frieman:2008}. For three months a year (Sep-Nov) over a three year period (2005-2007), the SDSS telescope \citep[]{Gunn:2006,gunn:1998a} performed multi--color $ugriz$ imaging \citep{fukugita:1996a,Gunn:2006,Ivezic:2007,Doi:2010} of this area of sky, with a cadence of a few times per week. The SDSS uses asinh magnitudes \citep{lupton:1999a}, although SDSS is on the AB system after applying small offsets (see Appendix~\ref{subsec:additional_ia_candidates}). The analysis of SDSS astrometry is described in \citet{pier:2003a}.

This multi-epoch data was then used to identify SN~Ia candidates in real-time for further spectroscopic observations \citep{sako:2008a}, resulting in over 500 spectroscopically-confirmed SNe~Ia \citep{Zheng:2008,ostman:2010a,konishi:2011a,foley:2012b}. Well-observed subsamples of these intermediate-redshift SNe have been used in a variety of studies, primarily focused on constraining cosmology \citep{kessler:2009a,sollerman:2009a,lampeitl:2010a}, the measurement of SN rates \citep{Dilday:2010,Smith:2012} and the study of host-galaxy properties and their correlations with SNe~Ia \citep{lampeitl:2010b,Gupta:2011,DAndrea:2011,galbany:2012a}. The full three-year SDSS-II SN Sample will be published in Sako et al. (in prep.).

The real-time spectroscopic sample of SDSS-II SNe is incomplete and potentially biased, as a function of redshift. This bias comes from a number of different, and sometimes competing, effects and is therefore hard to predict {\it a-priori} and thus correct for. First, the decisions made by observers following--up SN candidates was based on the local weather conditions (at a variety of telescopes), the position of the SN candidates on the sky, and the location of the SN candidate in the host galaxy. This is illustrated in Table~2 of \citet{Smith:2012} where the spectroscopic completeness of the SDSS-II SN Survey drops below 40\% at $z>0.4$ \citep[see also][]{kessler:2009a}. 

Second, our targeting of SN candidates gave priority to events in red elliptical host galaxies (see Eqn 7 of \citet{sako:2008a}), as these SNe are likely to be less affected by dust in their host galaxy. This prioritization is seen in the spectroscopically-confirmed SNe~Ia, which have lower reddening values than predicted from simulations without spectroscopic selection \citep[Figure~16,][]{kessler:2009a}. However, the rate of SNe~Ia in red elliptical galaxies is lower than seen in blue, star--forming spiral galaxies \citep{mannucci:2005a,wang:1997a,Smith:2012}, so this additional up--weighting given to the SNe~Ia in red ellipticals might be a subdominant effect.  As will be shown in Figure~\ref{host_galaxy} of this paper, the host galaxies of spectroscopically--confirmed SNe~Ia are, on average, representative of the whole population of host--galaxy colors studied in our BOSS sample (see Section \ref{subsec:host_bias}). However, it is clear that the spectroscopically-confirmed SNe~Ia from the SDSS-II SN Survey are biased in absolute magnitude with respect to the entire SN~Ia population. To avoid any such biases, we do not use spectroscopic SN information anywhere in our photometric classification (as discussed in Section~\ref{sec:algorithms}). Any spectroscopically-confirmed SNe~Ia that fail our photometric criteria are not included in our final sample to preserve selection consistency.

\subsection{BOSS Ancillary Targets}
\label{subsec:BOSS}

We undertook a BOSS ancillary spectroscopic program (see \citet{dawson:2012a} for details of these programs), to obtain host galaxy information for most SDSS-II SN candidates. We aim to obtain a larger, and more complete, SN~Ia sample from the SDSS-II SN Survey through photometric classification.  We also took the opportunity to target the host galaxies of a range of other, possibly interesting, transient events detected as part of the SDSS-II SN Survey.  Such a project was well-suited to the small number of ancillary targets available in each BOSS spectroscopic field and would have been impossible to achieve on a normal instrument (either through queue-scheduling or normal observer-mode), because of the combination of the low surface density of faint targets (between 5 to 25 targets per deg$^2$ to $r\sim22$) spread over a wide area ($\simeq300$ deg$^2$ of ``Stripe 82"). Table 1 of \citet{Frieman:2008} shows that the SDSS-II SN Survey used over 1000 hours (or $\sim100$ nights) of telescope time between 2005 and 2006 to spectroscopically confirm over 350 SNe~Ia, utilizing many of the larger optical telescopes in the world (HET, NTT, NOT, APO, Subaru, WHT, SALT, Keck, NOAO, TNG). It is hard to envisage how we could have used such resources to target thousands of host galaxies as presented herein.

In detail, we obtained spectroscopic observations using the BOSS spectroscopic system \citep{smee:2012a} of \notargeted\  galaxies spread almost evenly across the ``Stripe 82" region. These targets were chosen using two selection algorithms (described below) which were complementary, but different, in their scientific objectives. The first algorithm focused on improving cosmological constraints from SNe~Ia in the SDSS-II SN Survey by obtaining a sample free from the possible spectroscopic--selection biases discussed above. The other algorithm targeted interesting subsamples of transients detected as part of the SDSS-II SN Survey. In this paper we focus exclusively on the first of these objectives, but include in our analysis galaxies, and their associated SNe, observed by BOSS originating from either algorithm.

\subsubsection{Algorithm One: Additional Type Ia candidates}
\label{subsec:additional_ia_candidates}

We targeted galaxies that hosted a SN event of any type, based on object classifications using the ``Photometric SN IDentification" (PSNID) method of \sako, which we describe in detail in Appendix~\ref{Appendix_PSNID}. We applied PSNID to the multicolor ($ugriz$) light-curve data from SDSS-II created with the Scene-Modeling Photometry (SMP) method of \citet{Holtzman:2008}\footnote{Improvements in this photometry compared to that used in \citet{kessler:2009a} include the discovery of a small ($\sim10\%$) correction for an underestimate of uncertainties for low-flux data, and new (December 2011) AB offsets from the SDSS native magnitudes system of $\Delta u=-0.066, \Delta g=0.021, \Delta r=0.005, \Delta i=0.020,$ and $\Delta z=0.013$.  Details of this re-calibration can be found in \prepBetoule\ and \prepSako.}, assuming a flat prior on all SN parameters.  We do not include any spectroscopic information for these objects, and thus place a flat prior on the SN redshift as well. We select transient events that were classified as likely SNe of any type, i.e., probable SN Ia, SN II or SN Ibc.

For all these candidates, we visually inspected the SDSS images (from the DR7 Skyserver; \citealp{Abazajian:2009}) of the three nearest galaxies to each transient and manually assigned the most likely host galaxy for each candidate. In the majority of cases (95\%) the nearest galaxy on the sky (angular separation) to the candidate was classified as the host, but in some cases the nearest galaxy was either clearly a background object or a star misclassified as a galaxy.  In these cases we classified either the second (4\%) or third (1\%) nearest galaxy as the most likely host. Taking into account the observational limitations of the program, we gave priority to host galaxies with a fiber magnitude brighter than $r_{\rm fiber} = 21.25$, based on SDSS-I/II photometry. However, to fully utilise our allocation of BOSS fibers, we also include a small subsample of host galaxies fainter than this limit. Combined, these samples made up our main target list of \notargetSN\  galaxies.
 
In 66 cases, the nearest object to the SN event was classified as a faint star in the SDSS DR7 database, but was clearly a galaxy as seen in the SDSS images. For these objects, we targeted the nearest ``stellar" source in the DR7 database, which could have a fainter fiber magnitude than our main BOSS targets ($r_{\rm fiber} < 21.25$) and were given lower priority for BOSS observations. 

Finally, we cross-referenced the whole target list with the SDSS DR7 database and found 276 of these host galaxies had a spectrum already. For these cases we targeted the location of the SN event rather than placing the BOSS fiber at the centre of the galaxy, as was done in all other cases and in the original SDSS-I/II survey.  The motivation for this fiber placement is that galaxy spectra at the location of the SN could be useful for studies into correlations between SN properties and the environment in which the SN occurs \citep{gallagher:2005a, gallagher:2008a,Gupta:2011,DAndrea:2011,galbany:2012a}. As our focus here is only on obtaining redshifts to aid in classifying transients, these spectra (at the location of the SN event) serve our purposes equally well. In these cases, we ignored the fiber magnitude of the host galaxy (calculated at the center of the galaxy) and observed the SN location position regardless of our $r_{\rm fiber} = 21.25$ limit for the main sample. 
  
\subsubsection{Algorithm Two: Random sample of additional transients}
\label{subsubsec:algotwo}

The goals of our second sample were both to study our overall selection biases (e.g., determine the effect of AGNs) and to further study interesting and unusual variable objects observed by SDSS-II e.g., hydrogen--poor, superluminous supernovae \citep{Quimby:2011,leloudas:2012a}. Achieving the first of these goals requires an unbiased sample of non-SN transient host galaxies, created by choosing at random from the set of non-SN transients in the magnitude range of $19.5<r<21.5$.  We imposed no magnitude limit on the host galaxies of these targets, of which there were \notargetbias; however, we did require that the galaxy was detected in the DR7 galaxy catalogue. 

As with the SN host-galaxy sample in Section~\ref{subsec:additional_ia_candidates}, we visually inspected the three nearest galaxies in DR7 to each transient, selecting as our BOSS target the most likely host to the transient.  This proved to be the nearest galaxy (angular separation) in the vast majority (99.4\%) of cases.  The high percentage of host galaxies being matched to the nearest galaxy, compared to the previous sample in Section 2.2.1, is likely due to these transients being quasars that are located in the cores of galaxies, as opposed to SNe, which can be located throughout the galaxy.  The targets from this sample were given the lowest priority for observation when assigning BOSS fibers.

Another key difference between Algorithms One  and Two was that in Algorithm Two transient events were allowed to show variability over multiple years, or have light curves that failed in the initial photo-typing. Targets selected by Algorithm One were detected in only one season of the SDSS-II SN Survey.

\subsection{Reduction of BOSS Spectra} 
\label{subsec:BOSSred}

Our ancillary targets described above were merged with other ancillary BOSS targets on the ``Stripe 82" region and observed together as part of the normal SDSS-III observing program during 2009 and 2010 \citep[see][for a description of BOSS observations and programs]{Eisenstein:2011,dawson:2012a}.  Some of our targets were lost at this stage because of BOSS fiber collision issues, i.e., two objects within 55 arc-seconds of each other cannot be observed on a single BOSS spectroscopic plate.

By the end of 2010 all BOSS plates on ``Stripe 82", including most of our ancillary targets, had been observed.  For the purpose of this paper we are only interested in the galaxy redshift measurements (and their errors). The details of the spectral reductions of our sample of BOSS SN host galaxy spectra can be found in Olmstead et al. (in prep.), and the redshift and object classification of BOSS spectra in general is described in \citet{bolton:2012a}. In brief, our sample was processed using the standard BOSS spectroscopic analysis software, which is based on the original SDSS-I/II reduction pipelines. This pipeline has at least a 95\% success rate in obtaining redshifts for spectra from the primary galaxy sample \citep[e.g.,][]{anderson:2012a}. Olmstead et al. (in prep.) carried out a detailed comparison between the BOSS pipeline redshifts and those produced, and manually inspected, using the publicly available AAO {\sc runz} software. All the spectral data from our SN host galaxy ancillary program are now public as part of the SDSS DR9 data release \citep{ahn:2012a}.

After removing spectra with low redshift confidence, low signal--to--noise ($S/N$), and large redshift errors, our SDSS-II SN ancillary program on BOSS produced \nogood\ reliable redshifts.  This sample is composed of \nogoodSN\ likely SN host galaxies from the \notargetSN\ targets selected via Algorithm One in Section~\ref{subsec:additional_ia_candidates}, and \nogoodbias\ from the random sample of \notargetbias\ galaxies chosen in Section~\ref{subsubsec:algotwo}.  The spectroscopic target efficiency from Algorithm One is 86\%, which is approximately 10\% lower than for Algorithm Two. This difference is due to two subsamples in Algorithm One that have a lower efficiency, probably due to the lack of an imposed fiber-magnitude limit in these cases. The first of these subsamples targets the nearest photometric object to the SNe when there was some ambiguity about the star/galaxy separation; this had an efficiency of only 27.3\%, but was a very small subsample. The second subsample was created using less stringent cuts on the quality of the light curve (i.e., lower $S/N$) to provide an additional list of probable SN locations; this subsample has a lower average host galaxy fiber magnitude ($\langle r_{fiber}\rangle = 22.17$) than the main sample ($\langle r_{fiber}\rangle = 20.62$), which leads to a lower redshift efficiency (70.1\%). 

At this point, we checked the SDSS DR8 database for any host galaxies that failed to gain a redshift from our BOSS observations and reductions, based on either Algorithm One and Two as described above. We found an additional 178 host galaxies had a successful redshift measurement in DR8, and add them to our final BOSS sample.  We note that most of these galaxies were on our BOSS target list, but were not observed during our ancillary program. This combined sample (which, for simplicity, we will continue to call our 'BOSS' sample) forms the basis for our subsequent re-analysis of the SDSS-II SN light-curve data in Section~\ref{sec:algorithms}, now with the SN candidate redshift constrained to match the observed host galaxy. 

\begin{figure}[!t]
\begin{center}
\epsfig{file=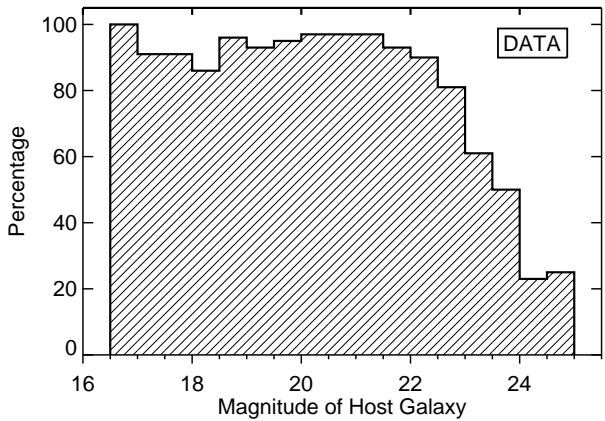,width=1.0\linewidth}
\caption{The percentage of well--measured (based on comparison with {\sc runz} and visual inspection) redshifts obtained as a function of the $r$--band fiber magnitude from the photometry of the SNe (Algorithm One) and general transients (Algorithm Two) host galaxies. We recover 3323 well--measured redshifts from the 3761 galaxies observed by our BOSS program.}
\label{fig:mag_dist_host_gal}
\end{center}
\end{figure}

\begin{figure}[!t]
\begin{center}
\epsfig{file=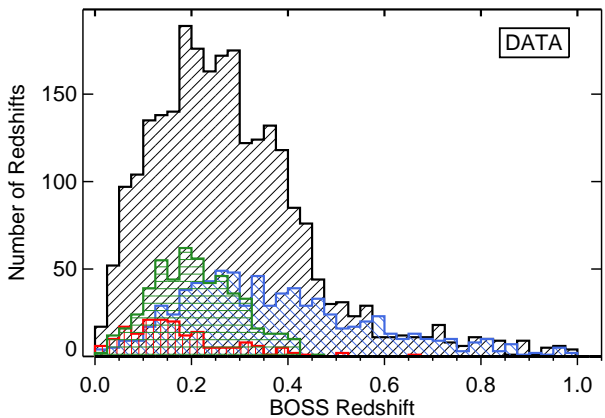,width=1.0\linewidth}
\caption{Redshift distribution of the targets observed with BOSS. The black histogram is the subset which were selected using Algorithm One (candidate SN host galaxies) and the blue are the from Algorithm Two (host galaxies for general transients) as discussed in Section~\ref{sec:Data}. The green histogram shows all spectroscopically--confirmed SNe~Ia from the SDSS-II SN Survey. The red histogram shows the additional 177 SDSS-II DR8 host galaxy redshifts.}\label{BOSS_z_dist_all}
\end{center}
\end{figure}

\begin{figure*}[!t]
\begin{center}
\epsfig{file=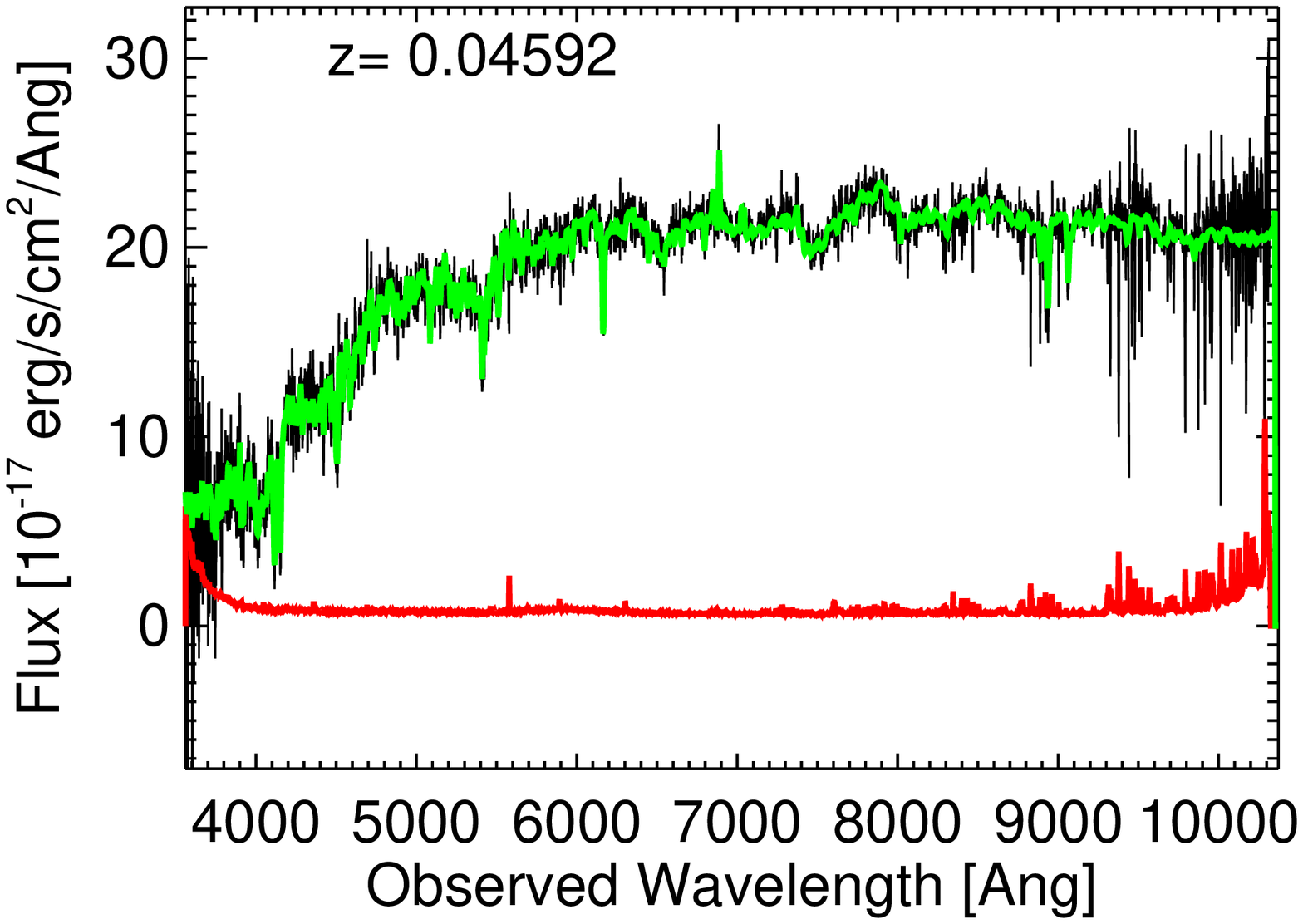,width=0.35\linewidth} \epsfig{file=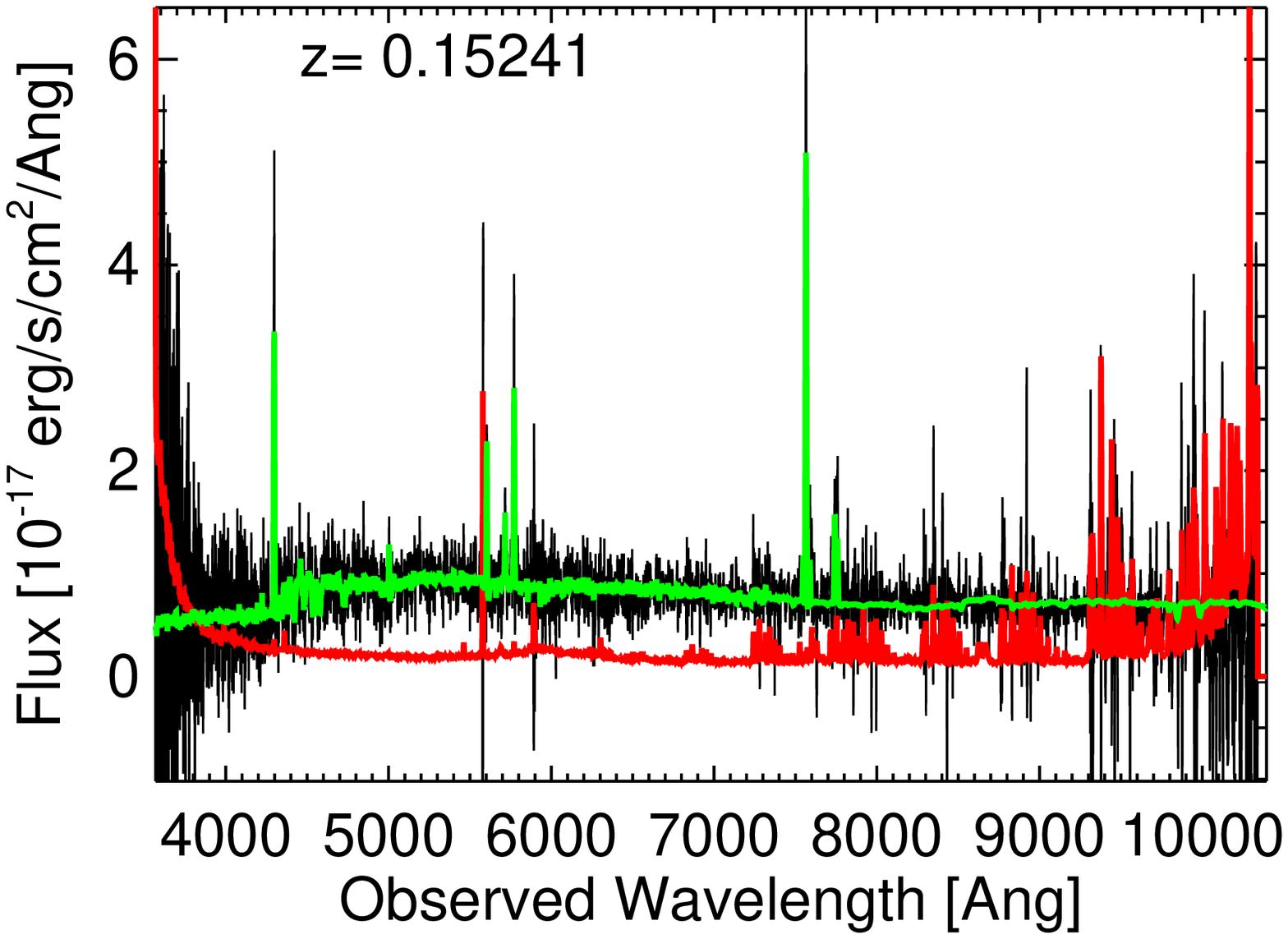,width=0.35\linewidth}
\epsfig{file=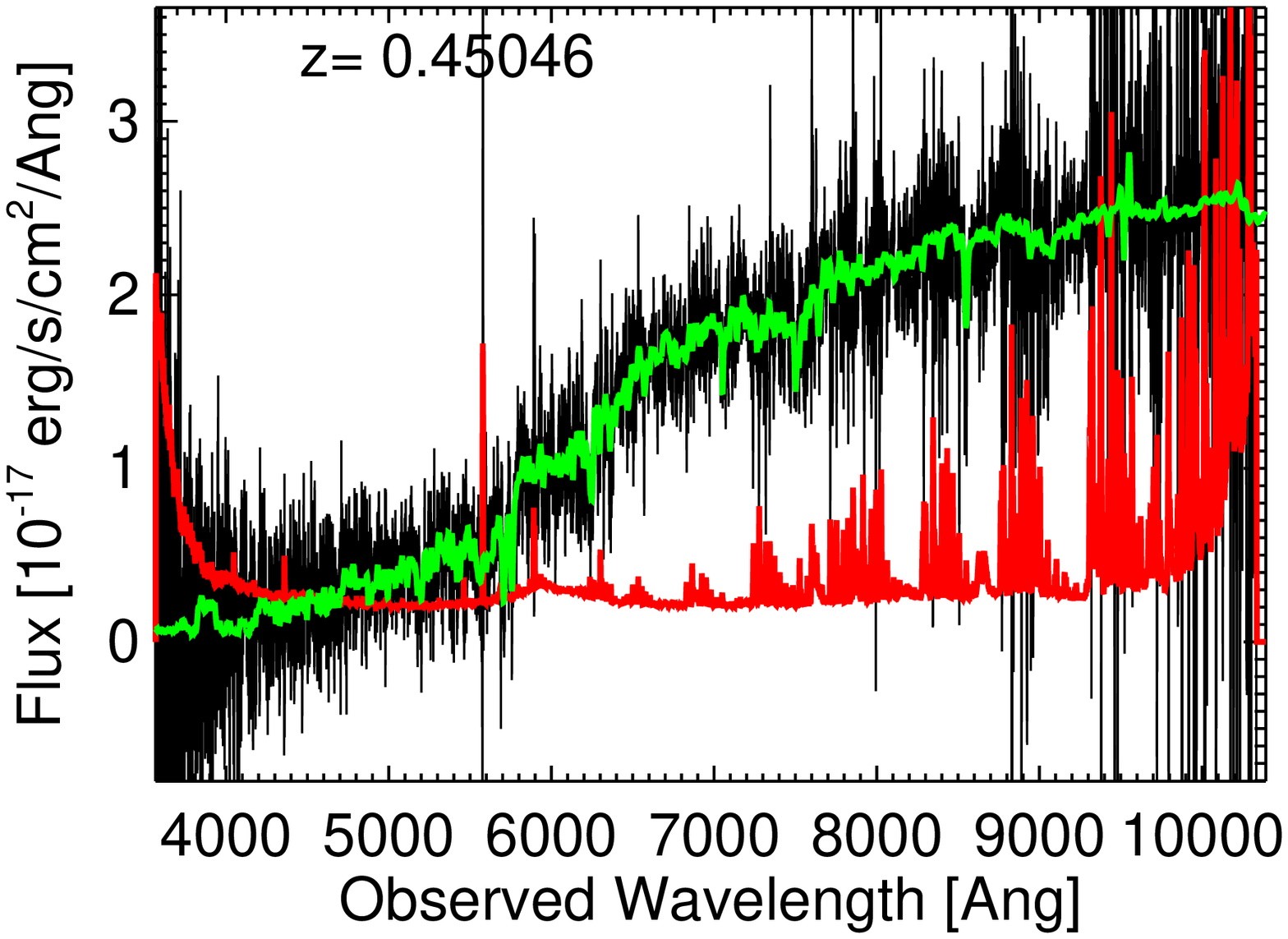,width=0.35\linewidth} \epsfig{file=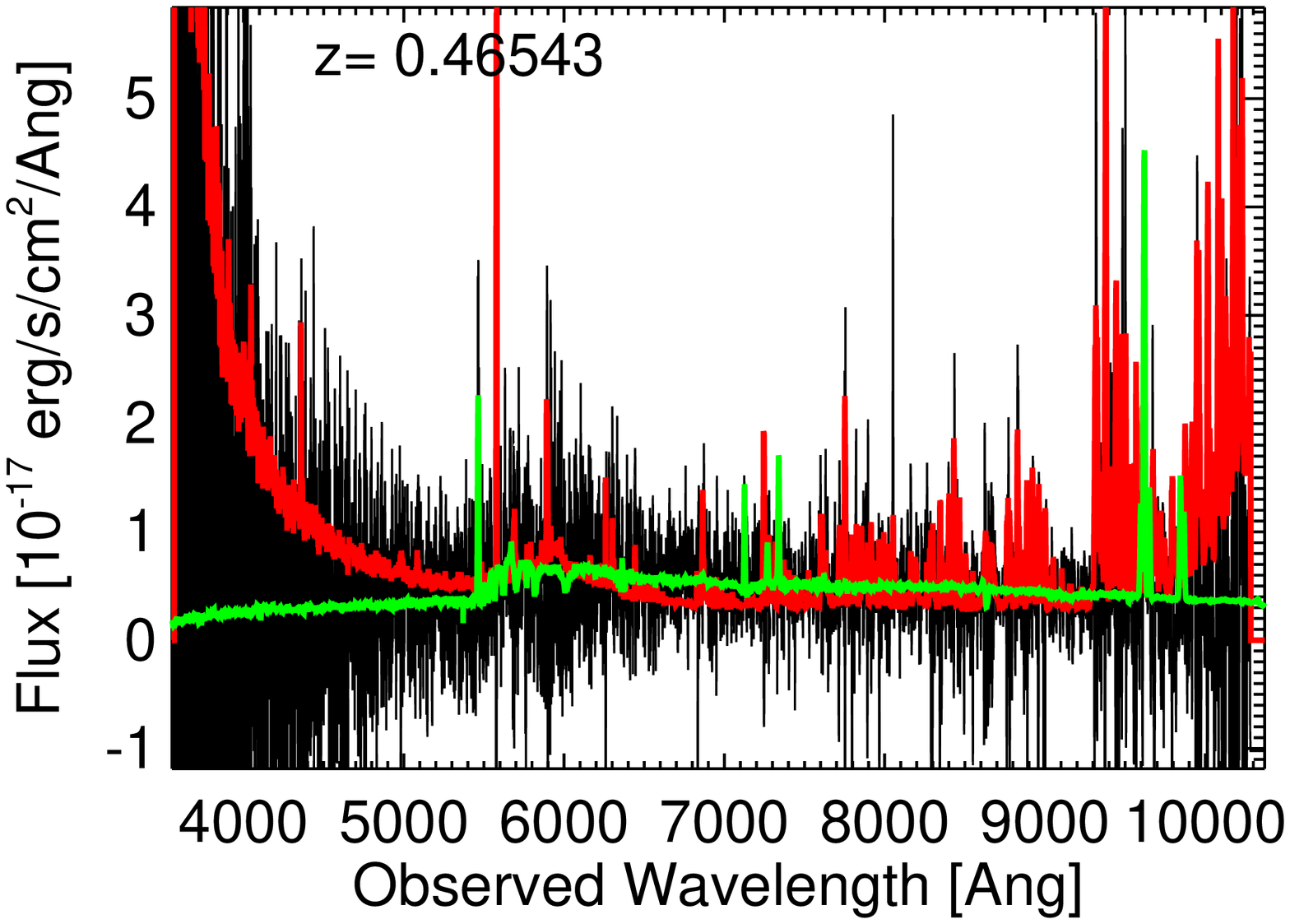,width=0.35\linewidth}
\caption{\label{spectra} Galaxy spectra from BOSS.  The top left panel is a low redshift galaxy with a high $S/N$ continuum, while the top right panel shows a galaxy with virtually no continuum, but several clear emission lines.  Both bottom panels are at the high end of our redshift range (z$>$0.4). The black lines show the data, the green is the best--fit eigenspectrum  spectra and the errors are in red (except masked points, which are set to zero). }
\end{center}
\end{figure*}

We present in Figure~\ref{fig:mag_dist_host_gal} the percentage of successful redshift measurements (from comparing with {\sc runz} and visual inspection) obtained from our BOSS--observed host galaxies, shown as a function of the host-galaxy $r-$band fiber magnitude.  Figure~\ref{BOSS_z_dist_all} shows the measured redshift distribution for our BOSS targets, grouped by their target algorithm. For comparison we also show in Figure~\ref{BOSS_z_dist_all} the spectroscopically--confirmed SNe~Ia from SDSS-II, which clearly peak at lower redshifts. There are a significant number of transient host galaxies with redshifts greater than 0.5, extending out to $z\approx1$, which are primarily quasars. In Figure~\ref{spectra} we display a few example BOSS spectra of our SN host galaxies, spanning the full redshift range of SDSS-II SNe.  Though there is a wide range of $S/N$ in these spectra, spectral features (particularly the 4000 \AA\ break and emission lines), which allow us to measure the galaxy redshift, are clearly visible. 

\section{Creating a Photometric Hubble Diagram}
\label{sec:algorithms}

In this section, we describe our construction of a photometrically-classified SN~Ia Hubble diagram. We provide details of the SN simulations used to determine the optimal selection criteria (Section~\ref{subsec:selection_cuts_sims}) and apply these to the data (Section~\ref{subsec:selection_cuts_data}). We describe the resulting Hubble diagram in Section~\ref{subsec:tests_photo}.

\subsection{Selection Criteria}
\label{subsec:selection_cuts_sims}

Here we define, using simulations, the criteria by which we construct our photometric SN~Ia Hubble diagram. Our primary focus is to minimize the contamination in our sample from non-Ia SNe; this is one of the major concerns associated with measuring cosmological parameters with photometrically-classified SNe, and has the potential to introduce significant systematic errors in the cosmological analysis (Section~\ref{subsec:cosmo_anal}). Given the large number of SN candidates included in our data-set, the conservative classification criteria that we seek will still result in statistical errors on our cosmology smaller than those due to systematics (Appendix~\ref{Appendix_OtherSys}). By focusing primarily on the purity of our sample we have sacrificed its overall completeness, and therefore we caution the reader against using this particular sample for analyses that require high completeness (e.g., SN rate measurements). 

There are three types of criteria that we apply to our sample.  First we apply ``light--curve quality" cuts which are criteria that ensure the SN light--curves (real or simulated) are of sufficient quality to be well-fit with SALT2 (see Appendix \ref{Appendix_SALT} for details) to give accurate distance moduli and provide meaningful typing constraints. Next we consider the optimal values for the PSNID classification parameters (see Appendix \ref{Appendix_PSNID} for details), which will differ from those used in \sako\ since we are both using an improved version of PSNID and wish to increase the recovered sample purity.  Finally, we can further improve the purity of our sample by applying ``color and stretch" criteria based on the derived SALT2 parameters for each light-curve and our knowledge of the likely acceptable range for these properties for SNe~Ia.  We outline each of these types of criteria below, following a brief overview of the simulations and Figure--of--Merit (FoM) we use to guide our criteria selection. In Table~\ref{cuts_contam}, we show the effect of each selection criterion as applied in the following sections.

\subsubsection{Supernova Simulations}
\label{subsubsec:sims}

To test the purity and completeness of our sample, and help define the best selection criteria, we use the publicly available simulations of the SDSS-II SN Survey created as part of the ``Photometric SN Classifier Challenge" of \kesslera. These simulations were made available via the SN challenge website$\footnote{http://sdssdp62.fnal.gov/sdsssn/SIMGEN\_PUBLIC}$ and were produced using the SNANA software \citep{kessler:2009a}, as described in \kesslerb. We decided to use these simulations to test our completeness and purity as they have been well-tested by many researchers, and were designed specifically for testing photometric classification of SNe. They also provide an accurate description of the conditions under which our data were acquired.

In detail, the simulations have ten times the number of SNe as the full three--year SDSS-II SN Survey. They are based on realistic weather, seeing, and photometric zero--point variations, and have a realistic mixture of different SN types out to a redshift limit of $z=0.45$.  The number of SNe~Ia created in these simulations is based on the observed SN~Ia redshift dependence from the SDSS-II SN Survey first--year rates analysis \citep{dilday:2008a}, and the non-Ia SNe redshift dependence is based on the core--collapse rate analysis from SNLS \citep{bazin:2009a}. The core--collapse contribution has been intentionally overestimated in the simulations in order to increase the statistics of non-Ia SNe that are misidentified as SN Ia, rather than underestimate the amount of non-Ia contamination.

For the analysis presented herein we only use the SALT2 simulated SN~Ia light-curves ($ugriz$) in this public data-set (the simulations also included MLCS generated light-curves).  Since this is only half of the generated SNe~Ia in the simulation, we only include half of the non-Ia SNe to keep the ratio of SNe~Ia to non-Ia SNe correct.  The SNe~Ia light--curves use the SALT2 standardization parameters of $\alpha=0.11$ and $\beta=3.2$, with an assumed intrinsic dispersion of $\sigma_{int}=0.12$ mag.  The intrinsic dispersion is included in the simulations of the SNe~Ia by adding random color variations for each SN in each passband, drawn from a Gaussian distribution with $\sigma_m=0.09$ mag, applied coherently to all SN epochs. The non-Ia~SNe in the simulations are based on 41 well-measured spectroscopically-confirmed non-Ia~SNe templates.  A flat $\Lambda$CDM cosmology with $\Omega_{\rm m}=0.3$ and $H_0=70\rm~km~s^{-1}~Mpc^{-1}$ was also assumed. 

We note that the simulations accurately model the SDSS-II SN Survey software detection pipeline, and thus any SNe that were too faint to be detected by the SDSS survey would not be included in the simulations. The simulations can also model spectroscopic selection, which we do not make use of in our main analysis but is included where we make checks against our spectroscopically--confirmed subsample.

\subsubsection{Defining the Figure--of--Merit}
As we know the true types of SNe in the simulations, we can estimate the efficiency and purity of our photometric classifications as a function of our selection criteria.  We use the definition of photometric typing efficiency $\epsilon_{\rm Ia}$ from \sako, 

\begin{equation}
\epsilon_{\rm Ia} = \frac{N^{\rm true}_{\rm Ia}}{N^{\rm CUT}_{\rm Ia}},
\label{efficiency}
\end{equation}

\noindent where $N^{\rm CUT}_{\rm Ia}$ is the total number of true SNe~Ia in the simulations that pass the light-curve quality cuts (see Section~\ref{dataqualitycuts}), and $N^{\rm true}_{\rm Ia}$ is the subsample of $N^{\rm CUT}_{\rm Ia}$ that are classified correctly (itself a function of additional cuts; see Table~\ref{cuts_contam}).  As noted in \sako, this equation measures the  efficiency of classification only for well-observed SNe~Ia, not the total efficiency of identifying all SNe~Ia in our simulated data. From here on in we refer to the parameter defined in Equation~\ref{efficiency} as simply our efficiency.

We additionally define the weighted purity as in \sako,

\begin{equation}
\eta_{\rm Ia}= \frac{N_{\rm Ia}^{\rm true}}{(N_{\rm Ia}^{\rm true} + W_{\rm Ia}^{\rm false} N_{\rm Ia}^{\rm false})},
\label{purity}
\end{equation}

\noindent where $N^{\rm false}_{\rm Ia}$ is the number of non-Ia SNe incorrectly classified as SNe~Ia and $W^{\rm false}_{\rm Ia}$ weights the contribution of misclassifications to the overall purity.  This definition has the usual meaning of purity for $W^{\rm false}_{\rm Ia}=1$, and for a given amount of contamination by non-Ia~SNe a higher (lower) value of $W^{\rm false}_{\rm Ia}$ results in a lower (higher) value of $\eta_{\rm Ia}$. 

The purity and  efficiency of the photometric classification can be combined to form a FoM.  As defined in \sako\  and \kesslerb, the FoM is simply the product of Eqns~\ref{efficiency} and \ref{purity},

\begin{equation}
FoM=\frac{N_{\rm Ia}^{\rm true}}{N_{\rm Ia}^{\rm CUT}} \frac{N_{\rm Ia}^{\rm true}}{(N_{\rm Ia}^{\rm true} + W_{\rm Ia}^{\rm false} N_{\rm Ia}^{\rm false})}.
\label{FoM}
\end{equation}

\noindent The FoM in Eqn~\ref{FoM} does not encapsulate information on the cosmology constraints, but rather is a simple metric that describes the broad merits of a classifier.  Nevertheless, we aim to create a classification that optimizes this FoM with a suitably chosen weighting factor ($W^{\rm false}_{\rm Ia}$) that represents our previously stated choice of prioritizing purity over efficiency.  The ideal choice for $W^{\rm false}_{\rm Ia}$ is a complicated function of the contaminating objects, sensitive to the redshift and magnitude distribution of each subtype.  This issue is not investigated in detail here; instead, we tested the effects of several different weighting values greater than one and empirically determined that $W^{\rm false}_{\rm Ia}=5$ is the best choice, as it produced the sharpest peak in the FoM in Figures~\ref{sim_fom_prob}, \ref{sim_fom_chi} and \ref{sim_x1_color_fom}. We therefore use this value in all subsequent analyses. For further discussion of the importance of purity in SN~Ia samples for cosmological analyses, see \citet{Bernstein:2011} and \citet{gjergo:2012}, who discuss this issue in the context of the Dark Energy Survey. We note that the  efficiency, purity and FoM are only calculated after we have applied our light--curve quality cuts, as this is our baseline for defining SNe~Ia that are potentially useful for cosmological constraints.

\subsubsection{Light--curve Quality Cuts} 
\label{dataqualitycuts}
We begin our analysis of the simulations by applying data quality cuts to all light--curves, removing SNe that have insufficient epochs to provide any useful measurement.  Defining $t$ as the rest-frame epoch (in days) of each SN relative to peak (determined from the best-fit PSNID SN~Ia model), we require at least one epoch of photometry near peak at $-5<t<+5$ and at least one additional epoch at $t>15$, as in \sako.  However, we do not apply the $S/N$ criteria outlined in \sako\ ($S/N>5$ in at least two of the $gri$ bands), as this could remove many high-redshift or under-luminous SNe that may be identified through photometric classification.  We note that an implicit $S/N$ limit does in fact exist, as the difference imaging software ({\tt sdssdiff}) in the SDSS-II SN Survey requires multiple detections at $S/N>3$ for an object to be labelled a SN candidate in the first place \citep{sako:2008a}. Finally, we check again that none of our SN candidates were detected in more than one of our SDSS-II SN search season.  

Since we have no hard $S/N$ limit, it is not strictly true that our light--curve quality cuts remove all objects that are incapable of providing useful cosmological constraints from their light curves.  The simplicity of our criteria is an attempt to balance the necessity of sufficiently useful data with the desire to be unbiased against faint objects.

\subsubsection{PSNID Criteria}
\label{psnidcriteria}

We run PSNID (described in Appendix~\ref{Appendix_PSNID}) on all simulated light--curves that pass our light-curve quality cuts (Section 3.1.3), placing flat priors on $A_V$ (the host-galaxy extinction), $T_{\rm max}$ (the time of peak brightness), and $\Delta m_{\rm 15}$ (the stretch parameter). We use the true redshift of each SN as a prior, with an uncertainty on $z$ of the measured error. 

\begin{figure}[!t]
\begin{center}
\epsfig{file=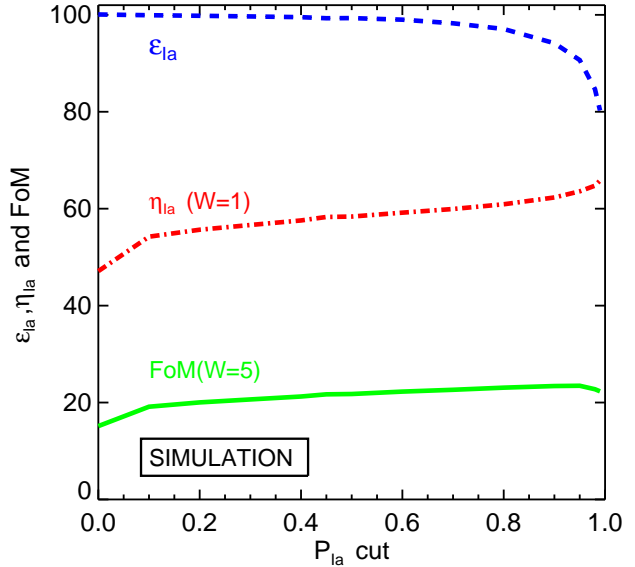,width=1.0\linewidth}
\caption{\label{sim_fom_prob} The  efficiency (blue dashed line), purity (red dot-dashed line), and FoM (green solid line) for the simulated sample as a function of the position of the PSNID $P_{\rm Ia}$ probability cut.  We plot the true purity ($W^{\rm false}_{\rm Ia}=1$), and only change the weighting factor to $W^{\rm false}_{\rm Ia}=5$ in the FoM.}
\end{center}
\end{figure}

We investigate PSNID criteria for removing non-Ia SNe using our simulations, examining their effect on the  efficiency, purity, and FoM.  First we optimize the cut on the PSNID probability of being a SN~Ia ($P_{\rm Ia}$). In Figure~\ref{sim_fom_prob} we show the  efficiency, purity, and FoM as a function of $P_{\rm Ia}$.  Due to the general behavior of PSNID, which tends to cluster values of $P_{\rm Ia}$ around zero or one (demonstrated in Figure~7 of \sako), these functions are relatively flat.  As $P_{\rm Ia}$ does not provide much discriminating power beyond these extreme values, the FoM has little sensitivity to the $P_{\rm Ia}$ cut value.  We thus require $P_{\rm Ia}> P_{\rm Ibc}$ and $P_{\rm Ia}> P_{\rm II}$ for inclusion in our SN~Ia sample, which combines high  efficiency with modest purity.  For reference, this is a less constrained criterion than \sako, which adopted $P_{\rm Ia}> $0.9. 

\begin{figure}[!t]
\begin{center}
\epsfig{file=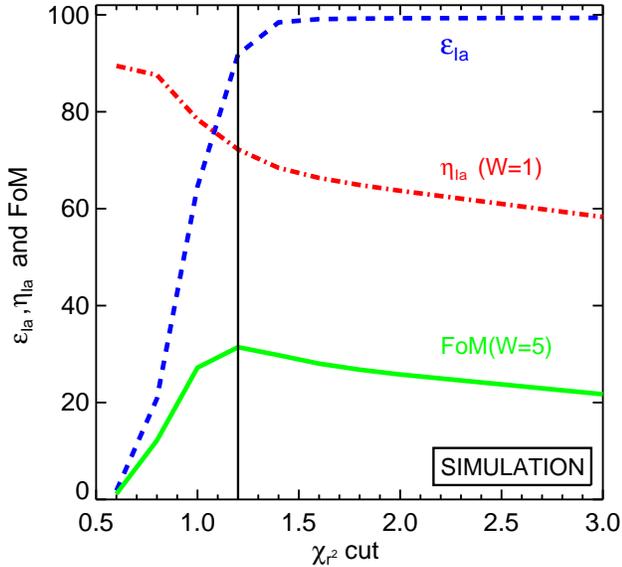,width=1.0\linewidth}
\caption{\label{sim_fom_chi} The  efficiency (blue dashed line), purity (red dot-dashed line), and FoM (green solid) for the simulated sample as a function of $\chi^2_r$, after the PSNID cut ($P_{\rm Ia}> P_{\rm Ibc}$ and $P_{\rm Ia}> P_{\rm II}$) was applied .  We plot the true purity ($W^{\rm false}_{\rm Ia}=1$), and only change the weighting factor to $W^{\rm false}_{\rm Ia}=5$ in the FoM. Our cut at $\chi^2_r=1.2$ is denoted by the vertical black line.} 
\end{center}
\end{figure}

We also use the reduced chi--squared ($\chi^2_r$) of the best-fit PSNID model as a discriminator of non-Ia SNe, and determine the optimal value for this cut after the PSNID cut ($P_{\rm Ia}> P_{\rm Ibc}$ and $P_{\rm Ia}> P_{\rm II}$) is applied.  In Figure~\ref{sim_fom_chi} one can see a clear peak in the FoM at $\chi^2_r\simeq1.2$, close to where the purity and  efficiency curves cross.  Thus we define a PSNID classified SNe Ia to be an object with  $P_{\rm Ia}> P_{\rm Ibc}$, $P_{\rm Ia}> P_{\rm II}$, and $\chi^2_r \le 1.2$ .

The value of our $\chi^{2}_r$ cut differs significantly from that in \sako, which can be seen in their Figure~10 to be located at a broad maximum of $\chi^{2}_r \simeq 1.8$. As noted in Appendix A, the version of PSNID used herein differs from that of \sako\ (which had larger model uncertainties), resulting in the optimal $\chi^{2}_r$ cut being smaller in this work. 

\subsubsection{SALT2 Criteria}
\label{subsubsec:SALT}
Despite our optimization of the PSNID classification criteria, over 25\% of our photometric SN~Ia sample remains non-Ia SNe (see Figure~\ref{sim_fom_chi} and Table~\ref{cuts_contam}). The purity of this sample is unsuitable for the cosmological analysis discussed in Section~\ref{sec:bias}, which is the goal of this paper.  We thus run the SALT2\footnote{We use only SALT2 throughout this paper.  As we are interested in the utility of photometric samples of SNe~Ia for cosmology, rather than presenting definitive cosmological constraints, we forego a detailed comparison of light-curve fitting algorithms.}  light--curve fitter (described in detail in Appendix~\ref{Appendix_SALT}) on all of our SNe (simulated and data) to obtain the best-fit light--curve parameters, and explore the effects of applying additional selection criteria to these parameters to further differentiate SNe~Ia from non-Ia SNe.

\begin{figure}[!t]
\begin{center}
\epsfig{file=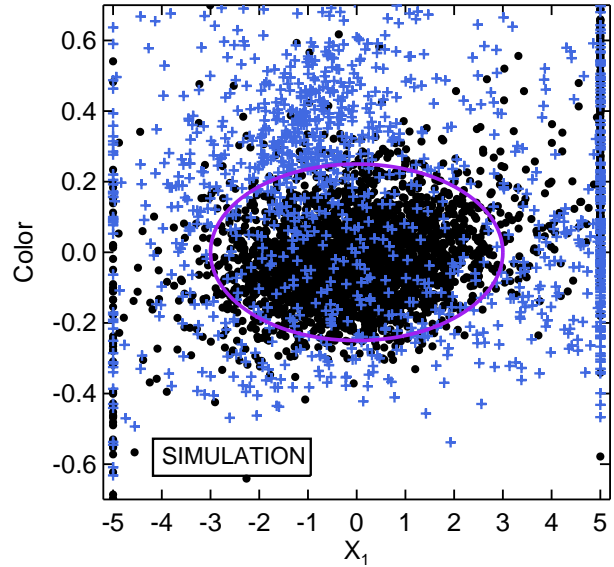,width=1.0\linewidth}
\caption{\label{sim_x1_color_ellipse} The $X_{\rm 1}$ and color ($c$) distributions for the simulated SNe remaining in our photometric sample after PSNID cuts with SNe~Ia  as black points and non-Ia SNe as blue cross symbols. The purple ellipse defines the area we keep for the photometrically-classified sample of SNe~Ia. The cutoff at $-5$ and $5$ in $X_{\rm 1}$ are hard limits set by SALT2.}
\end{center}
\end{figure}

\begin{figure}[!t]
\begin{center}
\epsfig{file=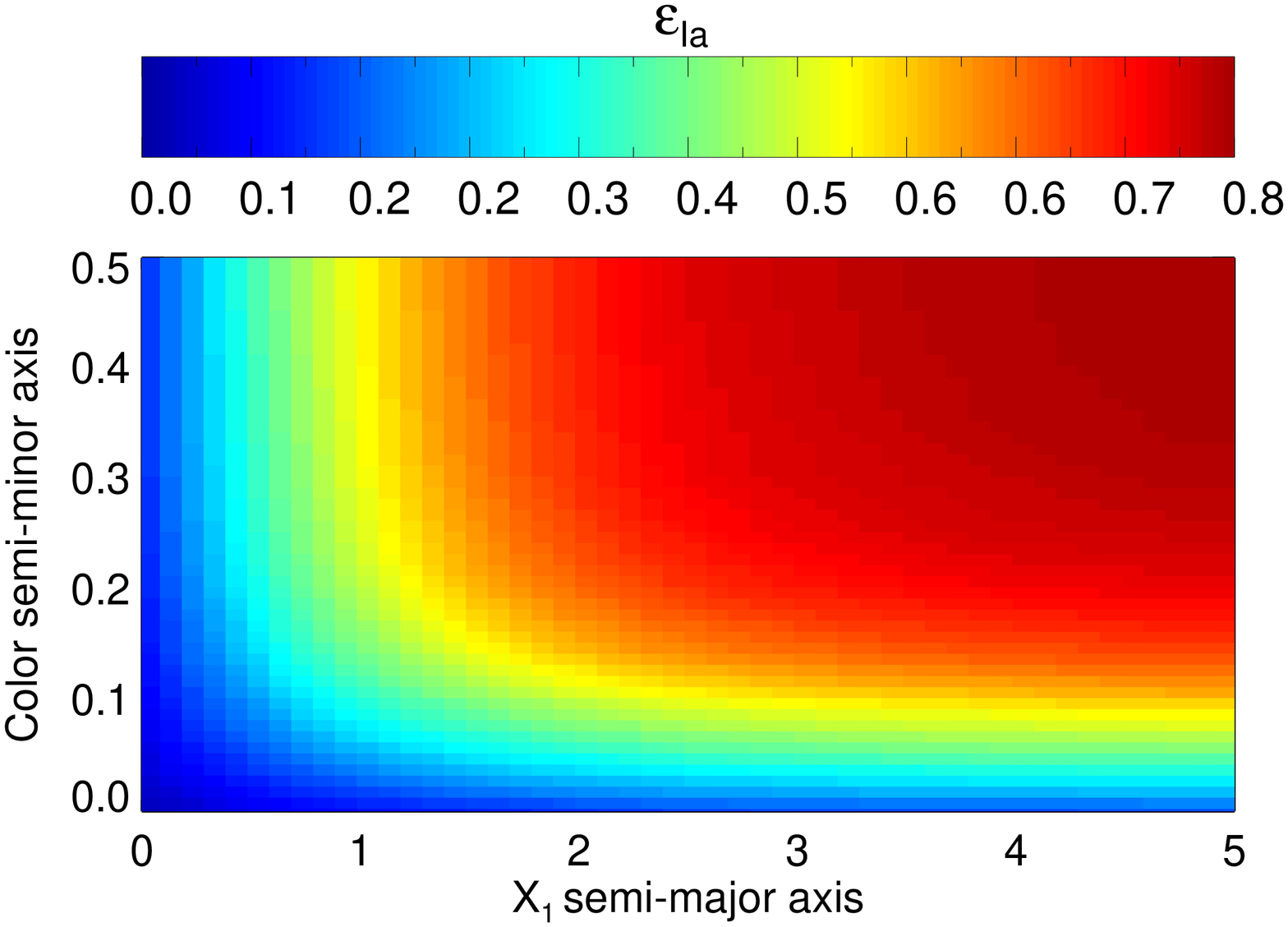,angle=90,width=1.0\linewidth} \epsfig{file=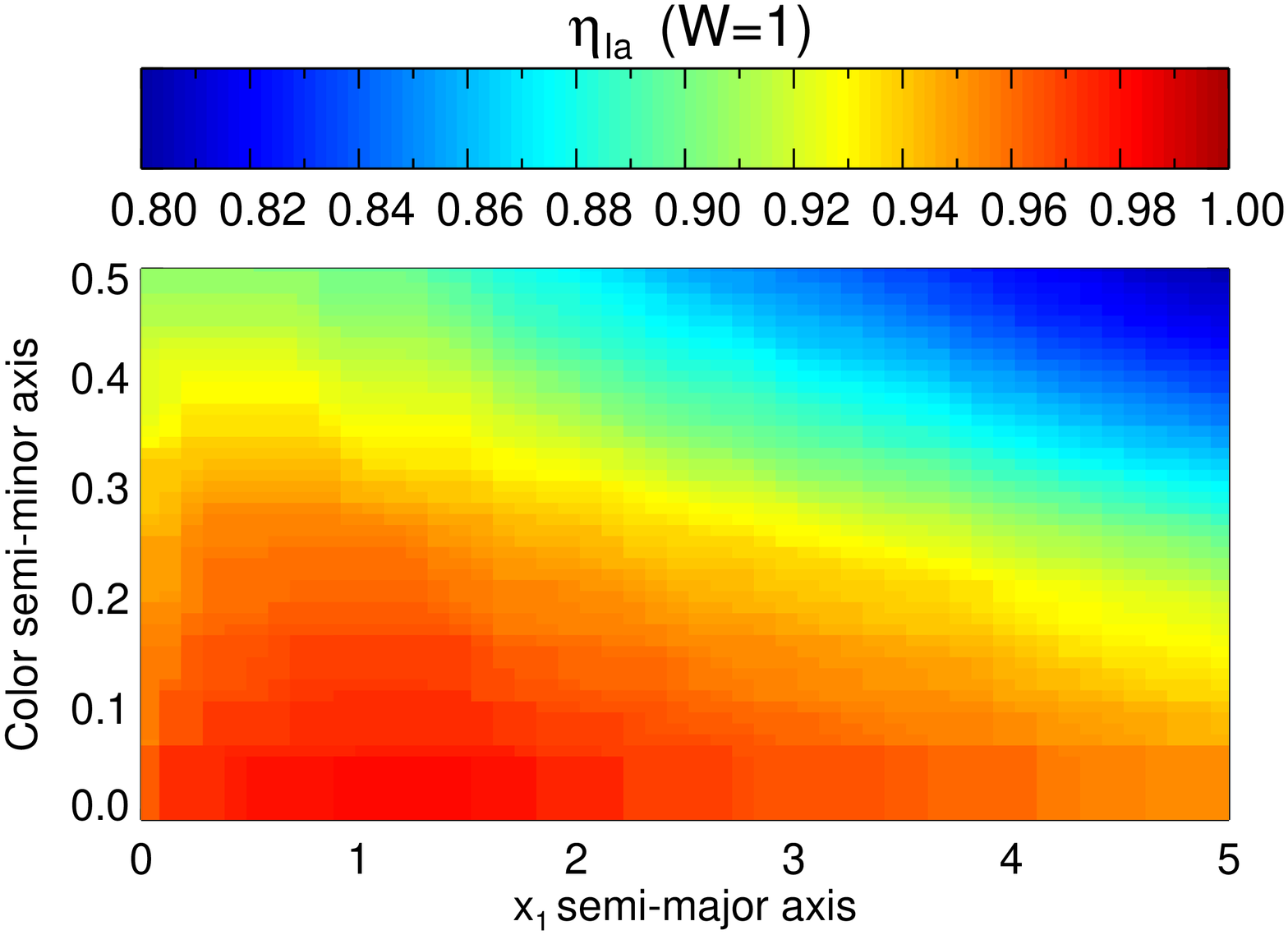,angle=90,width=1.0\linewidth}
\epsfig{file=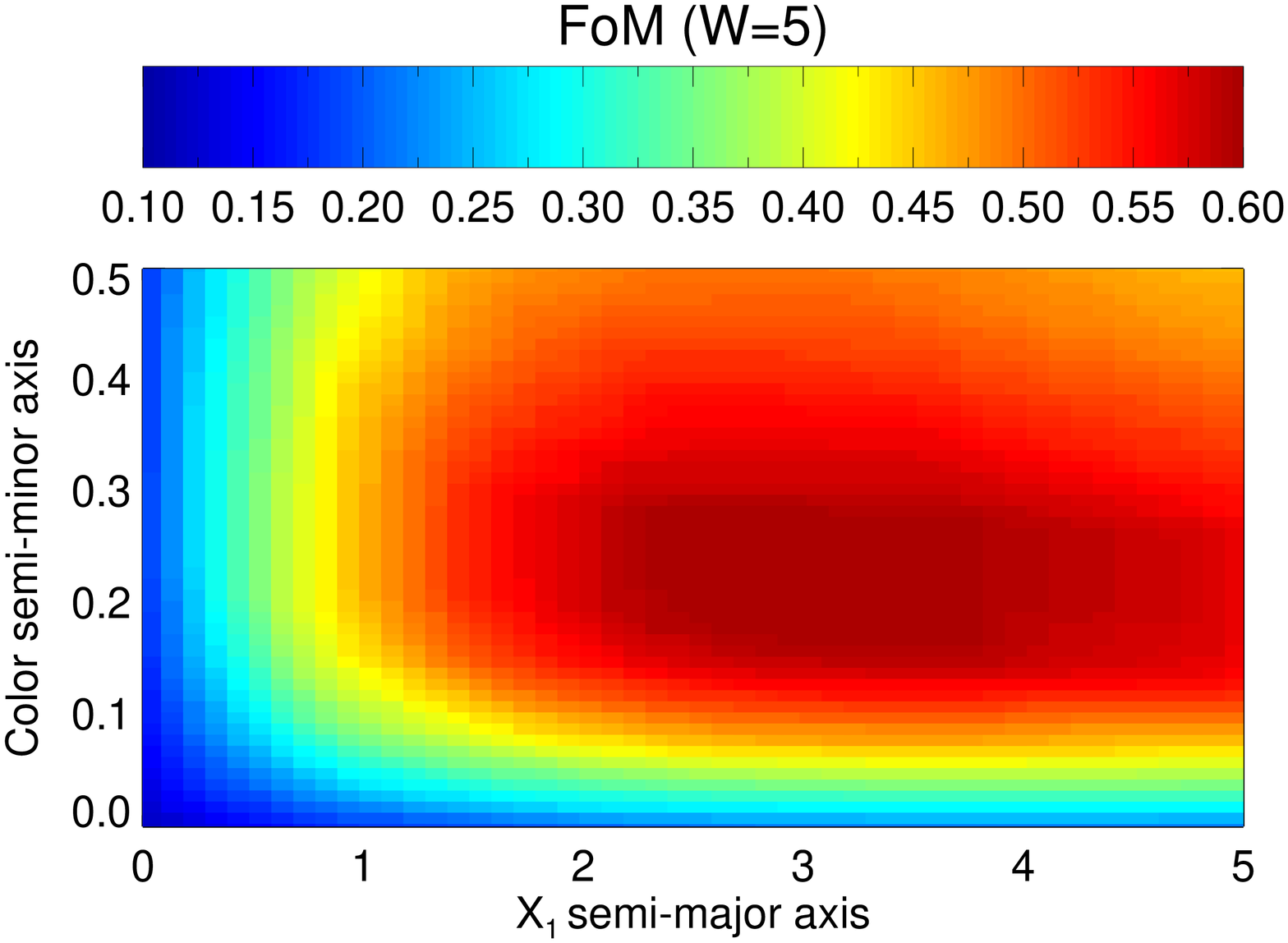,angle=90,width=1.0\linewidth}
\caption{\label{sim_x1_color_fom} The  efficiency (top), purity (middle), and FoM (bottom) plots from SN simulations, for changing the semi--major axis ($a_{X_{\rm 1}}$) and semi--minor  axis ($a_c$) of the ellipse in Figure \ref{sim_x1_color_ellipse}. }
\end{center}
\end{figure}

In Figure~\ref{sim_x1_color_ellipse} we show the distribution of the measured SALT2 parameters color ($c$) and ``shape" ($X_1$) for all SNe remaining in our photometric SN~Ia sample.  By definition, SNe~Ia form a well--defined cluster of points centered on zero in this parameter space, while non-Ia~SNe are more scattered. In \citet{lampeitl:2010a} independent limits were placed on $X_1$ and $c$, but it is clear that an ellipsoidal cut \citep[similar to the circular cut of][]{Bazin:2011} would yield a higher SN~Ia purity for a given  efficiency.  We determine the optimal lengths of the semi-major ($a_{x_{\rm 1}}$) and semi-minor ($a_c$) axes of this ellipse in Figure~\ref{sim_x1_color_fom}, which shows the  efficiency, purity, and FoM as a function of the ellipsoidal parameters.  For an ellipse centered at $(x_{\rm 1}, c)=(0,0)$, the FoM shows a clear peak at $a_{x_{\rm 1}}=3$ and $a_c=0.25$. We use this ellipse, shown in Figure~\ref{sim_x1_color_ellipse}, to further distinguish SNe~Ia from non-Ia SNe. As can be seen in Table~\ref{cuts_contam}, this procedure removes $\sim70$\% of our contaminating SNe at the expense of rejecting $\sim20$\% of the SNe~Ia. We investigated allowing the centre of the ellipse to vary, but found this did not significantly improve the purity or efficiency.

\subsubsection{Color--magnitude Cut}
\label{bazin}

\begin{figure}[!t]
\begin{center}
\epsfig{file=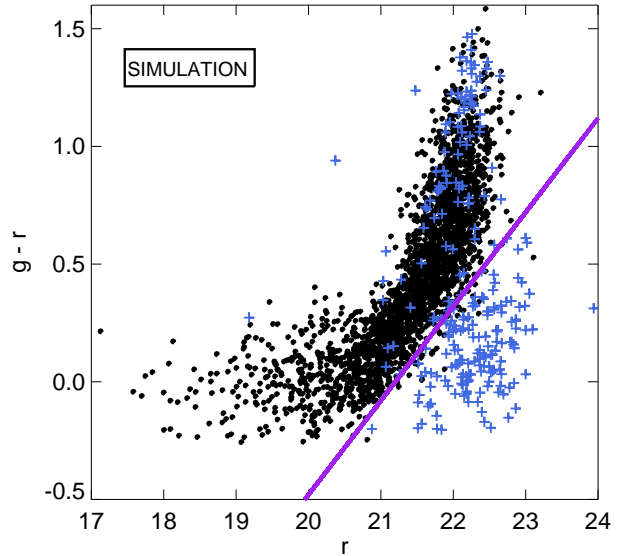,width=1.0\linewidth}
\caption{\label{sim_gr_g} The SN $g-r$ color as a function of peak SN $g-$band magnitude derived from the best--fit SALT2 model to our simulations.  The black dots are SNe~Ia and the blue cross symbols are non-Ia SNe. The purple line is our color--magnitude cut that best separates non-Ia SNe contaminants from our sample.}
\end{center}
\end{figure}

\citet{Bazin:2011} recently showed that a cut in observed color-magnitude space significantly improves SN~Ia sample purity by removing core-collapse SNe contaminants.  We investigated the effects of such a color-magnitude cut on our simulations using the $gri$ peak magnitudes from the best--fit SALT2 model for each SN. We found the most effective color-magnitude cut is in the $g-r$ versus $g-$band model magnitude plane.  The optimal orientation of this color--magnitude cut is described by $g-r < 0.3 \times (g-21.2)$, and is shown in Figure~\ref{sim_gr_g}. This diagnostic cleanly removes a population of non-Ia SNe in the simulations which are too ``blue"  at the given magnitude to be SNe~Ia.  We found that the application of additional color-magnitude constraints, e.g., using different filters, did not produce a significant reduction of our contamination beyond the first color--magnitude cut.

\subsubsection{Overall Contamination in the Simulations}
In Table~\ref{cuts_contam} we break down the effects of our selection criteria on our SN sample.  The first row of Table~\ref{cuts_contam} lists the number of SNe of each type that are included in the simulation; successive rows contain the number of SNe that remain in our SN~Ia sample after each of the criteria (as detailed in the previous sections) are applied.  We provide the contamination (1$-$purity, assuming $W^{\rm false}_{\rm Ia}=1$),  efficiency, and FoM (assuming $W^{\rm false}_{\rm Ia}=5$) after the application of each selection cut. These quantities are defined using the number of SNe~Ia that pass our light-curve quality cuts as the total number of SNe~Ia, i.e., $N^{\rm CUT}_{\rm Ia}$ in Eqn 1. 

Application of both the PSNID and SALT2 criteria significantly improves the purity of our sample, resulting in a SN~Ia sample with purity $>90\%$ and  efficiency $>70\%$.  The FoM has also increased significantly with these cuts, reflecting the weighting we have applied to purity. Interestingly, the inclusion of the color--magnitude constraints (Figure~\ref{sim_gr_g}) significantly improves our purity with almost no effect on our  efficiency.

\begin{figure}[!t]
\begin{center}
\epsfig{file=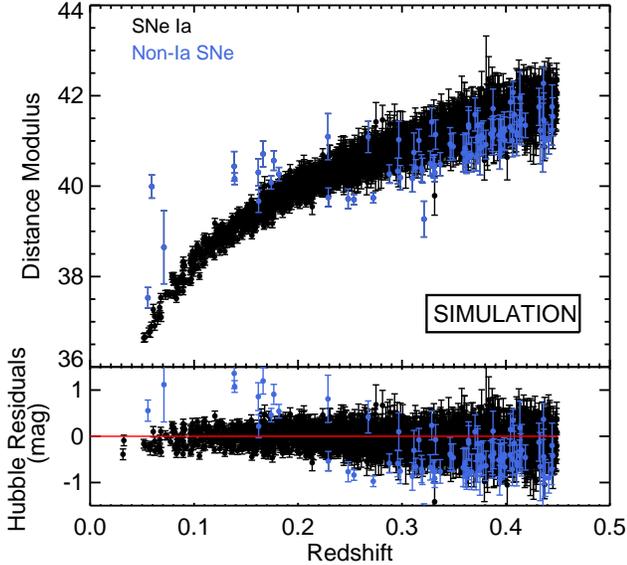,width=1.0\linewidth}
\caption{\label{sim_hubble_typed} Hubble diagram of our simulated SNe~Ia that pass all of our selection criteria. This includes 106 non-Ia SNe that have been misclassified (blue symbols) and 2644 correctly classified SNe~Ia (black symbols). The redshift limit of $z=0.45$ is artificial and set by the original limit in the public SNANA simulations. The plotted errors are a combination of the uncertainties on the SALT2 light--curve parameters and the redshift.  The SALT2 SN~Ia parameters have been fixed to the input simulation values of $\alpha=0.11$, $\beta=3.2$, and $M$=29.8. The bottom panel shows the Hubble residuals assuming the input cosmology.}
\end{center}
\end{figure}

Figure~\ref{sim_hubble_typed} is the Hubble diagram for our simulated photometrically--classified SN~Ia sample. The plotted errors are a combination of the uncertainties on the SALT2 light--curve parameters, the redshift and an assumed intrinsic dispersion of $\sigma_{int}$ = 0.12 magnitudes. The SALT2 SN~Ia parameters are fixed to the input simulation values of $\alpha=0.11$ and $\beta=3.2$.  

We show in blue in Figure~\ref{sim_hubble_typed} the 106 misclassified SNe that have passed all of our selection cuts.  This final simulated photometric SN~Ia sample has a contamination of \finalcontam\ and an efficiency of \finaleff.  Our purity is higher than that for any of the participating methods in the ``Photometric SN Classifier Challenge" (\kesslerb), although it is necessary to note that we have explicitly placed a higher priority on purity than was the stated goal in \kesslerb, and their analyses used a DES-like set of simulations that extend to higher redshifts than in our sample.  As expected, the source of our contamination is primarily SNe~Ib/c, as their light-curves most closely resemble those of a SNe Ia. 

\begin{figure}[!t]
\begin{center}
\epsfig{file=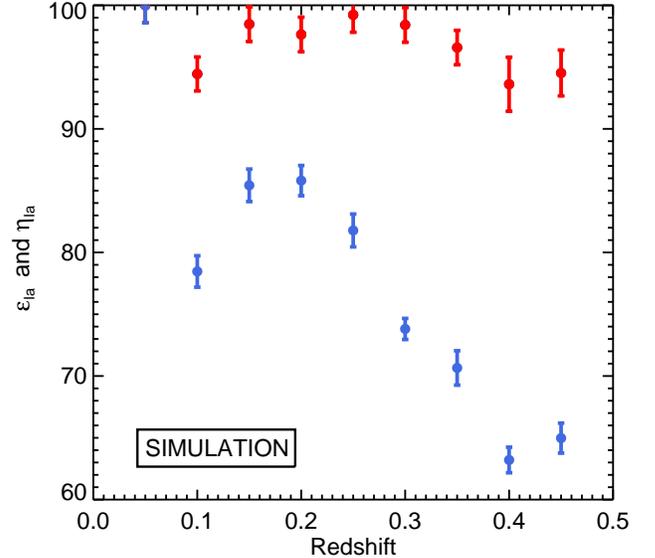,width=1.0\linewidth}
\caption{\label{purity_eff_z} The simulated efficiency (blue) and purity (red) for our final photometrically-classified sample. The error bars are determined from the propagation of errors using Eqns~\ref{efficiency} and~\ref{purity}.}
\end{center}
\end{figure}

Finally, in Figure~\ref{purity_eff_z}, we show the expected purity and efficiency of our sample (based on simulations) as a function of redshift. Within the errors, we see no significant redshift dependence in our purity, but we do witness a significant fall in our efficiency at higher redshifts. This is expected, as we have prioritised purity over efficiency for our cosmological study. Larger simulations (see Section 7.2) are required to probe more subtle effects with redshift.

\begin{center}
\begin{table*}
\begin{center}
\begin{tabular}{|l|c|c|c|c|c|c|}
\hline
Cut & \multicolumn{3}{|c|}{Classified SNe Ia}  & Contam-	 & SN~Ia   & FoM \\
  \cline{2-4}
  &   Total &  SNe Ia & non-Ia SNe & ination 	& Efficiency & ($W$=5)\\
  \hline
Number in simulation	  & 12203 & 5018 & 7185 &  &  &  \\  
Light--curve quality	 	& 9186 & 3734 & 5452 & 59.4$\%$ & 100$\%$  & 12.05  \\
P$_{\rm Ia}>P_{\rm non-Ia}$ (PSNID)		& 6354 & 3701  & 2653  & 41.8$\%$ & 99.1$\%$ & 21.6 \\
$\chi^2_r<$1.2 (PSNID)           & 4737 & 3420 & 1317 & 27.8$\%$ & 91.6$\%$ & 31.3   \\
$X_{\rm 1}$ and $c$ cut (SALT2)	  & 2918 & 2675 &  243 & 8.3$\%$ & 71.6$\%$ & 49.2  \\ 
Color--magnitude criteria (SALT2)	&  2750  & 2644 & 106 & \finalcontam  & \finaleff & 59.0  \\
\hline
\end{tabular}
\end{center}
 \caption{Number of SNe in our simulated sample as a function of the selection criteria applied to the data. In each row we show the cumulative effect of all the previous criteria on the contamination,  efficiency and FoM (assuming $W^{\rm false}_{\rm Ia}=5$).}
\label{cuts_contam}
\end{table*}
\end{center}

\subsection{Application of selection method to the data}
\label{subsec:selection_cuts_data}

We now apply our analysis and the selection criteria defined using our SN simulations to our SDSS--II BOSS SN sample. In Table~\ref{cuts}, we provide a break-down of the number of SNe classified at each stage of our selection (as discussed in Section~\ref{subsec:selection_cuts_sims}).  The two PSNID-based criteria remove $\approx45$\% of the sample that remain after the data quality cuts we applied (comparable to the 48\% seen in our simulations), resulting in 1443 objects in our SN~Ia sample at this stage of analysis.

\begin{table*}[!t]
\begin{center}
\begin{tabular}{|l|c|c|c|c|c|}
  \hline 
  Selection Criteria & Removed SNe & Kept SNe & Spec SNe~Ia & Spec non-Ia~SNe \\
  \hline
  Accurate BOSS Redshifts & -  & \nogoodSDSS\   & 329 & 59 \\
  Light--curve quality & 874 & 2626 & 249 & 24 \\
  P$_{\rm Ia}>P_{\rm non-Ia}$ (PSNID) & 579 & 2047 & 247 & 6 \\
  $\chi^2_r<$1.2 (PSNID) & 604 & 1443 & 239 & 2\\ 
  $X_{\rm 1}$ and $c$ cut  (SALT2)	& 634 &  809 & 209 & 0 \\
  Color--magnitude criteria  (SALT2)	& 54 & 755 & 209 & 0 \\  
  Correct host galaxy & 3 & 752  & 208 & 0 \\
  \hline
\end{tabular}
 \caption{The cumulative effect of applying each selection criteria to our data, leading to a final sample of photometrically-classified SNe~Ia (\nofinal). The right--hand column shows the effect of these criteria on known (spectroscopically-confirmed) SNe~Ia in our sample. \label{cuts} }
\end{center}
\end{table*}

\begin{figure}[!t]
\begin{center}
\epsfig{file=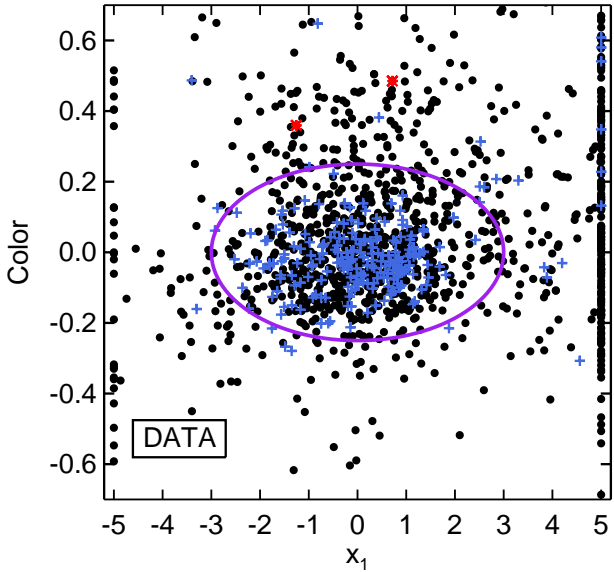,width=1.0\linewidth}
\caption{\label{boss_x1_color} The $X_{\rm 1}$ and $c$ distributions for PSNID--classified SNe~Ia from our SDSS-II SN candidate samples. The blue (red) cross symbols denote the sub-set of these candidates that have been spectroscopically  confirmed as SNe~Ia (non-Ia SNe).  The purple ellipse is our SNe~Ia boundary taken from Figure~\ref{sim_x1_color_ellipse}.}
\end{center}
\end{figure}

We use SALT2 (Appendix~\ref{Appendix_PSNID}), applied to the SDSS SMP $griz$ light-curve data, to measure the light-curve parameters and distance moduli for all \nogoodSDSS\ (\nogoodSN\ + \nogoodbias\ + \noSDSS ) SN candidates with host galaxy redshifts (see Section~\ref{subsec:BOSSred}). Figure~\ref{boss_x1_color} shows the distribution of our PSNID-classified SN~Ia sample in $X_{\rm 1}$--color space and the ellipsoidal criteria we derived from simulations.  This cut removes 634 SN candidates (44\%) from our photometric sample.  This is a higher percentage than predicted by simulations  in Section~\ref{subsec:selection_cuts_sims} (39\%), which could be attributed to the fact that non-SN contaminating sources (e.g., quasars, which are known to be in our BOSS target list) are not modeled in our simulations.  Many of the discarded SN candidates have poorly--fit templates, with $X_1=\pm5$, where SALT2 is driven to the extremes of its self-imposed (i.e., hard-coded) $X_1$ range.  This constraint also removes many ``red" objects with large $c$ values, which may include highly reddened SNe~Ia. However, we would prefer to exclude these red SNe~Ia from our cosmology analysis, as \citet{foley:2010} showed they potentially bias cosmological constraints. Our SALT2 criteria leave us with 809 photometrically--classified SNe~Ia candidates. 

\begin{figure}[!t]
\begin{center}
\epsfig{file=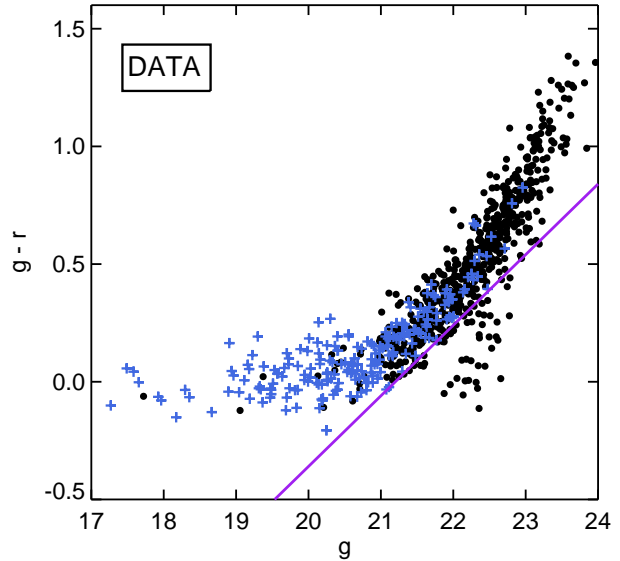,width=1.0\linewidth}
\caption{\label{boss_gr_g} The $g-r$ color as a function of $g$-band magnitude for our SDSS-II data.  The blue cross symbols denote the sub-set of these candidates that have spectroscopic confirmation as SNe~Ia. The location of the purple line is our SN~Ia boundary taken from Figure~\ref{sim_gr_g}.}
\end{center}
\end{figure}

In Figure~\ref{boss_gr_g} we show the application of the color--magnitude cut in the $g-r$ versus $g-$band model magnitude plane, which removes a sample of ``blue" SN candidates.  Both in the data and the simulations this cut removes $\approx7$\% of the sample that passed the SALT2 cut, leaving a photometrically-classified sample of 755 SNe~Ia candidates.  The agreement between the simulations and data is reassuring, and it should be noted that these simulations were made prior to the construction of our BOSS SDSS SN sample, so there has been no fine-tuning of the simulations to match our BOSS sample.

A potential source of error that is not included in our simulations is the misidentification of the host galaxy of our SN candidates, as this can result in an incorrect classification due to an incorrect redshift prior.  We have identified three cases in our sample where we have evidence, described in Appendix~\ref{Appendix_z_compare}, that either the galaxy observed by our BOSS program is not associated with the SN to which it was assigned, or the derived BOSS redshift is erroneous.  We remove these three SN candidates from our sample.

Our final sample contains \nofinal\ photometrically--classified SNe~Ia, of which \nofinalSpec\ have been spectroscopically confirmed as SNe~Ia as part of the original SDSS-II SN Survey (Table~\ref{cuts}).

\subsection{Tests of our photometric sample}
\label{subsec:tests_photo}
We present here a basic examination of our final sample of \nofinal\ photometrically--classified SNe~Ia before embarking on a cosmological analysis (Section~\ref{subsec:cosmo_anal}). First, we study the effect of our selection criteria on the subset of existing spectroscopically--confirmed SNe~Ia in our sample. Unlike \sako, we have not used this subset of known SNe to refine our selection criteria as there are concerns about potential bias in this sample (see Section~\ref{subsec:SDSSSN}).  As can be seen in Table~\ref{cuts}, we start with 329 spectroscopically--confirmed SNe~Ia that have had host-galaxy redshifts obtained by BOSS or SDSS. Of these, 249 SNe~Ia passed our data-quality criteria applied in Section~\ref{dataqualitycuts}.  After applying all of our additional selection criteria we are left with \nofinalSpec\ spectroscopically--confirmed SNe~Ia, resulting in an efficiency of 84$\%$ (\nofinalSpec\ /249) in classifying our SDSS-II spectroscopic SN~Ia sample. 

\begin{figure}[!t]
\begin{center}
\epsfig{file=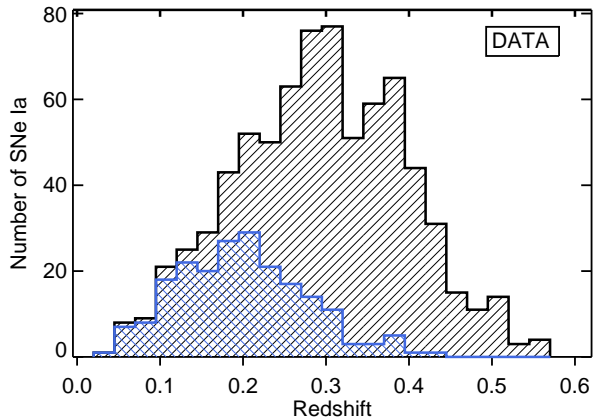,width=1.0\linewidth}
\caption{\label{BOSS_z_dist} The redshift distribution of our \nofinal\  photometrically--classified SNe~Ia (black), and the subset of \nofinalSpec\ that have spectroscopic classification (blue).}
\end{center}
\end{figure}

This  efficiency is higher than the predicted value from our simulations (71.6\%; see Table~\ref{cuts_contam}).  This difference can be mainly attributed to the spectroscopic sample probing a lower redshift range than the photometric sample. In Figure~\ref{BOSS_z_dist} we show the redshift distribution for our full sample of photometrically--classified SNe~Ia (black) compared to the subsample of spectroscopically--confirmed SNe~Ia (blue).  While the spectroscopic sample peaks at $z\sim0.2$ and drops to zero by $z>0.4$, the photometrically--classified SNe~Ia extend out to $z\simeq0.55$, with a median redshift of $z=0.30$. This explanation is checked by studying the "spectroscopically--confirmed" subset of SNe~Ia provided as part of the public SNANA simulations discussed in Section 3.1.1. The  efficiency of photometrically classifying this simulated spectroscopic SNe~Ia subset is 83.2\%, in reassuring agreement with the 84\% efficiency we find for the spectroscopic SNe~Ia subset in our data. 

We reiterate that we do not include in our final sample any of the spectroscopically confirmed SNe~Ia that were removed during our photometric classification procedure.  Our intention is to construct a sample of SNe~Ia based purely on their photometric properties and host galaxy redshifts, and examine the cosmological constraining power of such a sample. This procedure mimics the challenges facing the next generation of SN surveys.  

All spectroscopically--confirmed non-Ia~SNe in the original SDSS-II SN sample were removed by our selection criteria, resulting in 100\% purity for the spectroscopic subset of our photometric sample.  Of the 59 known non-Ia~SNe with host-galaxy redshifts obtained by BOSS, only two were misclassified as SNe~Ia by PSNID, and both were subsequently removed by the $X_{\rm 1}-c$ criteria.  There are two factors that lend themselves to this subset achieving higher purity than would otherwise be expected from our simulations.  First, the SDSS-II spectroscopic program was intentionally biased against targeting non-Ia~SNe; in our BOSS sample there is only one spectroscopically-confirmed non-Ia~SN for every six SNe~Ia, whereas nearly $60$\% of the SNe in our simulations are non-Ia.  Second, PSNID was tested, and optimized, on the SDSS-II spectroscopic SNe, so its efficiency at identifying non-Ia~SNe from this sample is unsurprising. 

\begin{figure}[!t]
\begin{center}
\epsfig{file=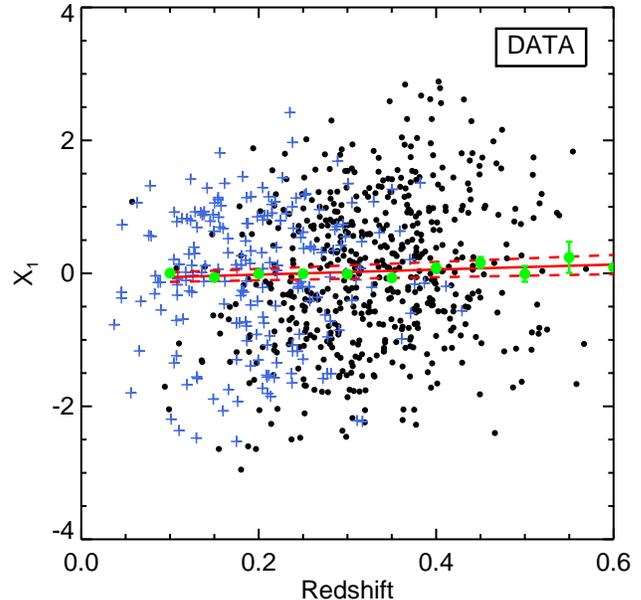,width=1.0\linewidth}
\caption{\label{x1_z} The SALT2 $X_1$ parameter as a function of redshift for the photometrically--classified SNe~Ia (black dots) and the subset of spectroscopically--confirmed SNe~Ia (blue crosses). The green points show the mean $X_1$ (and the error on the mean) in bins of redshift. The red solid line is the best-fit linear relation to the average $X_1$, and the red dashed lines are the error on the fit.} 
\end{center}
\end{figure}

\begin{figure}[!t]
\begin{center}
\epsfig{file=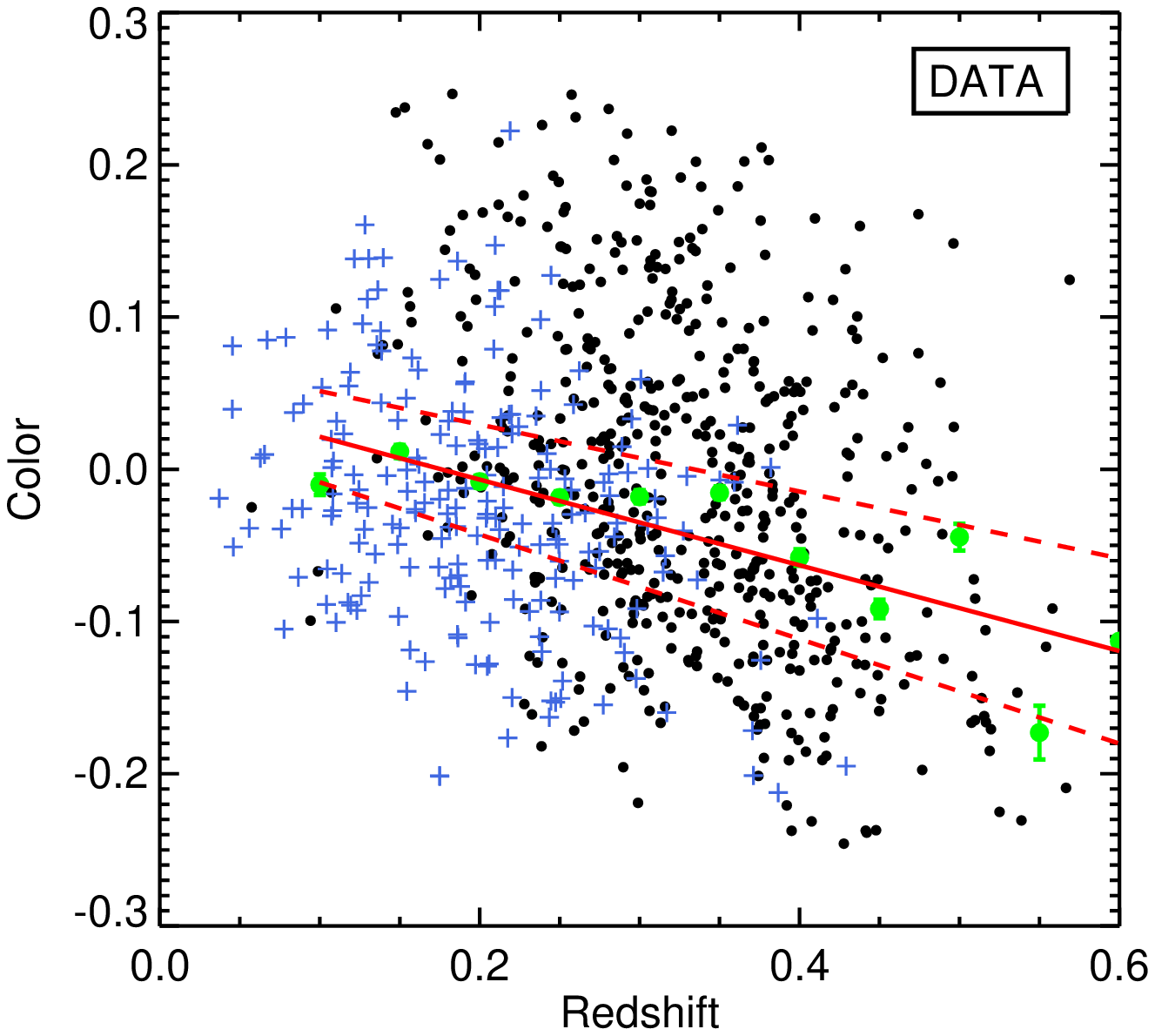,width=1.0\linewidth}
\caption{\label{color_z} The SALT2 color parameter as a function of redshift for the photometrically--classified SNe~Ia (black dots) and the subset of spectroscopically confirmed--SNe~Ia (blue crosses). The green points show the mean color (and the error on the mean) in bins of redshift. The red solid line is the best-fit linear relation to the average color, and the red dashed lines are the error on the fit.} 
\end{center}
\end{figure}

In Figure~\ref{x1_z}, we show the SALT $X_1$ values as a function of redshift for the photometrically classified SNe~Ia sample. We also show the mean $X_1$ in redshift bins and the best-fit linear relation to the binned data. We see no change in the distribution of $X_1$ values with redshift, as the best-fit slope is consistent with zero within the errors on the fit. In Figure~\ref{color_z}, we show the SALT $c$ (color) values as a function of redshift, again with their mean values, binned by redshift, and the best-fit linear relation. A correlation is evident in this case, with higher redshift SNe skewed towards bluer (negative) colors. However, this trend is driven by SNe at $z>0.4$, and if we limit the sample to redshifts below this value we find the best-fit slope is consistent with zero within errors. We may be seeing evidence of a color-dependent Malquist bias, i.e., bluer SNe~Ia are brighter and thus easier to detect at higher redshifts (more so than SNe with higher $X_1$ values).

\begin{figure*}[!t]
\begin{center}
\epsfig{file=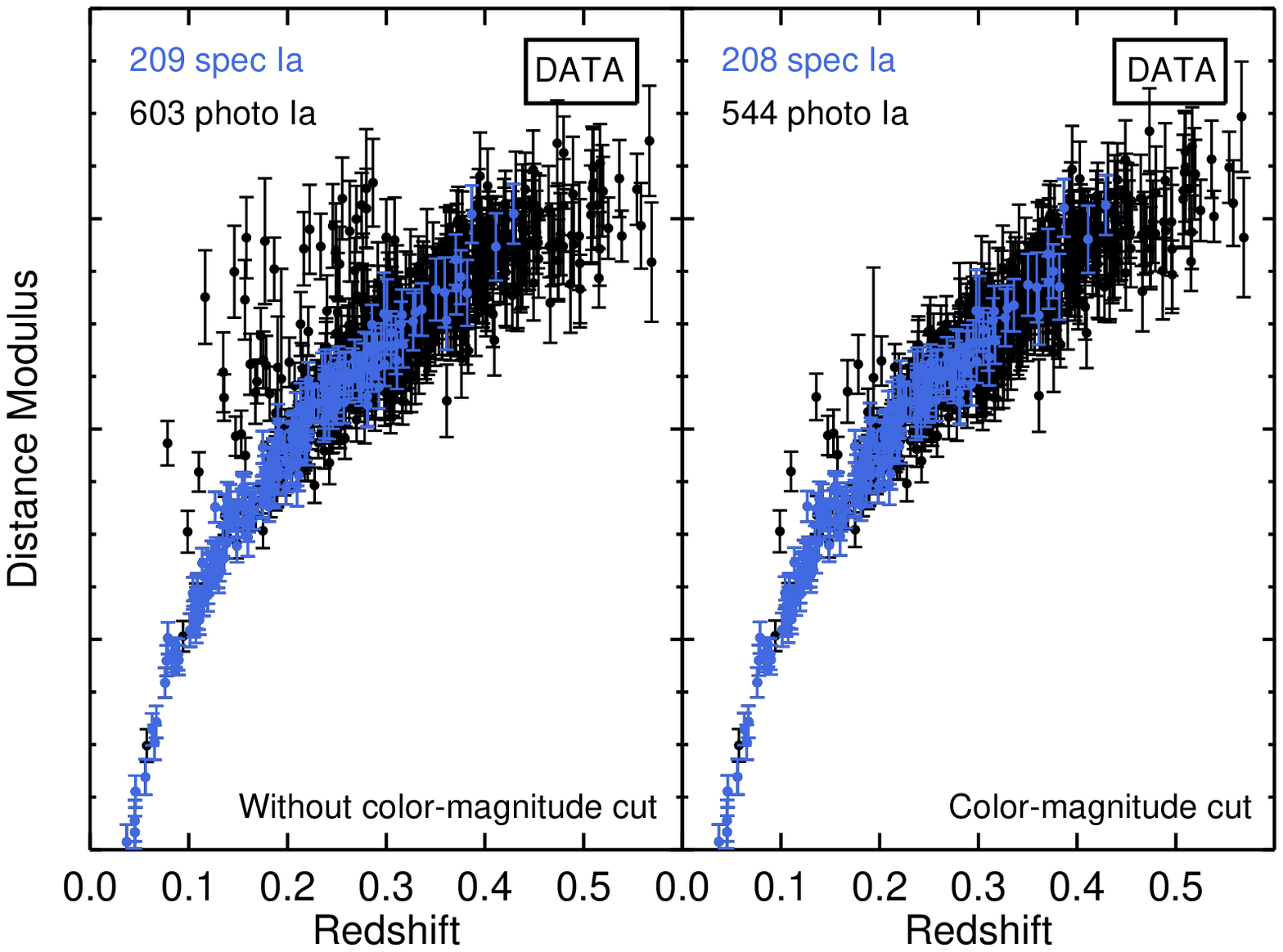,width=0.8\linewidth}
\caption{\label{hubble_diagram} Hubble diagram for our photometrically--classified SNe~Ia sample.  Blue points are the subsample of spectroscopically--confirmed SNe~Ia, while black points are new SNe~Ia. These Hubble diagrams are created using the best-fit $\alpha$ and $\beta$ from the full cosmological fit (Section \ref{subsec:cosmo_anal}), and include the intrinsic dispersion in the uncertainties.  Our sample before (left) and after (right) the color--magnitude criteria are applied demonstrates the utility of this criterion. The SALT2 SN~Ia parameters have been set to the best-fit values of $\alpha$ and $\beta$ from the cosmological fit, and as our cosmological fit analytically marginalizes over $M$ we use the same value here as in the simulations ($M$=29.8).}
\end{center}
\end{figure*}

In Figure~\ref{hubble_diagram} we present two Hubble diagrams for our photometrically--classified sample of SNe~Ia. On the left is the sample before our color--magnitude cut is applied, while on the right we show our sample after making this cut.  This criterion removes 54 SNe~Ia candidates, which are clearly offset (too faint) from the main Hubble sequence and thus likely to be non-Ia SNe (as seen in our simulations). These removed candidates are also clustered around $z\approx0.2$, again consistent with our simulations.

There are still ten possible outliers, around $z\approx0.2$, to the main Hubble diagram in Figure~\ref{hubble_diagram} (right-hand side).  We have studied the photometric data for these HR outliers individually (including visually inspecting the images) and can find no obvious reasons to remove them from our sample. These outliers could be Type Ibc SNe, which possess similar colors and light-curve shapes in the SDSS filters at these redshifts, or possibly odd SNe~Ia like PTF10ops \citep{maguire:2011a} and SN 2006bt \citep{foley:2010a}.  However, this small amount of contamination does not bias our cosmological results, as shown in Section~\ref{subsec:cosmoMC}, where we apply at 3$\sigma$ clipping to the Hubble diagram and find consistent cosmological results.
 
\section{Sources of Bias}
\label{sec:bias}
Before we undergo a detailed cosmological analysis of our sample (Section~\ref{subsec:cosmo_anal}), we investigate possible sources of bias in our photometrically--classified sample. 

\begin{figure}[!t]
\begin{center}
\epsfig{file=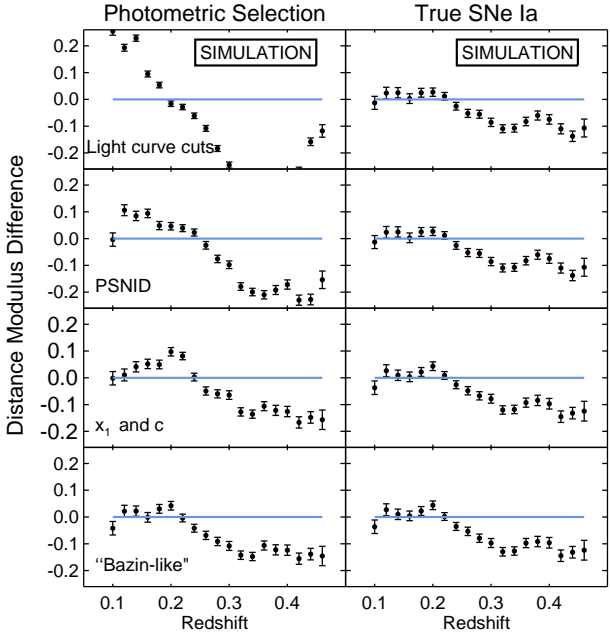,width=1.0 \linewidth}
\caption{\label{con_bias_plot} The weighted mean and uncertainty for Hubble residuals in our simulations in bins of $\Delta z=0.02$.  The left--hand panels show the photometrically--classified SNe~Ia, while the right-hand panels are for true SNe~Ia. Each row shows the cumulative effect of applying the various selection criteria discussed in Section~\ref{subsec:selection_cuts_sims}.}
\end{center}
\end{figure}

In Figure~\ref{con_bias_plot} we show the residuals about the Hubble diagram from our simulations (Section~\ref{subsec:selection_cuts_sims}); the Hubble residual (HR) is the difference between the observed and input distance modulus for each SN~Ia.  We have subtracted the true distance modulus, based on the input cosmology, from the measured distance modulus for each SN in the simulation, and binned the results in redshift.  The left panels of Figure~\ref{con_bias_plot} show the SNe~Ia photometrically--classified using the methodology outlined in Section~\ref{sec:algorithms}, while the right-hand panels shows residuals for the true SNe~Ia in the simulations.  Each row demonstrates the effect of adding one of our selection criteria, described in Section~\ref{subsec:selection_cuts_sims}. In the absence of any systematic bias, we would expect these residuals to be scattered about zero (blue line).

There are two sources of bias whose effects are easily seen in Figure~\ref{con_bias_plot}.  First, after making only the data-quality and PSNID cuts, our SN sample (second row, left panel) shows a strong bias towards positive HRs (under-luminous objects) at $z<0.2$.  We have shown in Table~\ref{cuts_contam} that this sample has high contamination, i.e., at $z\leq0.25$ non-Ia~SNe, though fainter than SNe~Ia, are still bright enough to be detected as SN candidates in the SDSS SN Survey. At higher redshifts core collapse SNe are too faint to be observed, and thus their contamination is naturally curtailed.  This bias at low-$z$, due to contamination, is effectively eliminated once the $X_1$--color cut is introduced. After the final color--magnitude cut (bottom panel) the HRs are consistent with zero bias at low redshifts.

\begin{figure}[!t]
\begin{center}
\epsfig{file=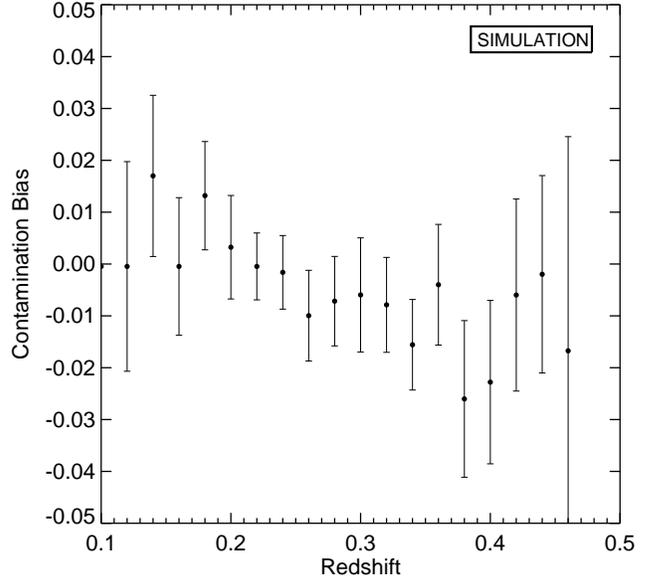,width=1.0 \linewidth}
\caption{\label{con_bias_photo_true} The difference between the weighted mean of the Hubble residuals from the photometrically~classified SNe~Ia and the true SNe~Ia, i.e., the difference between the bottom two panels in Figure 17. The error bars account for the correlations between the photometrically~classified SNe~Ia and the true SNe~Ia.} 
\end{center}
\end{figure} 

At high redshifts ($z>0.25$) the HRs in all panels of Figure~\ref{con_bias_plot} are biased low; i.e., SNe are, on average, brighter than expected given the input cosmology.  This selection effect is a combination of the classic Malmquist bias \citep{ malmquist:1936, wood-vasey:a} and other SALT2--related selection effects.  We have only analysed objects that would be classified as SN candidates in the SDSS-II SN Survey (survey pipeline cuts and post-survey cuts are applied within the simulations), so under-luminous objects at high-$z$ are never output from the simulations.  Additionally, though we place no implicit $S/N$ cut on our data, low $S/N$ SNe that are poorly fit by SALT2 are more likely to fail the $X_1-c$ cut, and our original data-quality cut, and are thus preferentially excluded (see Section~\ref{spec_subsample}).  

For clarity, we show in Figure~\ref{con_bias_photo_true} the difference between the recovered distance moduli for the photometrically--classified SNe~Ia and the true SNe~Ia as a function of redshift, i.e., the bottom left-hand panel in Figure~\ref{con_bias_plot} minus the bottom right panel in Figure~\ref{con_bias_plot}. This removes the Malmquist bias effect, allowing us to see the residual contamination bias as a function of redshift.  As one would infer from Figure~\ref{purity_eff_z}, the redshift-dependent bias of our HRs due to contamination is seen in Figure~\ref{con_bias_photo_true} to be a small effect. We find, using Akaike information criterion \citep[AIC; ][]{akaike:1974} to compare the two models, that a linear fit is favored over a redshift independent model, with a slope of -0.021 best describing the linear bias in the measured redshift range. This potential redshift dependence deviates from null by only $-0.01$ mag out to $z=0.5$. Thus the effect of the contamination is subdominant to the Malmquist bias, which has modeled uncertainties that are larger than the possible redshift dependence of the contamination bias.

The combination of these selection effects (the classical Malmquist bias and SALT2 effects) are by far the biggest source of bias affecting photometric SN sample if left uncorrected, which is discussed in detail in Section \ref{subsec:mal_bias}. However, we note that such effects, especially the classical Malmquist bias, are also present in spectroscopic samples where they are more difficult to correct for, as this bias will depend on the details of the spectroscopic program \citep{sako:2008a,kessler:2009a}. 

\subsection{Selection effects}
\label{subsec:mal_bias}

In this section we correct for the known selection bias at $z>0.25$ seen in Figure \ref{con_bias_plot}. This bias is a combination of the classical Malmquist bias and SALT2--related effects. For a magnitude limited SNe survey, the classical Malmquist bias results for a given $X_1$ and color in the preferential detection of SNe that appear brighter due fluctuations caused by Poisson noise and intrinsic scatter. We do not attempt to study these different effects separately in this paper and only provide a correction for their combined effect on the distance moduli of our photometric SNe~Ia sample. For this reason, we will from hereon refer to this combined selection effect as just the ``Malmquist bias", but ask the reader to recognise that it is a combination of magnitude (or $S/N$) effects, of which the classical Malmquist bias is likely the most important.

Before we present the details of our correction to the Malmquist bias, we first demonstrate the importance of this effect in our cosmological analysis.  We compute the equation-of-state of dark energy ($w$) for the entire sample of true SNe~Ia that pass data-quality cuts (top right panel of Figure~\ref{con_bias_plot}) using the publicly available software package {\it CosmoMC} \citep{Lewis:2002a}, including priors on the cosmological parameters (we detail our further usage of {\it CosmoMC}~in Section~\ref{subsec:cosmo_anal}).  We find the best-fit cosmology for this sample to be $w=-0.90\pm0.05$, which is inconsistent with the input cosmology of $w=-1$ (though only at the two--sigma level), and may indicate a bias.  We stress that this sample is, by definition, 100\% pure and 100\% efficient, yet still produces biased cosmological parameters.  

Therefore, to investigate the selection effects  bias, we have undertaken a new set of SNANA simulations that span a wider range in redshift than those used in our analysis of the sample contamination (Section~\ref{subsubsec:sims}).  We also increase the number of SNe in the simulation, as the Malmquist bias is small compared to the size of the error-bars and requires sufficient $S/N$ to characterize it. In detail, we use SNANA \citep[version v9$\_$97;][]{kessler:2009a} to create thirty thousand SNe~Ia over the redshift range $0.01\leq\ z \leq\ 0.5$. For consistency, the SALT2 SN parameters and the assumed cosmology are the same as those used for the public SNANA simulations described in Section~\ref{subsubsec:sims}. As we are characterizing an effect for SNe~Ia, we do not include core-collapse SNe in the simulation.  

\begin{figure}[!t]
\begin{center}
\epsfig{file=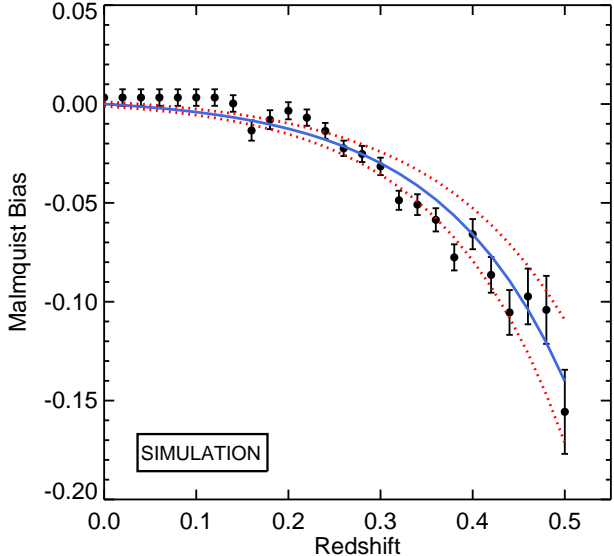,width=1.0 \linewidth}
\caption{\label{man_bias} The difference between the observed ($\mu_{\rm obs}$) and expected ($\mu_{\rm cosmo}$) distance modulus as a function of redshift for the simulations described in Section~\ref{subsec:mal_bias}. We show the best--fit exponential function to this data (blue line) and the one--sigma errors on the fit (red dashed line).}
\end{center}
\end{figure} 

In Figure~\ref{man_bias}, we show the weighted mean difference between the ``observed" ($\mu_{\rm obs}$) and true ($\mu_{\rm cosmo}$) distance modulus for our simulated SNe~Ia after applying our photometric--classification criteria to the data$\footnote{We apply all our data cuts to the simulated data except for the PSNID cuts, which require significant computational time for such a large suite of simulations. We have confirmed that excluding this cut does not effect our results using the public SNANA simulations since we are only interested in using true SNe~Ia for the Malmquist bias correction. Only an additional 7$\%$ of SNe~Ia in these simulations are removed by this cut. We have also demonstrated that the shape of the Malmquist bias is no different whether or not the PSNID cut is included.}$.  We use Eqn~\ref{mu} to calculate $\mu_{\rm obs}$, fixing $\alpha$ and $\beta$ to the values used in the simulation\footnote{Allowing $\alpha$ and $\beta$ to vary during the computation of $\mu$ does not significantly change our Malmquist bias correction, but does increase the statistical error.}.  There is little bias out to $z\simeq0.3$, but at higher redshifts a significant offset appears, growing with increasing redshift and becoming $\gtrsim0.1$ magnitude over-luminous.

We also show in Figure~\ref{man_bias} the best-fit analytical function to the Malmquist bias, which is an exponential of the form, 

\begin{equation}
\mu_{\rm corr}(z) = a e^{bz}+c,
\label{malmquist_corr}
\end{equation}

\noindent where $a=-0.004\pm0.001$, $b=7.26\pm0.31$, and $c=0.004\pm0.006$. We include this Malmquist bias correction in our distance modulus calculation (Eqn~\ref{mu}), resulting in a better estimate of the average distance modulus with redshift. This correction is based on simulations of the SDSS-II SN Survey, and thus is only valid for these data samples and selection criteria.

We note that the differences between the plot of $\Delta\mu$ in the bottom right panel of Figure~\ref{con_bias_plot} and that in Figure~\ref{man_bias} are the reflection of more than simply an increase in statistics. In the public SNANA simulations of K10a there existed an offset of 0.27 mag between the brightness of SNe~Ia created in the simulations and those observed in \citet{sullivan:2011a}, resulting in fainter SNe~Ia in the public simulations than are used in Figure~\ref{man_bias}. Thus the public simulations, while useful for our contamination analysis, should not be used for detailed cosmological calculations, although we have confirmed (with newer simulations) that Figures~\ref{sim_fom_prob} and \ref{sim_fom_chi} remain unaffected by this offset.  

The Malmquist bias we find is larger than has previously reported by other surveys.  This is primarily because we are pushing the SDSS-II SN survey to its limit of low $S/N$ observability; by $z=0.5$ only the very brightest SNe are observed. We note that the ESSENCE survey also found a significant Malmquist bias via their simulations, which they corrected for by adjusting the prior on the host galaxy extinction ($A_v$) as a function of redshift \citep{wood-vasey:a}.  This is an alternative method to adding an average redshift--dependent correction to each SN~Ia distance modulus, presented here.

\subsubsection{Testing the Malmquist bias}

The above Malmquist bias correction is computed using a particular set of cosmological parameters.  To determine whether this assumption biases our results, we have re-run our simulations with a set of widely varying input cosmologies, including $(\Omega_{\rm m},\Omega_{\Lambda},w)=(1,0,-1), (0.5,0.5,-1.5),$ and $(0.2,0.8,-0.8)$.  We find that the parameters describing the Malmquist bias correction in Eqn~\ref{malmquist_corr} do not change beyond their quoted statistical uncertainties. 

The value for $\alpha$ we have used in the simulations for deriving the Malmquist bias correction ($\alpha$=0.11) is much lower than what we recover from the data ($\alpha$=0.22; Section~\ref{subsec:SN+H0}).  We examine the effect this has on our results by determining the Malmquist bias correction from a new set of simulations that uses $\alpha$=0.22 in the input model.  We find the resulting Malmquist correction is consistent with what we previously derived; the $b$ parameter of Equation~\ref{malmquist_corr} is only changed by 0.14 (less than half the error), and the $a$ and $c$ parameters are unchanged.

To demonstrate the expected effect of the Malmquist correction on our photometrically-classified SN~Ia sample,  we draw ten subsamples of the same size and redshift distribution as our real data at random from photometrically--classified SNe~Ia in our simulations.  We derive the best-fit cosmology for these samples both with and without the Malmquist bias correction, again using {\it CosmoMC}.  For the uncorrected case we again find a best--fit value of $w=-0.87\pm0.03$, which is the weighted mean and uncertainty from the 10 randomly--drawn samples.  This result is significantly different from the input cosmology of $w=-1$, demonstrating the importance of the Malmquist correction, the application of which produces the expected best--fit value of $w=-1.00 \pm0.03$. 
 
\subsection{Host Galaxy Followup Bias}
\label{subsec:host_bias}
An additional source of potential bias is due to the spectroscopic follow--up program of SN host galaxies.  Our ancillary BOSS spectroscopy of likely SN~Ia host galaxies (Section~\ref{subsec:additional_ia_candidates}) had an apparent (fiber) magnitude limit of $r_{\rm fiber} = 21.2$. This should not cause an additional Malmquist bias, as the target selection is biased against fainter galaxies, not fainter SNe.  It is well known, though, that faint galaxies preferentially host only luminous (high $X_1$) SNe~Ia \citep{hamuy:1996b,gallagher:2005a}, and thus the target selection would appear to favor detection of {\it fainter} SNe~Ia.  However, it has recently been shown that SNe in massive galaxies tend to be {\it over-luminous} for their light-curve shape and color \citep{gallagher:2008a,kelly:2010,sullivan:2010a,lampeitl:2010b}.  These effects combine with our host-galaxy magnitude limit in a complicated manner that is not captured in our simulations, but which merits further study in the future.

\begin{figure}[!t]
\begin{center}
\epsfig{file=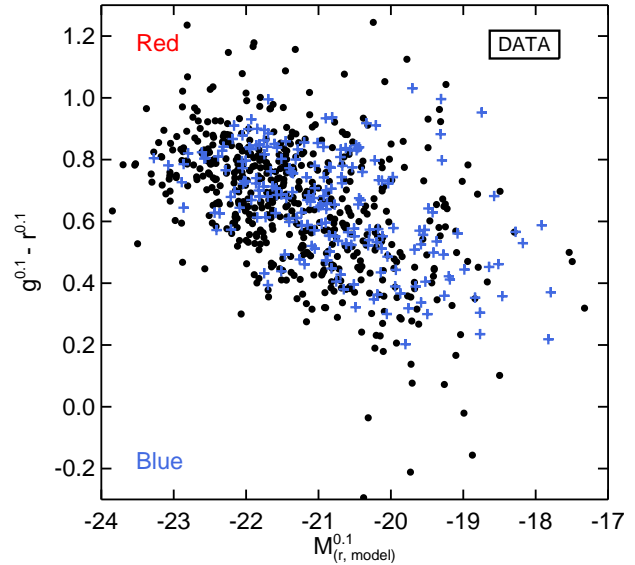,width=1.0 \linewidth}
\caption{\label{host_galaxy} The host--galaxy $g-r$ color as a function of absolute $r-$band model magnitude (both quantities k-corrected to $z=0.1$) for the host galaxies of the photometrically--classified SNe~Ia (black dots) and the subsample of SNe~Ia that have been spectroscopically--confirmed (blue cross symbols).}
\end{center}
\end{figure}

It is important to remember that despite these potential biases, photometric classification yields a less biased host-galaxy sample than our spectroscopic sample. In Figure \ref{host_galaxy} we show the color--magnitude diagram for the host galaxies in our BOSS sample, with the subsample of spectroscopically--classified SNe~Ia shown in blue. For each galaxy, we plot the $g-r$ color and the absolute $r$-band model magnitude, both of which have been k-corrected using the standard SDSS software \citep{blanton:2007}. As default, we quote all absolute magnitudes and the $g-r$ color at $z=0.1$. The host galaxies of the spectroscopic subset are, on average, fainter than the whole population of BOSS host galaxies. A Kolmogorov-Smirnov (KS) test of the k-corrected model $r-$band absolute magnitude distributions of the two galaxy samples (spectroscopically confirmed SN hosts and photometrically confirmed SN hosts) confirms the two distributions are not the same at a statistical significance of 99.9\%. The photometric sample includes intrinsically brighter host galaxies, which may be due to the increased volume sampled by the photometric sample as such luminous, massive galaxies are rare. It could also be a product of the SN spectroscopic follow-up avoiding the brightest hosts, as in these cases it is more difficult to separate the SN from the host galaxy light. However, there does not appear to be a bias in the $g^{0.1}-r^{0.1}$ model colors, which is reassuring.

\section{Hubble Diagram}
\label{sec:results}
\begin{figure*}[!]
\begin{center}
\epsfig{file=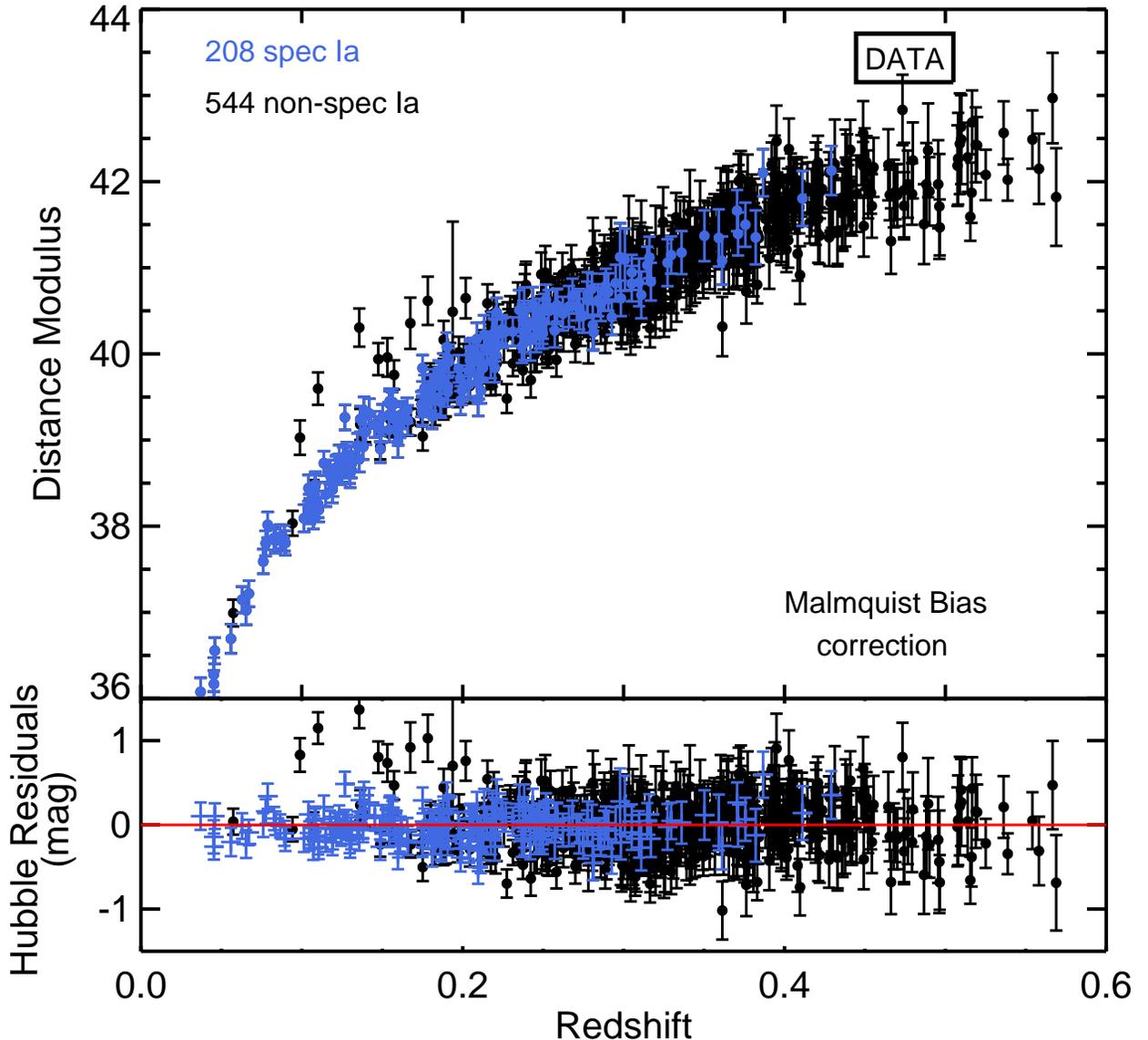,width=1.0\linewidth} 
\caption{\label{hubble_diagram_corrected} The Hubble diagram of the photometrically--classified SDSS-II SN~Ia sample. We have corrected for the Malmquist bias as discussed in Section~\ref{subsec:mal_bias}. We use the best--fit values of $\alpha$ and $\beta$ (see Section~\ref{subsec:cosmo_anal}) and assumed the same $M$ as in the simulations ($M$=29.8) when creating this Hubble diagram.  The SN intrinsic dispersion has been included in the error bars shown.  Blue points show the subsample of SNe~Ia that have been spectroscopically--confirmed as part of the SDSS-II SN Survey, while the black points only possess a photometric classification. The bottom panel shows the Hubble residuals of the data from the best--fit cosmology model (Section~\ref{subsec:cosmo_anal}).}
\end{center}
\end{figure*}

We show in Figure~\ref{hubble_diagram_corrected} the Hubble diagram for our final sample of \nofinal\ photometrically--classified SNe~Ia, derived from the SDSS-II SN Survey photometry and SDSS-III BOSS host--galaxy spectroscopy (Section~\ref{subsec:selection_cuts_data}).  In contrast to Figure~\ref{hubble_diagram}, we have now applied our Malmquist bias correction to the Hubble diagram as derived in Section~\ref{subsec:mal_bias}. We have not corrected our sample for host--galaxy mass correlations, as it is beyond the scope of this paper (see Section \ref{subsec:phot_sys}).

For comparison, we highlight in Figure~\ref{hubble_diagram_corrected} the subsample of \nofinalSpec\ SNe~Ia in our photometric sample that have spectroscopic confirmation from the SDSS-II SN Survey (shown in blue), and label this subsample ``spec Ia".  Therefore, \nofinalnoSpec\ of our photometrically--classified SNe~Ia are have no spectroscopic information at all, comprising 72.2\%\ of the sample. We note that only \nofinalphotoSpec\ of these \nofinalnoSpec\ SNe~Ia have been previously photometrically--classified, using host--galaxy spectra from the SDSS-I/II surveys \citep[\sako;][]{hlozek:2011a}.  The data for all SNe~Ia in our final sample can be found in Appendix~\ref{Appendix_data}. 

\subsection{Increased scatter}
\label{spec_subsample}

\begin{figure}[!t]
\begin{center}
\epsfig{file=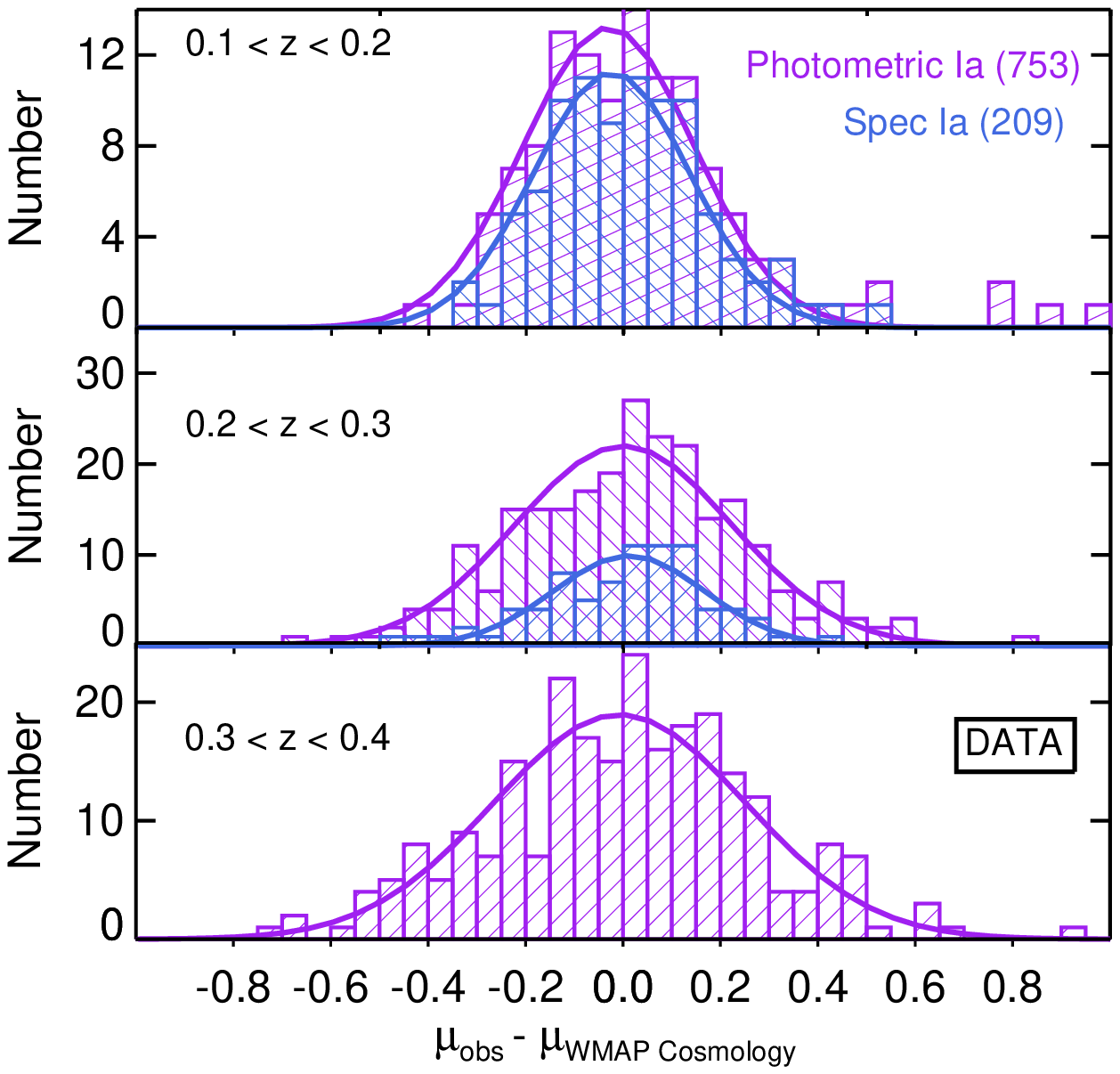,width=1.0 \linewidth}
\caption{\label{hubble_diagram_residuals_multi} The distribution of Hubble residuals as a function of redshift for the ``spec Ia'' subsample (blue) and full photometric sample of \nofinal\ SNe~Ia (purple).}
\end{center}
\end{figure}

The bottom panel of Figure~\ref{hubble_diagram_corrected} appears to show an increase in the scatter of the Hubble residuals for the photometrically--classified SNe~Ia (black points) compared to the spectroscopically--confirmed subsample (blue points). In Figure~\ref{hubble_diagram_residuals_multi}, we study this apparent increased scatter by comparing the distribution of Hubble residuals ($\Delta\mu=\mu_{\rm obs} - \mu_{\rm{WMAP}}$) in the ``spec Ia" subsample to our full photometric sample, assuming the latest WMAP+BAO+$H_0$ best-fit cosmological model \citep{Jarosik:2011}.  We show these residuals in three redshift bins of width $\Delta z=0.1$ over the redshift range $0.1<z<0.4$, which corresponds to the range of redshifts where these two sets of SNe~Ia significantly overlap. The blue histograms show the ``spec Ia" subsample (\nofinalSpec\ ), while the purple histograms includes the full \nofinal\ photometrically--classified SNe~Ia, i.e., blue plus black points from Figure~\ref{hubble_diagram_corrected}.

To quantify the trends we see in Figure~\ref{hubble_diagram_corrected} (increased dispersion of the photometric SNe~Ia with redshift), we fit the HR distributions with Gaussians, and report their full width at half maximum (FWHM) and centroids in Table~\ref{hubble_diagram_residuals_table}. We fit Gaussians to avoid our analysis being adversely affected by the small but noticeable tails in these distributions, which are likely non-Ia~SN contaminants and are clearly offset from the Hubble diagram (Figure~\ref{hubble_diagram_corrected}) at z$\simeq0.15$. 

In Table~\ref{hubble_diagram_residuals_table}, we see that the centroids of the best-fit Gaussians to both samples of SNe~Ia are consistent with zero, showing no bias in their HR distributions as a function of redshift.  However, the FWHM of the best-fit Gaussian does increase with redshift for the full photometric sample, and is additionally larger than the FWHM of the ``spec Ia" subsample. The quoted errors in Table~\ref{hubble_diagram_residuals_table} are given by {\tt GAUSSFIT} in IDL, and have been confirmed through bootstrap resampling of the distributions (with replacement). We do not report the FWHM for the high redshift bin of the ``spec Ia" subsample as it is unreliable due to small number statistics.

\begin{table*}[pt]
\begin{center}
\begin{tabular}{|c|c|c|c|c|}
  \hline
Redshift & Sample &\multicolumn{2}{c|}{Gaussian Fit}   & Number \\
  Bin & Type & FWHM & Centroid   & \\
\hline
0.1$<$z$<$0.2   & spec Ia  	 & 0.377$\pm$0.006 & -0.008$\pm$0.008  & 91\\
                         	  & photo Ia & 0.413$\pm$0.008  & -0.012$\pm$0.013  & 124\\
\hline
0.2$<$z$<$0.3	 & spec Ia 		& 0.366$\pm$0.010  & 0.010 $\pm$0.012  &  80\\
             		 & photo Ia 	 & 0.524$\pm$0.010  & 0.000$\pm$0.012  & 249\\ 
\hline
 	0.3$<$z$<$0.4		& photo Ia	& 0.610$\pm$0.016 & -0.005$\pm$0.016  & 251\\ 
			\hline
\end{tabular}	
\caption{Parameters of the best-fit Gaussian distributions to the data shown in Figure~\ref{hubble_diagram_residuals_multi}.}
\label{hubble_diagram_residuals_table}
\end{center}
\end{table*}

\begin{figure}[!t]
\begin{center}
\epsfig{file=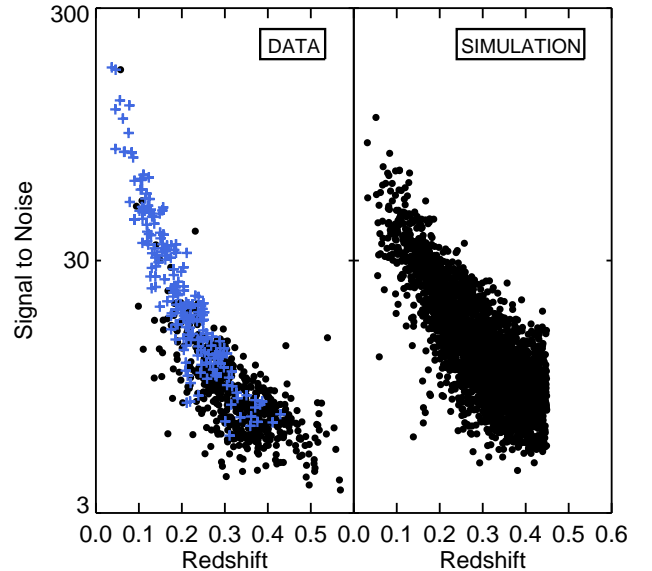, width=1.0 \linewidth}
\caption{\label{BOSS_SN_s_n_vs_z_log} Maximum observed $r$-band S/N (at any epoch) from both the observed (left--hand panel) and simulated (right--hand panel) light--curves, as a function of redshift.  The data sample is divided into SNe~Ia with (blue) and without (black) spectroscopic confirmation.}
\end{center}
\end{figure}

We investigate why this trend appears in the full photometric sample, but not in the ``spec Ia" subsample.  We plot in Figure~\ref{BOSS_SN_s_n_vs_z_log} the maximum $r$--band $S/N$ (at any epoch) for each SN~Ia in our observed (left panel) and simulated (right panel) photometrically--classified samples.  It is clear from the left panel of Figure~\ref{BOSS_SN_s_n_vs_z_log} that SNe in the ``spec Ia" (blue) subsample, at a given redshift, possess systematically higher $S/N$ light--curves than photometrically--classified SNe~Ia that weren't spectroscopically observed (black). The average $S/N$ for the ``spec Ia" subsample is 27.4, whereas the SNe~Ia with only photometric classification have an average $S/N$ of 9.6. This is of course expected, as the SDSS--II SN spectroscopic follow--up observations preferentially selected SN candidates that were easier to observe, naturally leading to a bias in $S/N$ for the spectroscopic sample. Thus the ``spec Ia" subsample has a smaller scatter because it contains the brightest SNe~Ia from the whole population, which are then easier to fit and thus produce tighter distance modulus constraints. At $z>0.3$, we see in Figure~\ref{BOSS_SN_s_n_vs_z_log} the emergence of an apparent detection limit at $S/N\simeq4-5$.  We have determined that this limit is due to the $X_1$-color cut (Section~\ref{subsubsec:SALT}), as the SALT2 parameters are not well-determined for such noisy light curves, frequently returning unphysical derived parameters which we then exclude.  

\section{Cosmology Analysis}
\label{subsec:cosmo_anal}

\subsection{Fitting issues}
\label{subsec:cosmoMC}

To allow consistent comparisons with SNLS, we use the same two methods to perform cosmological fits to our data as used in \citet{Guy:2010} and \citet{sullivan:2011a}.  First, we use a grid-based search technique when fitting simple cosmological models to just our photometrically-classified SN~Ia data. Specifically, we use the $simple\_cosfitter$ \citep{conley:2011a} software package, which computes the $\chi^2$ at every point in a regular grid (101 by 101) and converts those measurements to a probability via $P \approx exp (-\chi ^2/2)$, where the proportionality is set by normalizing over the grid. As in the SNLS analysis of \citet{Guy:2010} and \citet{sullivan:2011a}, we then marginalize over the SALT2 SN parameters ($\alpha$, $\beta$, $M$) to generate confidence contours for the cosmological parameters of interest. 

We also use the {\it CosmoMC}  \citep{Lewis:2002a} software package when fitting more complex cosmological models to our data (and simulations; see Section \ref{subsubsec:sims}) in combination with other cosmological information. This package uses the Markov Chain Monte-Carlo (MCMC) technique to efficiently probe multi-dimensional parameter space, allowing one to quickly investigate a large number of different regions in the parameter space. We have slightly modified {\it CosmoMC} to allow for the simultaneous fitting of both the cosmological parameters and the SALT2 SN parameters $\alpha$ and $\beta$, which define the standardization of SNe~Ia.  We include in the distance modulus calculation the redshift-dependent Malmquist bias correction (Eqn~\ref{malmquist_corr}) and the full SALT2 light--curve parameter covariance matrix. Finally, at each point in the MCMC chain, we determine $M$ (absolute magnitude at peak for SNe~Ia) for the value of $H_0$ at that step in the chain. This approach is the same as analytically marginalising over $M$, as outlined in \citet{bridle:2001a}, and is a similar methodology as used by SNLS \citep{sullivan:2011a}. This method is used in all our cosmological fits.

We execute our modified {\it CosmoMC} code using six chains in parallel to facilitate quick coverage of the large, multi--dimensional parameter space.  Each chain is started at a random location within our defined parameter space and typically converges after 50,000 to 100,000 steps. We assume that the MCMC has converged when $R-1<0.1$, where $R$ is the Gelman and Rubin statistic, i.e., when the variance within the chains is equal to the variance between chains \citep{an:1998a}.  

We provide in Table 4 a summary of the different combinations of priors we assume during our different cosmological fits discussed below. Throughout all our analyses we assume flat distributions for the priors on our SN parameters, i.e., $\alpha=(0.01,0.5)$ and $\beta=(1.0,5.0)$. We also assume flat distributions for the priors on $\Omega_{\rm dm}$ (density of dark matter) and $\Omega_b$ (density of baryonic matter) when these cosmological parameters are allowed to vary.  

Finally, we set the value of the intrinsic dispersion of SNe~Ia ($\sigma_{\rm int}$) to $\sigma_{\rm int}=0.12$ mag, which is then added in quadrature to all our SN errors.  Although this results in a reduced $\chi^2$ close to unity for all our cosmological fits, the best--fit value (i.e., delivers the reduced $\chi^2$ closest to one) for all our data is 0.16 mag. However, our simulations show that this larger value of $\sigma_{\rm int}$ is consistent with our small level of non-Ia~SN contamination$\footnote{We find that the intrinsic dispersion of our full sample drops to $\sigma_{int}=0.1$, and our cosmological results remain consistent, if we simply remove the 25 SNe~Ia that are located $>3\sigma$ away from the best-fit cosmological model. We do not recommend such a ``sigma-clipping" technique when using SNe~Ia to test cosmological models, but this test does illustrate the sensitivity of $\sigma_{int}$ to such outliers.}$, i.e., we input $\sigma_{\rm int}=0.12$ mag in our simulation, but measure $\sigma_{\rm int}=0.16$ mag for a photometrically--classified sample like ours. As $\sigma_{int}$ is used to explain unknown residual scatter in the SNe~Ia population, we feel it is appropriate to remove any extra scatter caused by non--Ia SN contamination from our `measured' value of $\sigma_{\rm int}$. We stress that this statement is not in contradiction with Section~\ref{spec_subsample}, where we state that the larger scatter we observe in our sample could be caused by lower S/N light curves, as such lower--quality data would increase both the scatter in the population and the observed errors bars on our SN distance moduli. However, contamination would increase the overall scatter in the SN Hubble residuals without necessarily increasing the distance uncertainties. It is worth noting that we also ran our cosmology fits with the $\sigma_{\rm int}=0.16$ mag and found no significant difference to our cosmological results.

We note that our assumed value of $\sigma_{\rm int}=0.12$ mag is still higher than that found in \citet[$\sigma_{\rm int}$=0.088 mag]{lampeitl:2010a} and in \citet[$\sigma_{\rm int}$=0.08 mag, SALT2 and SDSS-II SNe]{kessler:2009a}. These values were computed for the first--year SDSS SN Survey spectroscopic sample; limiting ourselves to the spectroscopically--confirmed subsample of our full photometric sample (Section 5), we still find a larger intrinsic scatter, $\sigma_{\rm int}=0.11$ mag. However, \citet{conley:2011a} found $\sigma_{\rm int}=0.10$ mag for the SDSS spectroscopically confirmed data, in agreement with our value. The apparent discrepancy may be caused by K09 and \citet{lampeitl:2010a} using an older version of SALT2 \citep{guy:2007a}, as \citet{conley:2011a} also uses the same newer version of SALT2 \citep{Guy:2010} used here.  As our simulations have shown that our selection criteria do not result in a bias between the input and best--fit value of $\sigma_{\rm int}$ for a pure (i.e., spectroscopic) sample of SNe~Ia, we have confidence in and continue to use throughout our value of $\sigma_{\rm int}=0.12$ mag. We stress that the assumed value of $\sigma_{\rm int}$, within the range discussed above, has little effect on the cosmological fits presented in this paper as the fits also have the freedom to adjust the values of other SN parameters, such as $\alpha$ and $\beta$.

\begin{table}
\centering
\begin{tabular}{|c|c|c|c|c|}
  \hline
    &  \multicolumn{2}{|c|}{ \it{Simple$\_$cosfitter}} &  \multicolumn{2}{|c|}{\it{CosmoMC}} \\
  \hline
  Parameter  & Set I & Set II & Set III  & Set IV \\ 
  \hline
  $w$ & -1 & -1 & -3,3 & -3,3 \\
   $\Omega_{\rm m}$ & 0.0, 1.5 & 0.0, 1.5 & - & - \\
$\Omega_{\Lambda}$ & -  & -0.5,2.5 & - & - \\
 $\Omega_{\rm k}$ & 0  & - & 0 & -1.5,1.5 \\
  $\Omega_{\rm dm}$  & -  & - & 0.0, 1.2 & 0.0, 1.2 \\ 
  $\Omega_{\rm b}$ & - & - & 0.0458  & 0.015, 0.200 \\
  $H_0$ & - & - & 50,100 & 50,100 \\

\hline
\end{tabular}
 \caption{Priors imposed on the fitted cosmological parameters in four different combinations (sets). The dash symbol in the table represents where we do not need to set priors for parameters, as they are constrained by a combination of other priors (e.g., $\Omega_{\rm m}$ is restricted by the priors on  $\Omega_{\rm dm}$ and $\Omega_{\rm b}$).}
\label{cosmo-priors}
\end{table}

\subsection{Our constraints on $\Lambda$CDM}
\label{subsec:SN_only}

We begin by studying the cosmological constraints obtained using only our SN~Ia sample, before combining with other data. For completeness, we provide a summary of all our cosmological fits, including the different combinations of data-sets and priors, in Table~\ref{data_priors_cosmo}.

\begin{figure*}[!t]
\begin{center}
\psfig{file=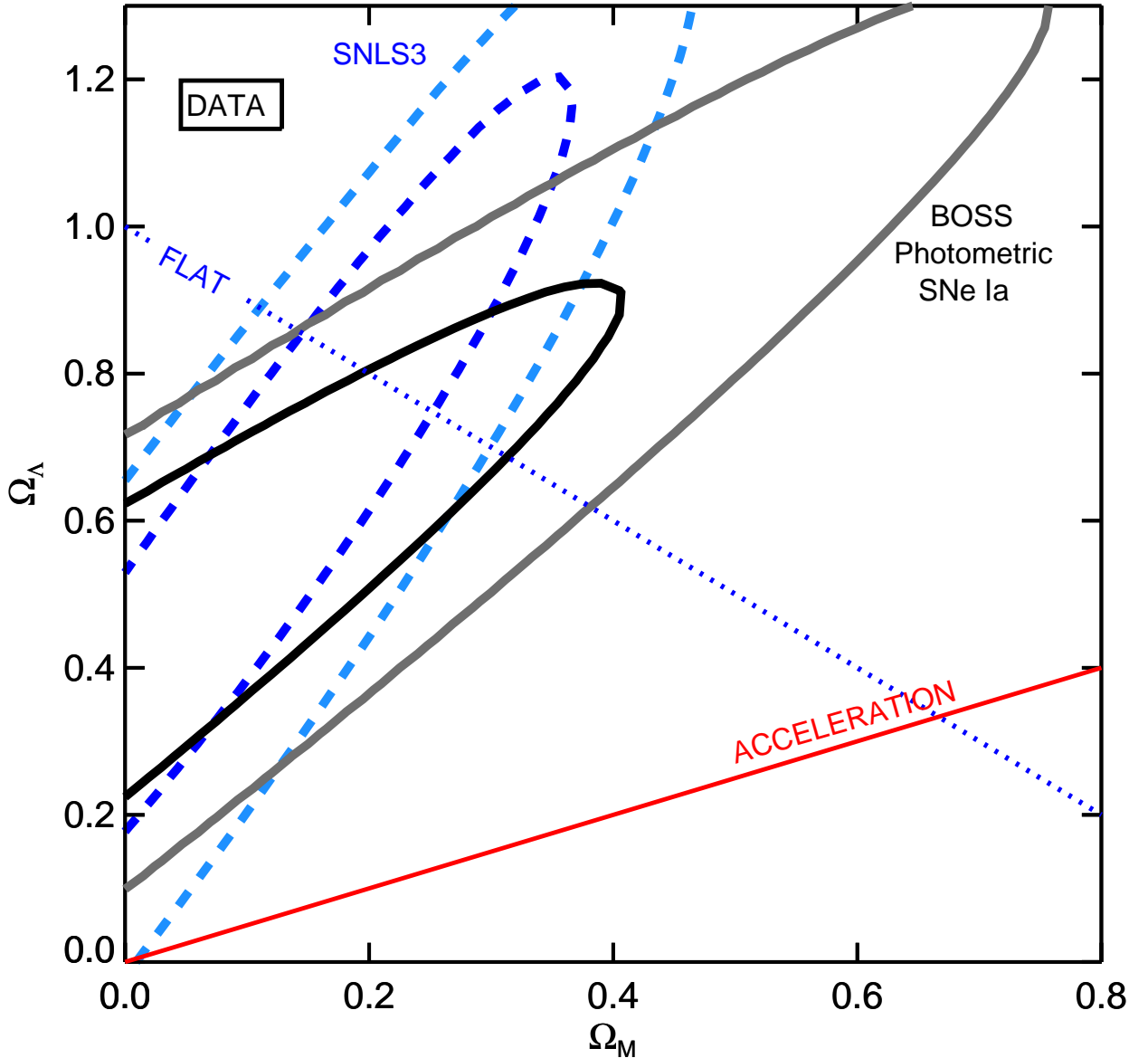,width=1.0 \linewidth}
\caption{\label{contour_cosmo_om_ol} The 68$\%$ and 95$\%$ confidence contours of $\Omega_{\rm m}$ versus $\Omega_{\Lambda}$ for a $\Lambda$CDM model using only our photometrically--classified SNe, with prior Set II (Table 5), and allowing curvature to vary. Only statistical errors in the contours are shown. The blue dashed contours show the comparable SNLS3 constraints taken \citet{Guy:2010}. }
\end{center}
\end{figure*}

\begin{table*}
\centering
\small
\begin{tabular}{|c|c|c|c|c|c|c|c|c|} 
\hline 
 \multicolumn{5}{|c|}{Data}  & \multicolumn{4}{|c|}{Results} \\
\hline
SNe & $H_0$ & CMB & LRGs & Priors & $\Omega_{\rm m}$ & $\Omega_{\Lambda}$ & $w$ & Figure\\ 
\hline
X & & &  & Set I &  $0.24^{+0.07}_{-0.05}$  & - & - & - \\
X & X & &  & Set III & 0.27$_{-0.16}^{+0.15}$ & 0.73$_{-0.15}^{+0.16}$ & -0.95$_{-0.32}^{+0.31}$  & \ref{contour_cosmo} \\
X & X & X & X & Set IV & 0.29$_{-0.02}^{+0.02}$ &0.71$_{-0.02}^{+0.02}$ & -0.96$_{-0.10}^{+0.10}$ & \ref{sullivan}\\
\hline
\end{tabular}
 \caption{Summary of the cosmological fits presented in Section \ref{subsec:cosmo_anal}}
\label{data_priors_cosmo}
\end{table*}

We begin by fitting the $\Lambda$CDM cosmological model ($w=-1$) to our photometrically--classified SNe~Ia using $simple\_cosfitter$ and prior Set I in Table~\ref{cosmo-priors}. Under the assumption of flatness we obtain a best--fit value of $\Omega_{\rm m}=0.24^{+0.07}_{-0.05}$ (statistical errors only). When we relax the prior on flatness (prior Set II), we obtain the (grey) confidence contours for $\Omega_{\rm m}$ and $\Omega_{\Lambda}$ in Figure~\ref{contour_cosmo_om_ol}. For comparison, we show similar constraints on these cosmological parameters using the three-year SNLS data (SNLS3) from \citet{Guy:2010}, which only includes 242 spectroscopically--classified SNe~Ia from SNLS.

Figure~\ref{contour_cosmo_om_ol} also demonstrates that our photometrically--classified SN~Ia sample alone is able to detect an accelerating universe (i.e., $\Omega_{\rm \Lambda} > \Omega_{\rm m}/2$).  Integrating over the whole parameter space in Figure~\ref{contour_cosmo_om_ol} (see table~\ref{cosmo-priors} for parameter ranges), we compute the probability of an accelerating universe given our data to be 99.96\% (statistical uncertainties only).

\subsection{Our constraints on $w$CDM}
\label{subsec:SN+H0}

We next fit for a flat, $w$CDM cosmological model using CosmoMC and the prior Set III in Table~\ref{cosmo-priors} \footnote{All other cosmological parameters are left at their default values (i.e., re-ionization optical depth, the primordial super--horizon power in the curvature perturbation on 0.05 Mpc$^{-1}$ scales, and the scalar spectral index).}. We fit this model to our sample of photometrically--classified SNe~Ia and the $H_0$ measurement of $73.8\pm2.4~\rm~km~s^{-1}~Mpc^{-1}$ from the recent ``Supernovae and $H_0$ for the Equation of State" sample \citep[SH0ES;][]{riess:2009,riess:2011}. This Gaussian prior on $H_0$ does not impact the cosmological constraints due to our marginalization over the absolute magnitude of SNe~Ia, but rather ensures that {\it cosmoMC}, which assumes a minimum age to the Universe, performs in a well--behaved manner. We note that within {\it cosmoMC}, $\Omega_{\rm m}$ is a parameter comprised of two components: $\Omega_{\rm b}$ and $\Omega_{\rm dm}$. Since the ratio of these two components is not constrained by SNe~Ia alone, we fix $\Omega_{\rm b}$ to the WMAP7 value of 0.0458 \citep{Jarosik:2011}, reducing the number of parameters to be constrained by the SNe data to two: $\Omega_{\rm dm}$ and w.

\begin{figure}[!t]
\begin{center}
\psfig{file=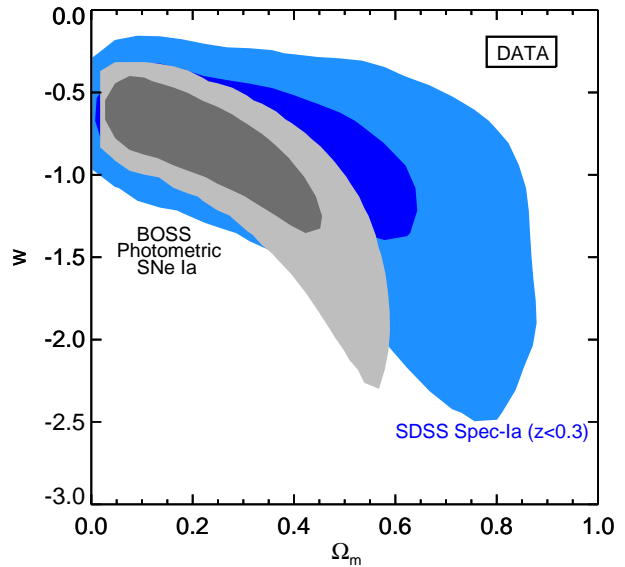,width=1.0 \linewidth}
\caption{\label{contour_cosmo} The 68$\%$ and 95$\%$ confidence contours of $w$ and $\Omega_{\rm m}$ in a flat $w$CDM model (assuming prior Set III in Table 5) for a combination of our photometrically--classified sample of \nofinal\ SNe~Ia and the SH0ES measurement of $H_0$ (only the statistical errors shown, grey). The blue contours show the equivalent constraints but using the 3-year sample of spectroscopically--confirmed SNe~Ia from the SDSS-II SN Survey (confined to z$<$0.3).}
\end{center}
\end{figure}

We find the best--fit values for this model to be $w=-0.95_{-0.31}^{+0.32}$ and $\Omega_{\rm m}$= $0.27_{-0.16}^{+0.15}$ (statistical errors only), and show in Figure~\ref{contour_cosmo} the joint 68$\%$ and 95$\%$ confidence intervals for this data. We also show the equivalent contours for the three--year sample of spectroscopically--confirmed SNe~Ia from the SDSS-II SN Survey. For this comparison, we include all 306 confirmed SNe~Ia (regardless of whether they are part of our photometric sample or not) at a redshift of $z<0.3$, below which any selection bias in the spectroscopic sample should be minimized. This comparison demonstrates that using photometric instead of spectroscopic classification, which increases the size of the sample by a factor of 2.5 and extends the redshift range, results in a reduction of the area of the confidence contours by a factor of 1.6. We stress that this simplistic comparison does not constitute a detailed analysis of the full 3--year spectroscopic sample of the SDSS-II SN Survey, which will be presented elsewhere. We also find that limiting our photometric SNe~Ia sample to $z<0.3$ gives consistent cosmological constraints with the samples plotted in Figure~\ref{contour_cosmo}, but with slightly larger uncertainties.

During this analysis, we simultaneously solve for the best--fit values of the SALT2 SN parameters. In our fit of the $w$CDM model, we find $\alpha=0.22_{-0.02}^{+0.02}$ and $\beta=3.12_{-0.12}^{+0.12}$. Our fitted value for $\beta$ is in agreement with previous analyses of the SDSS data \citep{lampeitl:2010a,marriner:2011a,conley:2011a}. However our value of $\alpha$ appears higher than previous analyses; \citet{lampeitl:2010a} found $\alpha$=0.16$\pm$0.03, and \citet{marriner:2011a} found $\alpha$=0.131$^{+0.05}_{-0.04}$. One potential explanation of this difference could be the non--Ia SNe contamination in our photometric sample, but this has been tested in our simulations:  we recover the input $\alpha$ and $\beta$ both with and without the expected level of contamination detailed in Section~\ref{sec:algorithms}. A more plausible explanation appears to be the higher average $S/N$ of the spectroscopically--confirmed SNe~Ia; using only the subset of spectroscopically--confirmed SNe~Ia in our photometric sample, we find $\alpha$=0.16$\pm$0.02, in agreement with previous results. 

We explored whether HR outliers (which we do not automatically clip) might be affecting the derived value of $\alpha$.  We subtracted $\alpha*X_1$ from the HR of each SN in our sample and plotted these as a function of $X_1$, finding four clear outliers. Removing these objects from our sample and refitting the cosmology, we found the resulting cosmological contours to be unchanged but the derived SALT2 parameters to both be lower ($\alpha$=0.18, $\beta$=2.79).  As this $\alpha$ is still larger than what is found in previous studies, the outliers cannot be solely responsible for this discrepancy, and regardless does not result in biased cosmological constraints.

\subsection{Constraints from combining data-sets and comparison with SNLS}
\label{subsec:combined}

Finally, we determine cosmology constraints with our photometrically--classified sample of \nofinal\ SNe~Ia combined with cosmological information from the power spectrum of Luminous Red Galaxies (LRGs) in the SDSS DR7 \citep{reid:2010a}, the full WMAP7 CMB power spectrum \citep{larson:2011a}, and the SH0ES $H_0$ measurement.  The SH0ES $H_0$ measurement is partially determined using nearby SNe~Ia measurements, and thus to be fully consistent we would have to consider the covariance between this value of $H_0$ and our SNe~Ia measurements. However, as we are assuming no prior information on $M$ in our treatment of intrinsic SN parameters, these measurements can be considered independent. Furthermore, the uncertainty in $M$ is a subdominant systematic uncertainty to the derived value of $H_0$ \citep{riess:2011}.
	
We fit this combination of data using $CosmoMC$ for a non-flat $w$CDM cosmology, using the priors listed as Set IV in Table~\ref{cosmo-priors}. With the addition of these external data-sets, we can now relax our priors on the re-ionization optical depth ($\tau$=[0.00, 0.50]), the primordial super--horizon power in the curvature perturbation on 0.05 Mpc$^{-1}$ scales ($\log A$=[0,30]), and the scalar spectral index ($n_{\textrm{s}}$=[0,1.5]).

 We find the best--fit value for the equation--of--state of dark energy using these data is $w=-0.96_{-0.10}^{+0.10}$, with $\Omega_{\rm m}$= 0.29$_{-0.02}^{+0.02}$ and $\Omega_{\rm k}$=$0.00_{-0.01}^{+0.01}$ (statistical errors only). We also find a best--fit value of $H_0$=67.97$_{-2.25}^{+2.28}$ (stat) $\rm{km~s^{-1}~Mpc^{-1}}$. These cosmological constraints are summarized in Table~\ref{data_priors_cosmo}.

\begin{figure}[!t]
\begin{center}
\psfig{file=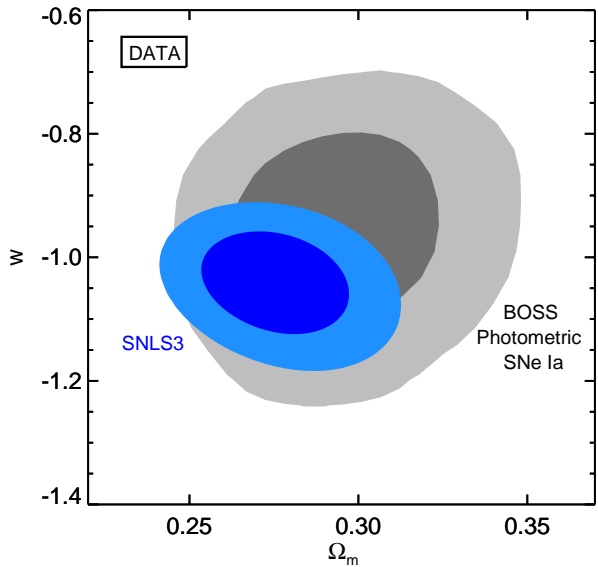,width=1.0 \linewidth}
\caption{\label{sullivan} The 68$\%$ and 95$\%$ confidence contours of $w$ and $\Omega_{\rm m}$ in a $w$CDM model from the spectroscopic SNLS3 sample \citep[][blue]{sullivan:2011a} and our photometrically--classified SDSS-II SN~Ia sample (grey) using prior set IV in Table 5. Both sets of contours include external data from the CMB, LRGs, and the SH0ES $H_0$ measurement.  Contours represent statistical uncertainties only.}
\end{center}
\end{figure}

Figure~\ref{sullivan} shows our joint confidence contours for $w$ and $\Omega_{\rm m}$ in comparison with similar SNLS3 constraints from \citet{sullivan:2011a} using the same combination of external data--sets (CMB, LRGs, and the $H_0$ SH0ES).  SNLS3 only uses spectroscopically--classified SNe~Ia, collected from the following SN data-sets: 242 SNe~Ia from SNLS; 123 low-redshift SNe~Ia from the literature \citep[primarily][]{hamuy:1996a,riess:1999a,jha:2006a,hicken:2009a, contreras:2009a}; 14 high redshift SNe~Ia from HST \citep{riess:2007a}; and 93 SNe~Ia over the redshift range $0.06\leq z \leq0.4$ from the first--year SDSS SN Survey \citep{kessler:2009a}. The inclusion of the SDSS SN data means there are some SNe~Ia in common between these analyses.  There appears to be good agreement in these two sets of constraints, although there is a small offset in the best--fit values for $\Omega_m$ between the two analyses: $\Delta \Omega_m$=0.018, significant at less than $1\sigma$. The best-fit values of $w$ differ by $\Delta w$=0.080 between the two analyses, which is again consistent within the quoted uncertainties. One possible explanation for these small differences is the fact that we have not corrected for the correlation with host--galaxy stellar mass, as discussed in Section 7. Overall this comparison is reassuring, considering the lower redshift leverage of the SDSS-II SN sample ($z<0.55$) and the lack of spectroscopic confirmation used herein.  These results demonstrate the potential for photometrically--classified SN~Ia samples to be used to improve cosmological constraints in the future.

\section{Discussion}
\label{sec:discuss}

\subsection{Systematic Uncertainties in Photometric Classification}
\label{subsec:phot_sys}

The goal of this paper has been to illustrate the power of photometrically--classified SNe~Ia in delivering competitive cosmological results when compared to spectroscopically--confirmed samples.  As such, we have paid particular attention to the sources of uncertainty unique to our methodology of SN classification, and to the SDSS-II SN Survey in general, i.e., contamination from non-Ia~SNe and Malmquist bias.  We have not undertaken an exhaustive study of systematic uncertainties as exemplified by \citet{kessler:2009a} and \citet{conley:2011a}.  

In Section~\ref{sec:bias}, we studied the effect of non-Ia~SN contamination of photometric samples on cosmological constraints.  The low predicted contamination rate (\finalcontam) in our sample has an insignificant effect on the best--fit cosmological parameters in simulations; compared to a completely pure sample of SNe~Ia we find a bias in the equation--of--state of only $w\simeq0.007$ (using prior Set III in Table~\ref{cosmo-priors}). 

We have also studied the systematic offset associated with the Malmquist bias effect (Section~\ref{subsec:mal_bias}). This correction, as a function of redshift, is included in all our cosmological fits.  We also investigated allowing the three parameters in the Malmquist bias exponential parameterization (Eqn 4) to vary in the cosmological fit within the error bars; the cosmological constraints remained identical. Correlations between the three parameters in the Malmquist bias fit have not been investigated in this paper.

We do not attempt to correct for the known correlation between the SN~Ia Hubble residuals and the properties of their host galaxies; SNe~Ia in massive galaxies are over-luminous even after light-curve corrections. Our sample of host galaxies span a range in absolute magnitude of approximately $-24 < M_r^{0.1} < -18$ (Figure 20), which corresponds to a stellar mass range of $\simeq10^9$-$10^{11}$ M$_{\odot}$, or a predicted difference of $\Delta\mu\simeq0.1$ \citep{lampeitl:2010b}.  We additionally expect that the magnitude limit imposed on our host--galaxy selection could cause a bias by preferentially selecting more massive (more luminous) host galaxies at higher redshift, thus preferentially selecting over-luminous SNe~Ia.  Unfortunately, the underlying physical mechanism that drives this correlation remains unclear \citep[e.g.,][]{DAndrea:2011,Gupta:2011}, and as such the mass-HR relationship could also be subject to a redshift dependence. Therefore, we do not correct for this effect here, but note that it will be essential for future studies.  We note that introducing a correction term for host-galaxy properties is no more difficult in photometrically-classified SNe~Ia samples than in those obtained from spectroscopic follow up, so it is a common problem for all future SN surveys.

A systematic uncertainty that is unique to photometrically--classified SN~Ia samples like ours, based on host galaxy spectroscopy, are errors in associating the SNe with the correct host galaxy. As described in Appendix~\ref{Appendix_z_compare}, we found only one mismatch between the SDSS-II SN spectroscopic subsample and our photometrically--classified SNe~Ia, one mismatch between the SDSS-II host galaxy spectroscopy and BOSS host galaxy spectroscopy, and giving an error rate of only 0.6$\%$.  We have removed this object, as well as another likely matching error discussed in Section~\ref{subsec:selection_cuts_data}, from our photometric sample.  We cannot rule out the possibility that other SN-galaxy pairs have been incorrectly matched, although based on the rate found in the spectroscopic subsample we expect the number to be low and thus have a negligible effect on our cosmological fits.

Finally, our estimation of systematic effects due to Malmquist bias and contamination relies on the assumption that our simulations accurately represents the final sample after selection cuts.  However, some implicit assumptions in the simulation may not be valid. For example, we assume that all candidates are SN~Ia or non-Ia~SNe, even though some photometrically--classified candidates may in fact be another type of transient (e.g., AGN). We also assume that the 41 non-Ia templates reflect a complete sample of non-Ia SNe, and that the non-Ia properties are redshift independent. These assumptions may be inadequate, as discussed further in Section~\ref{improvements} and \citet{kessler:2010a}.

The magnitude of these systematic biases on our results are either small compared to our statistical uncertainties or unknown, and as such we do not include them in the error budget of our derived cosmological parameters in Table~\ref{data_priors_cosmo}.

\subsection{Future Improvements to Photometric Classification}
\label{improvements}

We have provided in this paper a procedure for photometrically--classifying SNe.  Of course, other methods also exist, with their own relative advantages and disadvantages.  For example, rather than applying limits on light-curve properties in 2-D parameter space (e.g., Figure~\ref{sim_x1_color_ellipse}), one could apply a nearest-neighbor algorithm to look for clustering in higher dimensional parameter space.  Alternatively, we note that the analysis presented here excludes from our sample all transients that have $P_{\rm Ia}$ computed to be below a hard threshold (see Section~\ref{psnidcriteria}).  One could choose instead to retain all SN candidates in their cosmology analysis, weighting each candidate by its Bayesian probability ($P_{\rm Ia}$) of being a SN~Ia. This approach avoids the uncertainty of choosing the optimal $P_{\rm Ia}$ threshold to obtain a ``clean" sample of SNe~Ia, and prevents the removal of actual SNe~Ia and the information that they provide. These methods -- the nearest-neighbor algorithm and the fully Bayesian method using the BEAMS algorithm \citep{hlozek:2011a} -- are currently being investigated using the SDSS-II SN candidates (and their BOSS host-galaxy redshifts).

Although we have optimized the selection cuts implemented in this paper, utilizing higher-order criteria could further improve the accuracy and efficiency of photometric classification. For example, the ellipsoidal $X_1$--color cut could be derived with an additional rotation parameter that accounts for the correlations between the two SALT2 light--curve parameters. 

There is also room for improvement in the simulations we use to describe our observations and measure the efficacy of our selection criteria. The SN~Ia templates included in the simulations are missing a number of subclasses, such as 02cx-like SNe, which have similar light curves to but are fainter than normal SNe~Ia \citep{li:2003a, phillips:2007a, mcclelland:2010a}; 06bt-like SNe, which are particularly problematic \citep{foley:2010a, stritzinger:2011a}; and super-Chandrasekhar mass SNe~Ia \citep{howell:2006a, scalzo:2010a}. The simulations also include only a limited number of non-Ia~SN templates, a deficiency which is sure to be improved in the future as the quantity of observations of these objects begins to reflect their diversity.  

At present, the only transient objects included in these simulations are SNe (Ia, II, and Ibc). In the future we would like to include other transient objects that are known contaminants in SN surveys, such as AGN.  Therefore, when we quote an estimated contamination of \finalcontam, this could be underestimated as we have only included contamination from non-Ia~SNe. Furthermore, errors on assigning the correct SN host redshift were not modeled in our simulations.  This is a rather complicated effect, as it depends on both the distribution of SNe within galaxies and the luminosity function of galaxies (those with and without SNe) as a function of redshift.  While we believe we have removed possible mis-identified host galaxies from our data (Appendix~\ref{Appendix_z_compare}), it would be interesting to model whether the expected number of such misidentifications is consistent with our findings.

In this work we have assumed a constant value for the intrinsic dispersion of our sample. However, \citet{kessler:2012} have recently shown that the intrinsic dispersion of SDSS and SNLS SN data may be better described as a wavelength-dependent function. Therefore, future analyses will want to include a more complex model than is currently standard for the intrinsic dispersion to improve their cosmological constraints.

We note that the relative rates as a function of redshift of SNe~Ia \citep[based on][]{dilday:2008a} and non-Ia~SNe \citep{bazin:2009a} assumed in the simulations still have associated uncertainties.  It is certain that these constraints will be improved in the near future by the next generation of large, deep SN surveys.

Finally, in this paper we have used a simple redshift-dependent correction for the Malmquist bias. In the future, higher-order Malmquist bias corrections could be investigated, such as stretch-dependent or color-dependent corrections. The latter of these may be important, as Figure~\ref{color_z} shows there is a clear bias in the recovered color distribution with redshift. The ESSENCE survey used a color-dependent Malmquist bias correction, adjusting the prior on the host galaxy extinction ($A_v$) as a function of redshift \citep{wood-vasey:a}.  This method would be interesting to explore with our photometrically-classified SNe~Ia.

\subsection{Prospects for Future SN Surveys}
\label{future}
The DES Supernova Survey \citep{Bernstein:2011} should start by the end of 2012 and run for at least five seasons. It is expected to measure high-quality light curves for $\sim4000$ SNe~Ia out to a redshift of $z\approx1.2$, for which real--time spectroscopy of every SN~Ia candidate, as done previously, will be impractical. The SNLS, SDSS and ESSENCE surveys  combined used over a year of telescope time (4 and 8-m class) to spectroscopically confirm fewer than 1000 SNe~Ia \citep{foley:2009a, howell:2009a}.

Therefore, DES will need to use photometric classifications to reduce the burden of real--time spectroscopic confirmation and use allocated spectroscopic resources wisely, e.g., targeting hostless SNe for spectroscopic confirmation or building up training samples of SNe for photometric classifiers. Our work suggests the need for obtaining spectroscopy of SN host galaxies which, as was the case of SDSS-II, can be done over a longer period of time and can be coordinated with other science goals \citep{lidman:2012}.

Photometric classification is well suited for obtaining large, uniformly-selected samples of SNe~Ia.  In light of the systematics discussed in Section~\ref{subsec:phot_sys} and Appendix~\ref{Appendix_OtherSys}, there is a great scientific benefit in having such samples that can be subdivided and analysed a number of different ways to determine the magnitude of such systematic effects.  We describe below two examples of how these samples will be useful.

SN lensing is the increase in observed flux from a SN due to lensing by the structure the light passes through on its journey through the universe.  \citet{Clarkson:2011} has discussed how the number of galaxies we observe along the line of sight to a SN should be correlated with the over-luminosity of the SN (voids play a role in this determination as well).  Thus grouping large numbers of SNe~Ia by the amount of foreground structure is vital for measuring this effect; such a program is currently being investigated with this sample (Smith et al. in prep).

The correlation between SN~Ia host-galaxy properties (mass, star-formation rate, metallicity) and SN~Ia HRs {\rm after} light-curve and color corrections are made is an important discovery for SN~Ia cosmology \citep{gallagher:2008a,hicken:2009b,kelly:2010,sullivan:2010a,lampeitl:2010b,Gupta:2011,DAndrea:2011}. Using spectroscopically-confirmed SDSS-II SNe, \citet{lampeitl:2010b} showed that SNe~Ia are $\sim$0.10 mag intrinsically over-luminous in passive hosts, and also found that parameterization using stellar mass gives an improvement of $\sim$4$\sigma$ on the cosmological constraints. This effect, however, is most likely not driven by the host mass; using low-$z$ SDSS-II SNe~Ia, \citet{DAndrea:2011} found intrinsically over-luminous SNe occur in high-metallicity galaxies at $>$3$\sigma$. The recent SNLS3 cosmology analysis \citep{sullivan:2011a} is the first major SN study to include host-galaxy corrections in a full cosmology analysis. 

Our SDSS-II photometric SN~Ia sample is much larger than any previously analysed sample, and the improved statistics and reduced bias may lead to an improved understanding of this effect.  Correlations with photometrically-derived stellar mass and spectroscopically-derived metallicities and star-formation rates are typically degenerate, but a large sample will allow one to hold these parameters constant while allowing only one to vary.

\section{Conclusions}
\label{sec:conclusion}
In this paper we use the full three-year photometry from the SDSS-II SN Survey, together with BOSS spectroscopy of the host galaxies of transients, to create a photometrically-classified sample of SNe~Ia to be used for cosmology.  Our main results are: 

\begin{itemize}
\item We have created a homogeneous sample of \nofinal\ photometrically-classified SNe~Ia; the largest collection of SNe~Ia ever selected from a single photometric survey. Our sample spans a redshift range of $0.05<z<0.55$ and contains \nofinalnoSpec\ newly classified SNe~Ia; \nofinalSpec\ SNe~Ia in our sample had been previously spectroscopically confirmed and another \nofinalphotoSpec\ had been photometrically--classified using SDSS-II host galaxy spectra. Based on SNANA simulations, we estimate that this sample is 70.8\% efficient at detecting SNe~Ia, with a contamination of only 3.9\% from core--collapse SNe.  We demonstrate that this level of contamination is negligible when estimating constraints on cosmological parameters using this sample of SNe~Ia.

\item Malmquist bias and SALT2--related effects are the largest systematic selection uncertainties in our photometrically--classified SNe~Ia sample. We estimate the combined size of these effects using extensive SNANA simulations, which show that they can be $>0.1$ magnitudes at the high--redshift limit of our sample ($z=0.5$; Figure~\ref{man_bias}). We use our simulations to correct for these biases and show we can then recover the correct cosmological model input into the simulations. 

\item We show that the ``spec Ia" subsample of \nofinalSpec\ SNe~Ia is potentially biased, as they possess higher $S/N$ light curves than the majority of our photometrically--classified SNe~Ia (Figure~\ref{BOSS_SN_s_n_vs_z_log}). Furthermore, there is evidence in Figures~\ref{x1_z} and \ref{color_z} that the ``spec Ia" subsample is biased to brighter (higher stretch) and slightly bluer (lower color values) SNe~Ia than the whole population. The weighted means of the SALT2 parameters for the ``spec Ia" sample are $X_1= 0.033\pm0.015$ and $c=-0.021\pm0.002$, compared to $X_1=-0.017\pm0.013$ and $c=-0.018\pm0.002$ for the sample as a whole. It is additionally clear from Figure~\ref{host_galaxy} that the host galaxies of the ``spec Ia" subsample are biased towards fainter (in absolute magnitude) galaxies compared to the whole population of host galaxies. These biases may be related, but we have not investigated this possibly herein.

\item We present the corrected Hubble diagram for our photometrically--classified SN~Ia sample in Figure~\ref{man_bias}. The extra scatter seen in this diagram at high redshifts is likely caused by our sample including SNe~Ia with lower $S/N$ than are found in the spectroscopically--confirmed SNe~Ia subsample (based on our simulations). We then fit this Hubble diagram with a straightforward $\Lambda$CDM ($w=-1$) cosmological model to obtain $\Omega_{\rm m}=0.24^{+0.06}_{-0.05}$ (statistical errors only) for a flat Universe. If we relax the constraint on flatness we obtain the constraints on $\Omega_{\rm m}$ and $\Omega_{\Lambda}$ shown in Figure~\ref{contour_cosmo_om_ol}, where we have detected an accelerating Universe at the 99.96\% confidence level. This figure also shows that our statistical constraints on these important cosmological parameters are comparable to the recent SNLS three-year constraints published in \citet{Guy:2010}.

\item In Figures~\ref{contour_cosmo} and \ref{sullivan}, we show our constraints on the equation--of--state of dark energy ($w$) for our photometrically--classified sample on its own (with only H$_0$ data) and when combined with other cosmological information (CMB, LRGs, H$_0$). We find $w=-0.95_{-0.31}^{+0.32}$ and $w=-0.96_{-0.10}^{+0.10}$ respectively (statistical errors only), which are consistent with both the SDSS-II and SNLS three-year spectroscopic samples. These cosmological analyses illustrate that our photometrically-classified sample can deliver competitive constraints even though it lacks extensive SN spectroscopic follow-up and probes a smaller range of redshifts (compared to the SNLS3). 

\end{itemize}

Creating a photometric SN~Ia sample is a fundamentally different task from creating a spectroscopic sample.  There are various equally-valid approaches that can be taken into account in designing a SN~Ia photometric--classification algorithm depending on the purpose to be fulfilled, each resulting in a sample of different size and composition.  As our focus is on obtaining useful cosmological constraints, this required prioritizing a highly pure sample (few non-Ia~SNe included) over a highly efficient one (few SNe~Ia excluded).  Had our intention been to study, for example, SN rates, then the selection pressures driving our sample construction would have been different.  A focus on cosmology and thus purity will necessitate the sort of strict cuts on photometric properties that we apply.

Our classification is based on the technique of \sako, which computes the Bayesian probability of each SN subtype based on the fit of the data to templates and models (Appendix~\ref{Appendix_PSNID}).  However, on its own this method does not perform nearly as efficiently as it does when a prior on the redshift exists for each SN candidate (Olmstead et al. in prep.).  We undertook an ancillary program using BOSS to obtain host-galaxy redshifts for a large number of SN candidates; this forms a crucial part of our paper.  We have argued that obtaining the resources to spectroscopically observe the majority of SNe~Ia in future surveys is not feasible. This is because SN spectroscopy is a highly time sensitive (observing windows of a few weeks) and scattered (low density per solid angle per unit time) undertaking.  For this reason, it is also subject to significant selection biases.  Host-galaxy spectroscopy allows the observer to obtain the redshift of each object much more efficiently: multi-object spectrographs can be used to sample the higher spatial density of targets at scheduled dates long after the SNe have faded away.  For this reason we believe host spectroscopy will remain a vital component of SN surveys in the future.

It is not desirable for future surveys to abandon real-time spectroscopy completely; it will remain necessary to identify subtypes of SNe~Ia, train classifiers, and study detailed properties.  It is even possible for spectroscopic samples with sizes and redshifts ranges exceeding that of SNLS and SDSS-II to be created.  However, this method could not possibly achieve the volume of SNe~Ia identification possible with photometric classification.  And although statistical uncertainties today are quite small, sample size is very important for understanding systematic uncertainties (Appendix~\ref{Appendix_OtherSys}).  Future surveys will have to understand SN lensing, host-galaxy correlations, intrinsic color, evolution, and other effects that have the potential to bias cosmological constraints.  Large samples that allow a complicated parameter space to be explored will thus be necessary.  For all these reasons, photometric--classification as an underpinning of SN cosmology is here to stay.  

\acknowledgements We thank the referee, whose diligent but friendly review greatly improved the quality of this paper. We thank Mark Sullivan for many helpful discussions during the lifetime of this project and analysis. We thank Julien Guy for sending us the data plotted in Figure~\ref{contour_cosmo_om_ol}, and Alex Conley for his advice with the $simple\_cosfitter$. We thank Michael Wood-Vasey and Raffaele Flaminio for helpful discussions on earlier drafts of the paper. Additionally we would like to thank Alex Kim and Rick Kessler for their useful inputs on advanced draft of the paper. We thank Marisa March for her comments on grid versus MCMC search methods. We thank Emma Beynon for her advice with $ComsoMC$, Rob Crittenden and David Bacon for useful discussions on statistics.  Numerical computations were done on the Sciama High Performance Computer (HPC) cluster, which is supported by the ICG, SEPNet and the University of Portsmouth.  We thank Gary Burton for his help in using Sciama HPC.  

Funding for the SDSS and SDSS-II has been provided by the Alfred P. Sloan Foundation, the Participating Institutions, the National Science Foundation, the U.S. Department of Energy, the National Aeronautics and Space Administration, the Japanese Monbukagakusho, the Max Planck Society, and the Higher Education Funding Council for England. The SDSS Web Site is \verb9http://www.sdss.org/9.  The SDSS is managed by the Astrophysical Research Consortium for the Participating Institutions. The Participating Institutions are the American Museum of Natural History, Astrophysical Institute Potsdam, University of Basel, Cambridge University, Case Western Reserve University, University of Chicago, Drexel University, Fermilab, the Institute for Advanced Study, the Japan Participation Group, Johns Hopkins University, the Joint Institute for Nuclear Astrophysics, the Kavli Institute for Particle Astrophysics and Cosmology, the Korean Scientist Group, the Chinese Academy of Sciences ({\small LAMOST}), Los Alamos National Laboratory, the Max-Planck-Institute for Astronomy ({\small MPIA}), the Max-Planck-Institute for Astrophysics ({\small MPA}), New Mexico State University, Ohio State University, University of Pittsburgh, University of Portsmouth, Princeton University, the United States Naval Observatory, and the University of Washington.

Funding for SDSS-III has been provided by the Alfred P. Sloan Foundation, the Participating Institutions, the National Science Foundation, and the U.S. Department of Energy. The SDSS-III web site is \verb9http://www.sdss3.org/9.  SDSS-III is managed by the Astrophysical Research Consortium for the Participating Institutions of the SDSS-III Collaboration including the University of Arizona, the Brazilian Participation Group, Brookhaven National Laboratory, University of Cambridge, University of Florida, the French Participation Group, the German Participation Group, the Instituto de Astrofisica de Canarias, the Michigan State/Notre Dame/JINA Participation Group, Johns Hopkins University, Lawrence Berkeley National Laboratory, Max Planck Institute for Astrophysics, New Mexico State University, New York University, Ohio State University, Pennsylvania State University, University of Portsmouth, Princeton University, the Spanish Participation Group, University of Tokyo, University of Utah, Vanderbilt University, University of Virginia, University of Washington, and Yale University.

\appendix

\twocolumngrid

\section{A. Photometric Supernova Identification}
\label{Appendix_PSNID}

In this paper we make use of the Photometric SuperNova IDentification (PSNID; \sako) software to obtain a typing of SN candidates using only the photometry of each object.  PSNID has been shown to perform better than other photometric--classification algorithms, scoring the highest Figure-of-Merit in the recent classification challenge of \kesslera.  This methodology has been applied to the full 3-year data release of the SDSS-II SN Survey and will be presented in full in \prepSako.

We refer the reader to \sako\ for a full description of PSNID, but describe here some of the key features of the algorithm which are relevant to this paper. PSNID uses the observed multi-color light curve to calculate a reduced $\chi^2$ fit to a grid of SN~Ia light-curve models and non-Ia~SN templates, assigning probabilities for each SN using the Bayesian evidence criteria.  PSNID returns the SN~Ia Bayesian evidence ($E_{\rm Ia}$), given by,

\begin{align}
E_{\rm Ia} = \int &~&P(z,A_{\rm v},T_{\rm max},\Delta m_{\rm 15,B},\mu)  e^{-\chi^{2}/{2}}  \nonumber \\
&~& dz  dA_{\rm v}  dT_{\rm max}  d\Delta m_{\rm 15,B}  d\mu,
\label{evidence1}
\end{align}

\noindent where the redshift prior $P(z)$ for an externally constrained redshift $z_{\rm ext}$ is given by,

\begin{equation}
P(z) = \frac{1}{\sqrt 2 \pi \sigma_{\rm z}} e^{{ -(z-z_{\rm ext})^{2}}/{2\sigma^{\rm 2}_{\rm z}}},
\label{prob}
\end{equation}

\noindent and $\sigma^{\rm 2}_{\rm z}$ is the error on the observed (external) redshift.

The $E_{\rm Ia}$ is computed by marginalizing the product of the likelihood function and the prior probabilities over the model parameter space.  Five parameters are included in the model: redshift ($z$), host-galaxy extinction ($A_{\rm v}$), time of maximum light ($T_{\rm max}$), the amount by which a SN Ia declined in the B-band during the first fifteen days after maximum light ($\Delta m_{\rm 15}$, \citealp{phillips:1993a}), and the distance modulus ($\mu$). Flat priors are assumed on $A_V$, $T_{\rm max}$, and $\Delta m_{\rm 15}$. Galactic extinction is corrected for using $A_{\rm V}$ from \citet{Schlegel:1998}, assuming the \citet{Cardelli:1989} extinction law with R$_{\rm V}= A_{\rm V}/E(B-V)=3.1$.  The color law for SN host galaxies, however, is assumed to be steeper, with R$_{\rm V}=2.2$ \citep{kessler:2009a}. 

The non-Ia~SNe Bayesian evidences ($E_{\rm Ibc,II}$) are given by,

\begin{equation}
E_{\rm Ibc,II} = \sum\limits_{\rm templates} \int P(z)  e^{-\chi^{2}/{2}} dz dA_{\rm v} dT_{\rm max} d\mu.
\label{evidence_non}
\end{equation}

\noindent This quantity is the summation over a variety of Type Ibc and Type II templates, as given in Table~1 of \sako. PSNID returns a Bayesian probability for each of the three SN types considered herein, given by

\begin{equation}
P_{\rm type} = \frac{E_{\rm type}}{E_{\rm Ia}+E_{\rm Ibc} + E_{\rm II}},
\end{equation}
where, by definition, $P_{\rm Ia}+P_{\rm Ibc}+P_{\rm II}=1$. 

We use in this paper a version of PSNID that has incorporated several improvements over that presented in \sako.  The primary difference is in the light--curve template uncertainties (Equations~6~\&~7 from \sako); these have been changed to better match the observed distribution of errors in spectroscopically--confirmed SDSS-II SNe.  Specifically, the magnitude errors $\delta m$ on the SNe~Ia $gri$ model light curves are now 

\begin{equation}
  \delta m_{\rm{Ia}} = \left\{ \begin{array}{ll}
    0.06 + 0.04 \times (|t|/20) & |t| < 20~\mathrm{days}, \\
    0.10 + 0.18 \times ((|t|-20)/60) & |t| \ge 20~\mathrm{days},
  \end{array} \right.
\end{equation}

\noindent while the new magnitude errors for the non-Ia~SN light--curve templates are

\begin{equation}
  \delta m_{\rm{CC}} = 0.06 + 0.08 \times (|t|/60),
\end{equation}

\noindent where $t$ is the rest-frame epoch in days from B-band maximum.  These changes require us to re-determine goodness-of-fit thresholds used in \sako\ on PSNID, which we describe in Section~\ref{psnidcriteria}.

In Section~\ref{subsec:additional_ia_candidates} we obtain a preliminary classification of our SDSS-II SN candidates by running PSNID on the $ugriz$ light-curves while placing a flat prior on the redshift.  This classification shapes our subsequent target list for BOSS host-galaxy spectroscopy, and is the only occasion in this paper where we use a flat redshift prior. For both our simulated SNe (Section~\ref{psnidcriteria}) and our SN candidates that have host-galaxy spectra from BOSS we use the spectroscopic redshift as our prior (Section~\ref{subsec:selection_cuts_data}).  This additional prior on the redshift helps compress the parameter space being investigated and break degeneracies between non-Ia SNe at low-$z$, which could appear similar to higher redshift SNe~Ia.  As in \sako, we assume flat priors for all other PSNID model parameters.  

\section{B. Spectral Adaptive Lightcurve Template}
\label{Appendix_SALT}

We determine the distance moduli for our SN candidates (both data and simulated) using the SALT2 light-curve fitting software \citep[version 2.2;][]{Guy:2010}.  For each SN light--curve the SALT2 fitting program determines the best--fit value of three parameters ($X_0$, $X_1$, $c$), which describe the observed luminosity offset, stretch, and color of the SN, respectively \citep{guy:2007a,Guy:2010}. These fitted values are then used to ``standardise" the light curve, as the distance modulus to each SN is calculated using,

\begin{equation}
\mu = m_{\rm *B} -M +\alpha x_{\rm 1} -\beta c,
\label{mu}
\end{equation}

\noindent where $m_{\rm *B}$ is the $B$--band peak apparent magnitude and is defined as $-2.5\log_{10}(x_{\rm 0})$. Parameters $\alpha$, $\beta$ and $M$ (absolute B--band magnitude at peak) are constants that can either be derived for the whole sample simultaneously with the best--fit cosmology, or can be constrained from other data. In our cosmology analysis presented in Section~\ref{subsec:cosmo_anal} we allow $\alpha$ and $\beta$ to float within priors and analytically marginalize over $M$ (which is degenerate with $H_0$). 

We fit (in flux) SALT2 to the SDSS SMP light-curve data in the $griz$ passbands.  Although the SALT2 template does not extend to the rest-frame $z$--band, we include $z$--band data in our fits as it should help constrain the model at higher redshifts.  However, the $z$--band has low throughput in the SDSS, and we obtain consistent light--curve fits with and without including this information. We exclude $u$--band photometry due to its low $S/N$, but we note that this does not significantly effect the quality of our fits, as this data is of much lower quality than in $griz$. 

\begin{figure*}[!t]
\begin{center}
\epsfig{file=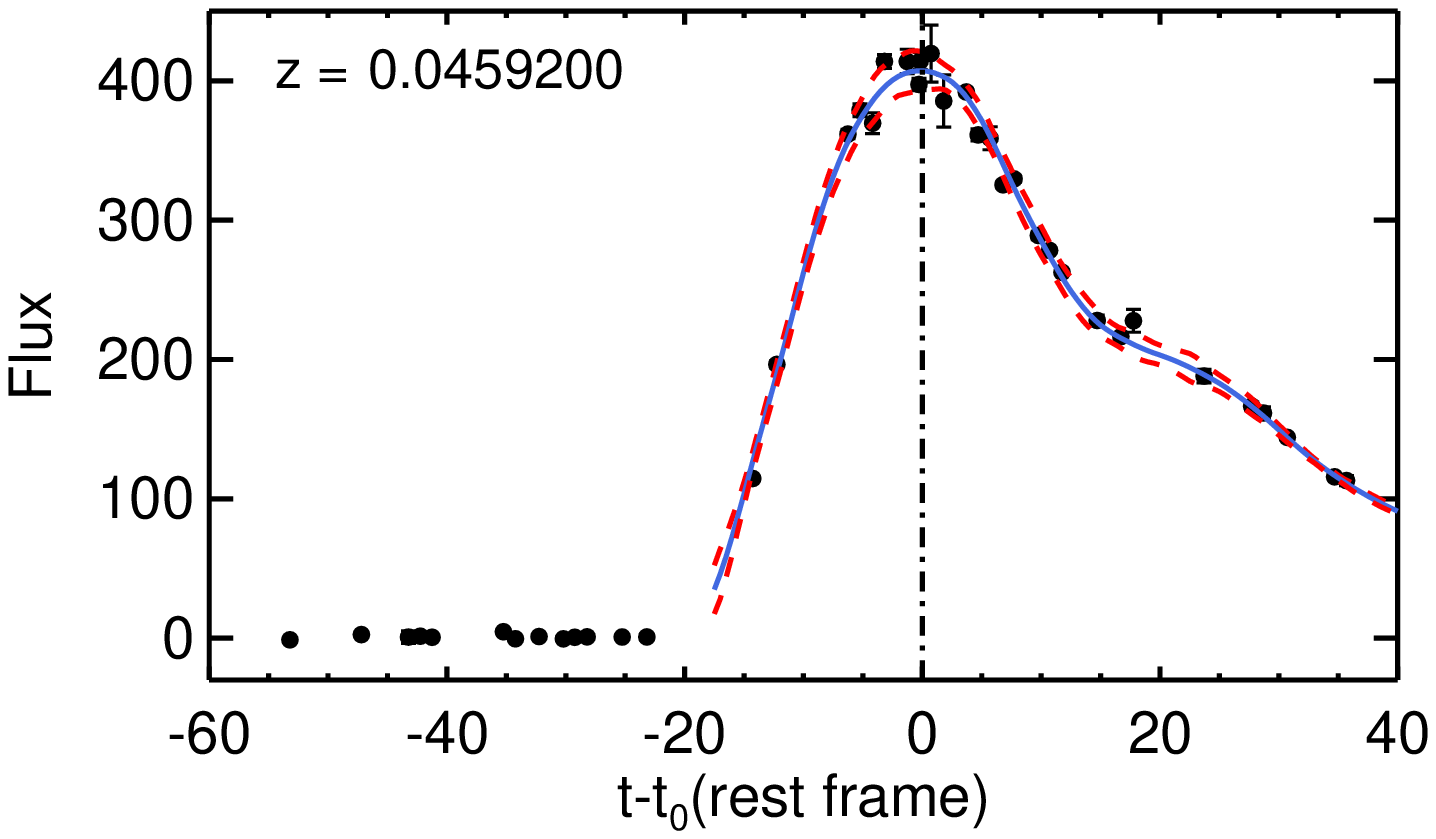 ,width=0.40\linewidth} \epsfig{file=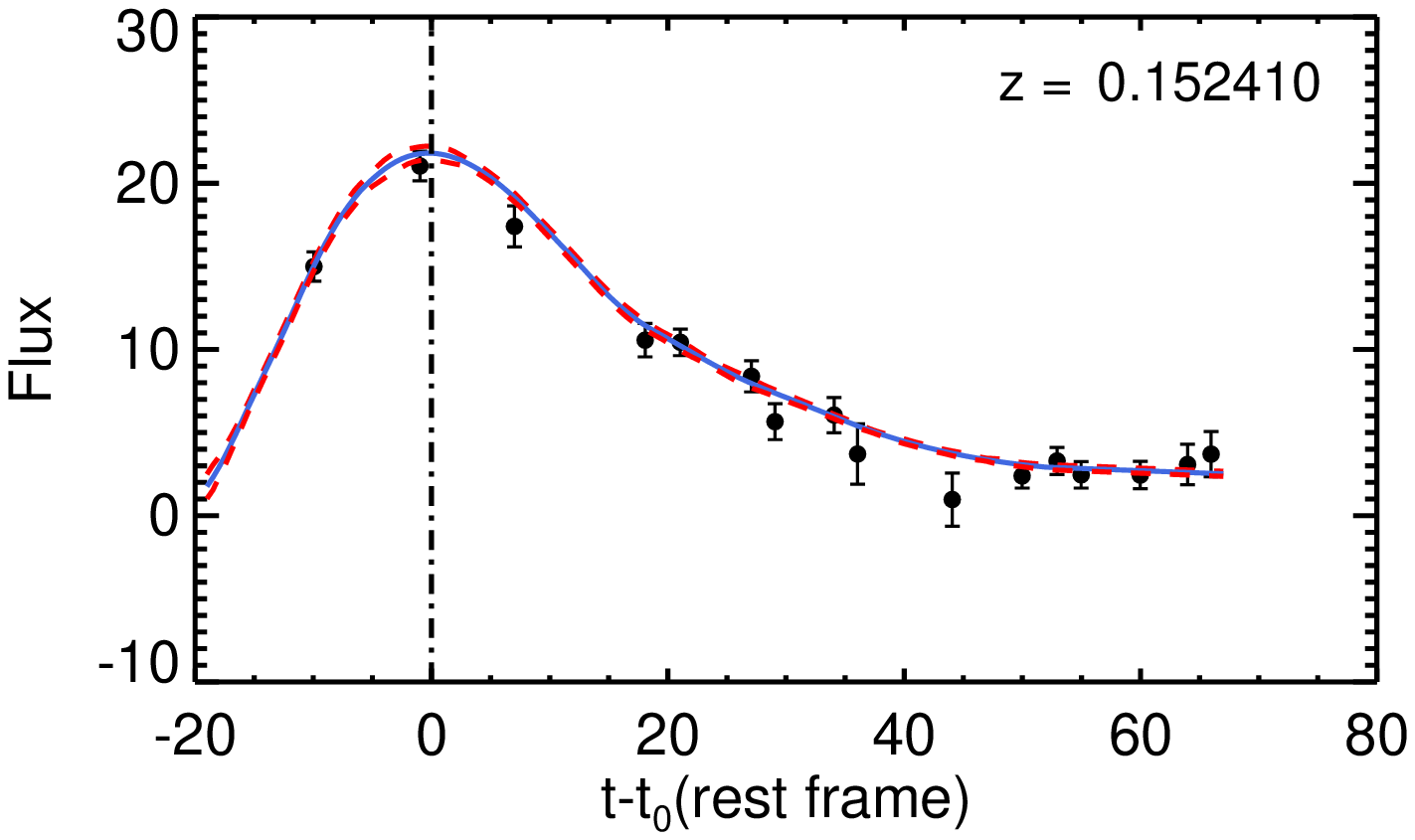,width=0.40\linewidth}
\epsfig{file=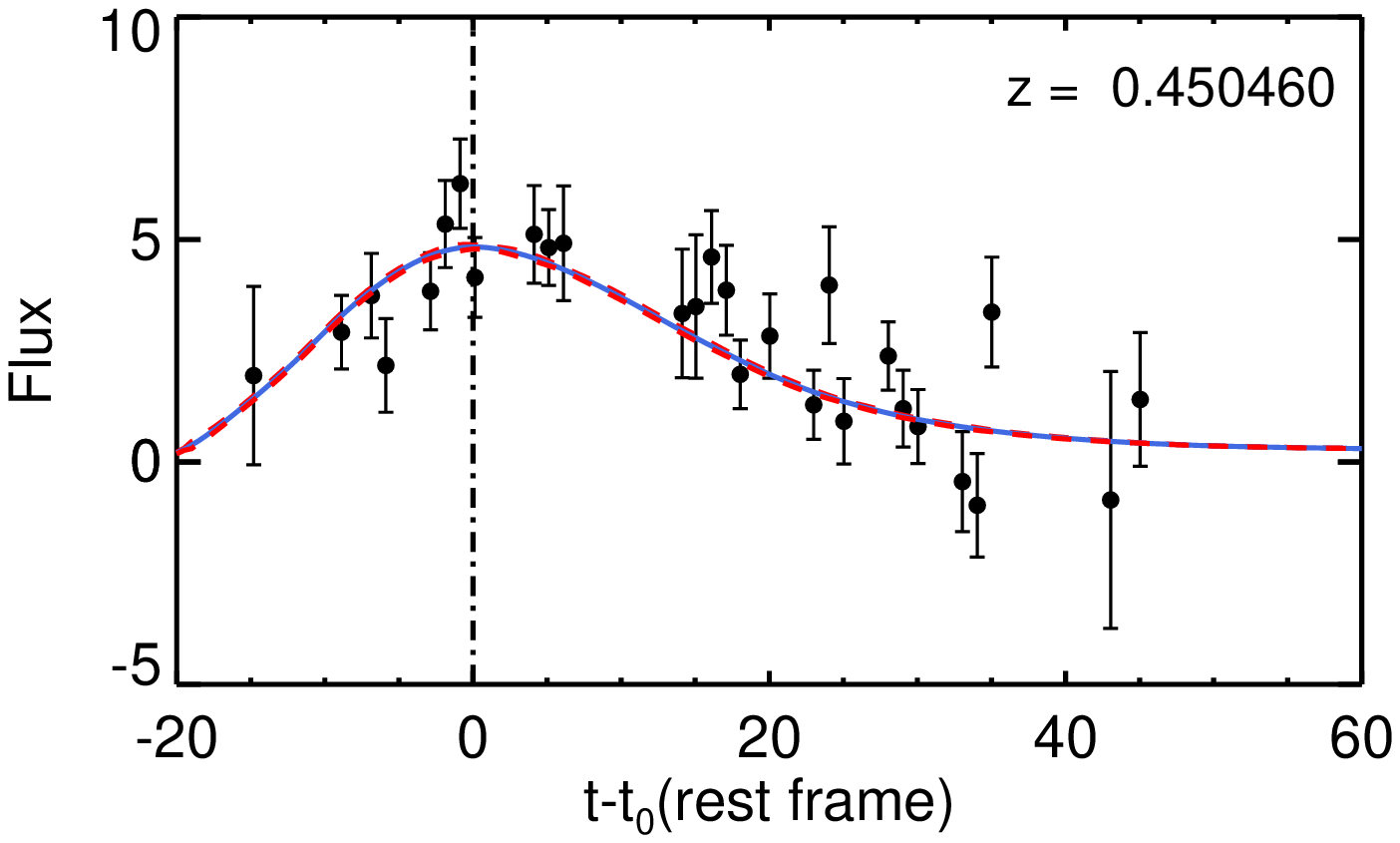,width=0.40\linewidth} \epsfig{file=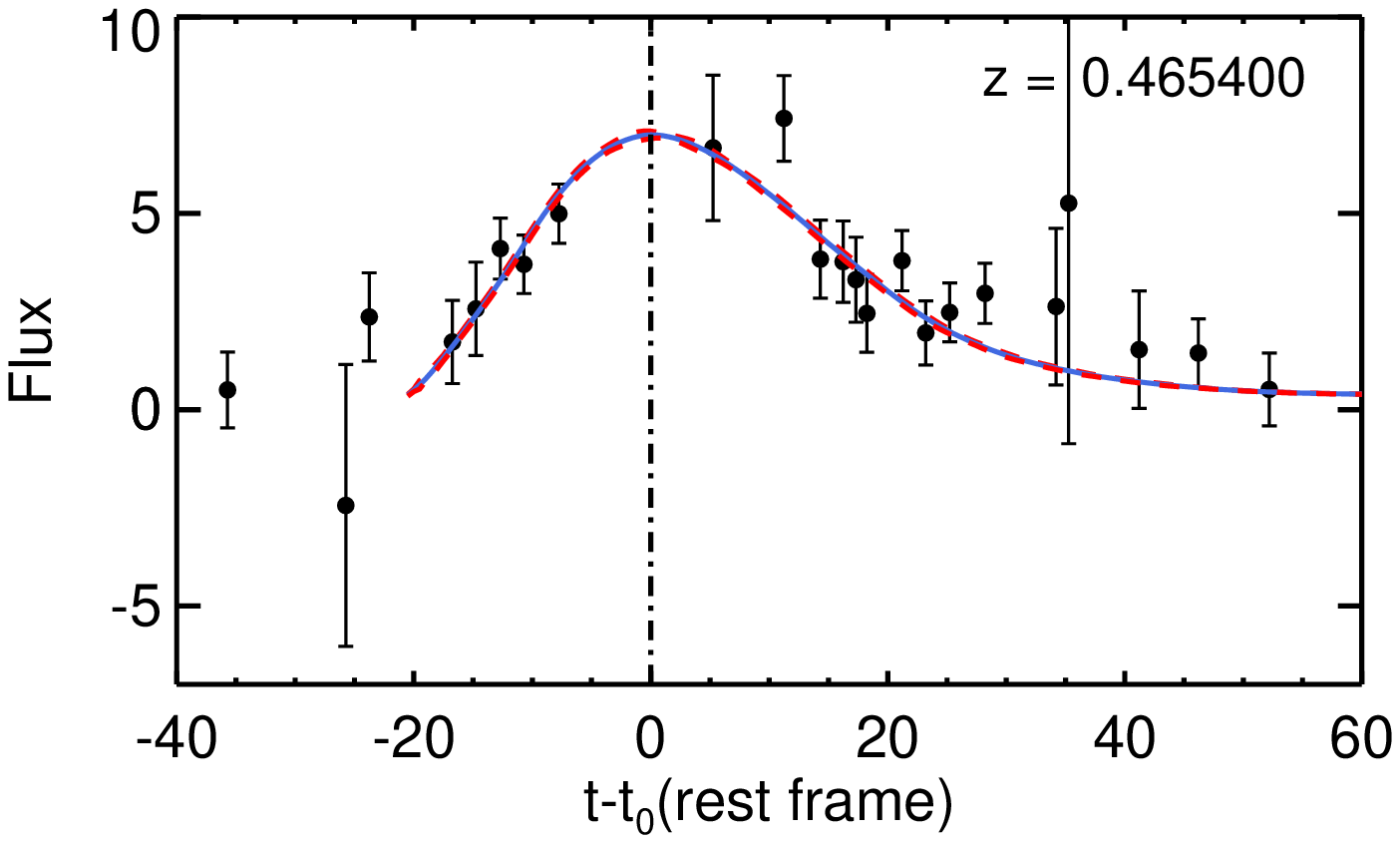 ,width=0.40\linewidth}

\caption{\label{lc} The $r$--band light-curves fitted using SALT2 for the SNe whose host galaxy spectra are shown in Figure~\ref{spectra}. The blue solid curve is the best--fit SN~Ia model light curve, and the red dashed lines represent the 1$\sigma$ uncertainties on this fit. The vertical dotted line shows the best--fit time $t_0$ of peak brightness. }
\end{center}
\end{figure*}

In Figure~\ref{lc}, we present the SDSS--II light curves for the four SN candidates whose BOSS host--galaxy spectra are shown in Figure~\ref{spectra}. We also present the SALT2 best--fit SN~Ia model and the one--sigma error on this fit provided by SALT2. 

\section{C. Redshift comparison}
\label{Appendix_z_compare} 

\begin{figure*}[!t]
\begin{center}
\epsfig{file=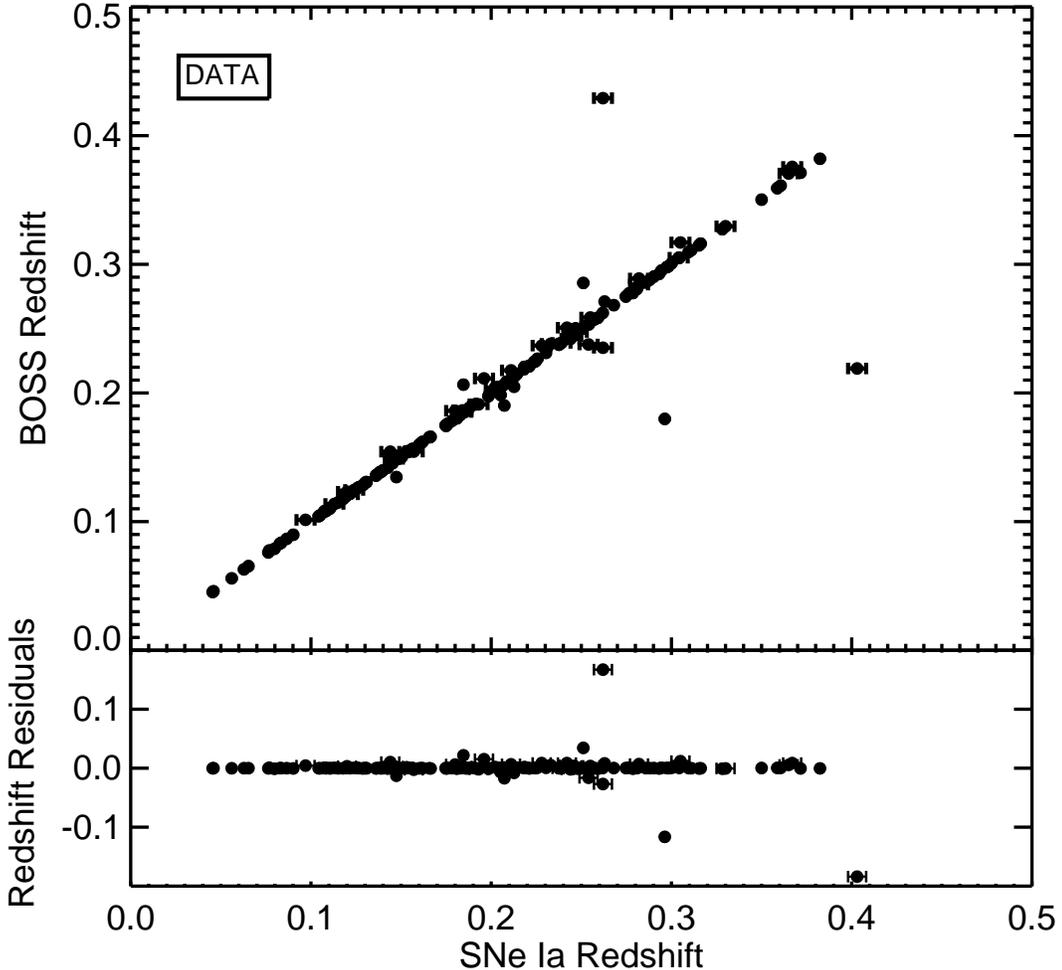 ,width=0.85\linewidth} 
\caption{\label{boss_sn_z} Comparison of the spectroscopic redshift for our spectroscopically--confirmed SNe~Ia and the corresponding host--galaxy redshift from BOSS. The bottom panel shows the redshift residuals (BOSS galaxy redshift - SNe Redshift) for this sample.}
\end{center}
\end{figure*}

In Figure \ref{boss_sn_z} we compare the BOSS host--galaxy redshift and the SN spectroscopic redshift for the 186 spectroscopically--classified SNe~Ia that pass our selection cuts.  In only four cases do these redshift measurements disagree significantly (SN3199; SN13956; SN15301; SN19757).  Based on visual inspection of the SDSS-III DR8 catalogue we identify and remove SN3199 from our sample, as the BOSS--targeted galaxy does not appear to be the most likely SN host galaxy. The remaining three SNe are retained in our sample as the identified host galaxy appears to be correct and they reside close to the Hubble diagram (small HRs) when using the BOSS host--galaxy redshifts rather than the SN--spectrum redshift.  Apart from this check, the purpose of which is to determine the likelihood of host mis-identification, we make no other use of SN spectroscopy in this paper.

We have also compared our BOSS host-galaxy redshifts with SDSS galaxy redshifts, where the latter are available.  We found one SNe~Ia (SN6491) hosted in a galaxy where the redshifts from these two surveys disagree, and removed this object from our sample.

Additionally, we have visually inspected the host galaxies of other photometrically--classified SNe~Ia candidates in our sample. First, we inspected a random subset of 70 photometric SNe~Ia, finding no obvious misidentification of the appropriate host. This is reassuring, as it confirms that the rate of misidentification of hosts must be low.  Next, we inspected the host association for SNe~Ia candidates that are clearly offset from the Hubble diagram in Figure \ref{hubble_diagram} to ensure the correct galaxy had been assigned during targeting. We found only one host galaxy that was likely to be incorrect (SN9052), being located 25 arcseconds from our photometrically--classified SN, and removed this object from our sample. In total we removed only two SNe~Ia from our sample because of likely host--galaxy mismatches.

\section{D. Other Systematic Uncertainties}
\label{Appendix_OtherSys}

There are a number of other systematic uncertainties that are likely to affect our SDSS-II SN sample beyond the photometric--classification specific uncertainties discussed in Section~\ref{subsec:phot_sys}. As outlined in \citet{kessler:2009a} and \citet{conley:2011a}, present SN samples have major uncertainties associated with their photometric calibration and the light--curve fitting technique used, as well as many astrophysical uncertainties such as correlations with host--galaxy properties, SN lensing, peculiar velocities, galactic dust, and SN evolution.  Though we do not address these important systematics in detail in this paper, we discuss the likely effect of these additional systematic uncertainties on our results. 

The optimal method of light--curve fitting is not known; there can be significant differences in the cosmological results obtained from using different algorithms \citep[e.g., SALT2 and MLCS2k2;][]{kessler:2009a, sollerman:2009a}.  Both \citet{Guy:2010} and \citet{conley:2011a} find consistent cosmological results between SALT2 and a different technique (SiFTO), with a possible systematic uncertainty of only $\Delta\mu\simeq0.02-0.03$ magnitudes between light--curve fitters. We have chosen to use only the SALT2 light--curve fitting algorithm \citep{Guy:2010} in this work.  

As discussed in \citet{conley:2011a} for SNLS, the most important systematic uncertainty in present SN surveys is the photometric calibration.  Therefore, \citet{conley:2011a} recommended that future SN surveys should be calibrated onto a {\it ``more modern, better understood photometric systems such as USNO/SDSS"}.  By using photometric data obtained wholly from the SDSS-II SN Survey, we believe we have minimized calibration uncertainties, as the SDSS photometric system is now mature and well--understood \citep{Ivezic:2007,Doi:2010}.  \citet{Mosher:2012} has recently compared the SDSS-II photometric system to that of the Carnegie Supernova Project (CSP) using light--curve data for nine SNe~Ia observed concurrently by the two projects.  They conclude that measurements from the two surveys agree in all bands at or better than 2\% in flux, and are consistent with no difference in $g-$ and $r-$band magnitude at the $2\sigma$ level. This is an indication of the relative calibration between the SDSS and other surveys (CSP in this case) and not a direct statement on the absolute calibration, which would require observation of a known source like NIST photodiodes, as discussed by \citet{stubbs:2012a}. Such techniques will be implemented in future surveys (e.g., DES). Based on these findings, we assume systematic uncertainty of only $\Delta\mu\approx0.02$ mags on the distance modulus of our SNe, relative to other surveys. 

The next largest systematic uncertainty at present is the recently discovered correlation between the (corrected) peak absolute magnitude of SNe~Ia and the stellar mass of the host galaxy \citep{kelly:2010, sullivan:2010a, lampeitl:2010b}. This relationship has been observed to have an effect as large as $\Delta\mu\simeq0.07$ mags in the SDSS-II SN sample \citep{lampeitl:2010b}. We do not attempt to make this correction here, as it is beyond the scope of this paper, but we note it is being investigated in Smith et al. (in prep.). There will likely be significant stellar population modeling uncertainties associated with determining reliable stellar masses of our faint host galaxies; the large magnitude errors on the photometric colors of our BOSS galaxies make this especially difficult at $z>0.3$. We will revisit the topic of SN~Ia correlations with host-galaxy properties using our photometrically--classified sample in a future paper.

There are a number of other astrophysical uncertainties that could be considered, especially SN lensing effects, peculiar velocities and possible SN evolution. These are all smaller in size compared to the systematic uncertainties discussed above, and should be further mitigated in our sample because of the lack of relatively high-redshift SNe~Ia where these effects are most prominent (especially lensing and evolution). 

\begin{figure}[!t]
\begin{center}
\epsfig{file=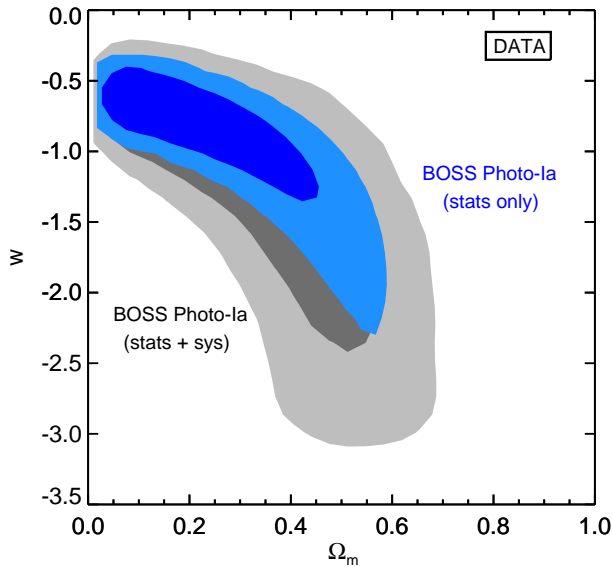,width=1.0 \linewidth}
\caption{\label{contour_cosmo_sys} The 68$\%$ and 95$\%$ confidence contours of $w$ and $\Omega_{\rm m}$ in a flat $w$CDM model. We assume a Gaussian prior on $H_0$; other priors are summarised in Table \ref{data_priors_cosmo}, using prior Set III. The blue contours show the statistics only contours and the grey contours include the estimate of the systematics, as well as statistical uncertainties.}
\end{center}
\end{figure}

In Figure~\ref{contour_cosmo_sys} we show an estimate of the effect of systematic uncertainties on our results, compared to Figure~\ref{contour_cosmo} in Section~\ref{subsec:SN+H0} which includes only statistical errors. These measurements are obtained by adding in quadrature an additional uncertainty of $\Delta\mu=0.1$ magnitudes to the distance moduli errors of our observed SNe~Ia; the same methodology as \citet{kowalski:2008a}.  This level of uncertainty is an estimate from the combination (in quadrature) of uncertainties associated with the light--curve fitter (0.03 mags), photometric calibration (0.02 mags), lensing and peculiar velocities (0.05 mags) and possible host galaxy correlations (0.07 mags).  When including our systematic errors in the cosmological analysis we fix the SALT2 SN parameters $\alpha$ and $\beta$ to the values found in Section~\ref{subsec:cosmo_anal}; this is done to prevent $\alpha$ and $\beta$ changing values to counteract the increase in dispersion that the systematic errors cause. The results do not change significantly with our inclusion of these estimates of the unknown systematics, giving a best-fit of $w=-0.93_{-0.49}^{+0.39}$ and $\Omega_m=0.314_{-0.06}^{+0.07}$. However, we stress again that this analysis is not comprehensive.

\newpage

\section{E. Photometrically--classified SNe~Ia data}
\label{Appendix_data}

In Table~\ref{DATA_table} we present the key information used in this paper for our sample of \nofinal\ photometrically--classified SNe~Ia. The full SDSS-II SN sample, including all the light-curve data, redshifts and classifications for all transients, will be published in Sako et al. (in prep.).  Table~\ref{DATA_table} includes a unique identification number for the whole SDSS-II SN Survey (CID; column 1), BOSS host--galaxy redshift and error (columns 2 \& 3), the RA and DEC (in degrees) of the SN event (columns 4 \& 5) and its host (columns 6 \& 7), a unique SDSS object identifier for the host galaxy from DR8 (column 8), the SALT2 parameters $X_0$ in flux units, $X_1$ and color both in magnitudes (columns 9, 10, 11), and finally the uncorrected and corrected distance modulus (columns 12 \& 13) with error (column 14), all in magnitudes.

The data in Table~\ref{DATA_table}, along with the SALT2 covariance matrices and the SN type probabilities, can be electronically downloaded from {\tt http://www.icg.port.ac.uk/$\sim$campbelh}. The probabilities listed there are those used in this paper (with the BOSS host galaxy redshift prior) and include the probability assigned to each object of being a SNe~Ia ($P_{Ia}$), a Type~II~SN ($P_{II}$), and a Type~Ibc ($P_{Ibc}$).

\begin{center}          
 
  \begin{sidewaystable*}[!tp] \centering

\vspace*{9.2cm}
\begin{tabular}{|c|c|c|c|c|c|c|c|c|c|c|c|c|c|}
\hline
CID & z  & z err & \multicolumn{2}{|c|}{SNe~Ia} & \multicolumn{2}{|c|}{Host galaxy} & Host objID & \multicolumn{3}{|c|}{SALT2}   & \multicolumn{3}{|c|}{$\mu$} \\
\cline{4-7} 
\cline{9-14}
 & & ($10^{-5}$) & RA & DEC & RA & DEC & (DR8) & $X_0$ ($10^{-5}$) & $X_1$ & color & uncorrected & corrected & err \\
  \hline
       703  &  0.2980   &     2.03   &  -23.7820   &   0.6508   &  -23.7821   &    0.6507   &  1237663544222483004   &    4.32   &   0.63   &  -0.01   &  40.69   &  40.72   &   0.14   \\  
 762  &  0.1914   &     2.41   &   15.5361   &  -0.8797   &   15.5354   &   -0.8790   &  1237666338114765068   &    9.93   &   1.15   &  -0.01   &  39.90   &  39.91   &   0.08   \\  

  779  &  0.2381   &     2.13   &   26.6738   &  -1.0207   &   26.6737   &   -1.0206   &  1237657069548208337   &    6.11   &   0.38   &   0.01   &  40.17   &  40.19   &   0.10   \\  
   822  &  0.2376   &     14.3   &   40.5608   &  -0.8622   &   40.5608   &   -0.8622   &  1237657584950379049   &    5.34   &  -0.52   &  -0.07   &  40.39   &  40.41   &   0.13   \\  
     859  &  0.2783   &     1.93   &   -9.4480   &   0.3867   &   -9.4483   &    0.3866   &  1237666408438301119   &    5.19   &   0.56   &   0.02   &  40.37   &  40.40   &   0.12   \\  
     893  &  0.1101   &     2.62   &    5.4942   &  -0.1317   &    5.4942   &   -0.1317   &  1237657190907641944   &    6.03   &  -1.05   &   0.11   &  39.59   &  39.60   &   0.11   \\  
    1112  &  0.2576   &     2.67   &  -20.9830   &  -0.3749   &  -20.9825   &   -0.3752   &  1237663478724428434   &    4.40   &  -0.41   &  -0.03   &  40.49   &  40.52   &   0.14   \\  

   1119  &  0.2978   &     1.16   &  -39.5870   &   0.8952   &  -39.5865   &    0.8946   &  1237663458851619714   &    4.17   &   0.55   &  -0.14   &  41.10   &  41.13   &   0.22   \\  

    1166  &  0.3821   &     9.98   &    9.3553   &   0.9740   &    9.3556   &    0.9733   &  1237663716555293384   &    2.74   &   1.36   &   0.00   &  41.29   &  41.35   &   0.20   \\  
    1241  &  0.0898   &     1.54   &  -22.3290   &  -0.7762   &  -22.3274   &   -0.7766   &  1237656567586226517   &   41.46   &  -0.55   &   0.04   &  37.80   &  37.81   &   0.06   \\  
     \hline
      \end{tabular}      
  
 \caption{Table containing data for 10 of the \nofinal\ photometrically--classified SNe~Ia presented in this paper. For the full table, which includes errors and covariance on the SALT2 parameters, see the electronic table at http://www.icg.port.ac.uk/stable/campbelh/SDSS\_Photometric\_SNe\_Ia.fits. The 15 SNe with entries of * in the electronic table are ones where there is no photometric object ID for the host galaxy in DR8; these galaxies do appear in the co-added images, and from this catalog we quote the HostID.}
        \label{DATA_table}
\end{sidewaystable*}
      \end{center}  

\bibliographystyle{apj}

\end{document}